\documentclass[twocolumn]{aastex63}
\usepackage{natbib}

\newcommand{\msun}{{M_{\odot}}}
\newcommand{\mstar}{{M_{\ast}}}
\newcommand{\ser}{S\'ersic }

\shorttitle{Galaxy sizes since $z=2$}
\shortauthors{Mosleh et al.}

\begin{document}

\title{Galaxy Sizes Since $z=2$ from the Perspective of Stellar Mass Distribution within Galaxies}

\correspondingauthor{Moein Mosleh}
\email{moein.mosleh@shirazu.ac.ir}

\author[0000-0002-4111-2266]{Moein Mosleh}
\affiliation{Biruni Observatory, School of Science, Shiraz University, Shiraz 71454, Iran}
\affiliation{Department of Physics, School of Science, Shiraz University, Shiraz 71454, Iran}

\author{Shiva Hosseinnejad}
\affiliation{Biruni Observatory, School of Science, Shiraz University, Shiraz 71454, Iran}

\author{S. Zahra Hosseini-ShahiSavandi}
\affiliation{Biruni Observatory, School of Science, Shiraz University, Shiraz 71454, Iran}

\author[0000-0002-8224-4505]{Sandro Tacchella}
\affiliation{Center for Astrophysics $\vert$ Harvard \& Smithsonian, 60 Garden St, Cambridge, MA 02138, USA}

\begin{abstract}

How stellar mass assembles within galaxies is still an open question. We present measurements of the stellar mass distribution on kpc-scale for $\sim5500$ galaxies with stellar masses above $\log(\mstar/\msun)\geqslant9.8$ up to the redshift $2.0$. We create stellar mass maps from Hubble Space Telescope observations by means of the pixel-by-pixel SED fitting method. These maps are used to derive radii encompassing 20\%, 50\%, and 80\% ($r_{20}$, $r_{50}$ and $r_{80}$) of the total stellar mass from the best-fit \ser models. The reliability and limitations of the structural parameter measurements are checked extensively using a large sample ($\sim3000$) of simulated galaxies. The size-mass relations and redshift evolution of $r_{20}$, $r_{50}$ and $r_{80}$ are explored for star-forming and quiescent galaxies. At fixed mass, the star-forming galaxies do not show significant changes in their $r_{20}$, $r_{50}$ and $r_{80}$ sizes, indicating self-similar growth. Only above the pivot stellar mass of $\log(\mstar/\msun)\simeq10.5$, $r_{80}$ evolves as $r_{80}\propto(1+z)^{-0.85\pm0.20}$, indicating that mass builds up in the outskirts of these systems (inside-out growth). The \ser values also increase for the massive star-forming galaxies towards late cosmic time. Massive quiescent galaxies show stronger size evolution at all radii, in particular the $r_{20}$ sizes. For these massive galaxies, \ser values remain almost constant since at least $z\sim1.3$, indicating that the strong size evolution is related to the changes in the outer parts of these galaxies. We make all the structural parameters publicly available. 
\end{abstract}


\keywords{galaxies: evolution -- galaxies: structural -- galaxies: star formation}
 

\section{Introduction}
\label{introduction}

An open challenge in the galaxy evolution is to understand how stellar mass assembles within galaxies. The stellar mass distribution of a galaxy contains a wealth of information on its past evolution and it is believed to be able to constrain physical processes that operate on spatially resolved scales. In this paper, we present detailed measurements of the stellar mass distribution on kpc-scale for a large sample of galaxies over the past 10 billion years (since redshift $z\approx2$), focusing specifically on evolution of galaxy sizes.

A wide range of physical mechanisms is believe to shape both star-forming and quiescent galaxies. While galaxies are star-forming, the gas and the resulting star-formation distribution are mostly driving the mass growth. This is particularly the case at early cosmic times, when the star-formation rates (SFRs) are high relative the stellar masses ($M_{\star}$), i.e. for galaxies with high specific SFRs (sSFR). A variety of processes can lead to rapid gas inflow to the central region \citep[summarily called ``gas compaction'';][]{Dekel2014, Zolotov2015, tacchella2016}, including disk instabilities, mergers, and migration of gas clumps \citep{Hernquist1989, Noguchi1999, dekel2009, Bournaud2011, Sales2012, Wellons2015}. At later cosmic times, when gas fractions and sSFRs are lower, secular processes related to spiral arms and bars (including stellar migration) may contribute to the spatially resolved stellar mass growth \citep[see, e.g., a review by][]{kormendy2004}. When galaxies are quiescent, gas-poor mergers and perturbations such as tidal interactions from neighboring galaxies are important in adding and re-distributing stellar mass \citep{naab2009, Bekki2011, oser2012}.

Different physical mechanisms have distinct imprints on the mass distribution and, hence, the morphological indicators. For example, gas-poor minor merger preferentially lead to mass growth in the outskirts, leading to an increase in the half-mass radius $r_{50}$ \citep{naab2009, bezanson2009}. On the other hand, gas compaction leads to mass growth in the center, decreasing $r_{50}$ and increasing the stellar mass density within the central kpc ($\Sigma_1$; \citealt{Bournaud2007, elmegreen2008, dekel2009, tacchella2016}). These and other processes lead to a large diversity of galaxies today with a structural dichotomy between star-forming and quiescent galaxies: star-forming galaxies are larger, where as quiescent galaxies have a more prominent bulge component \citep{cheung2012, fang2013, Bluck2014, Huertas2016, whitaker2017}. Nevertheless, it is not clear how much (or if any at all) morphological transformation takes place when galaxies cease their star formation \citep{tacchella2019}: today's quiescent galaxies were star forming in the past when gas fractions were much higher (relative to today's star-forming galaxies), leading to higher efficiency of bulge formation. 

\begin{figure*}
\centering
\includegraphics[width=\textwidth]{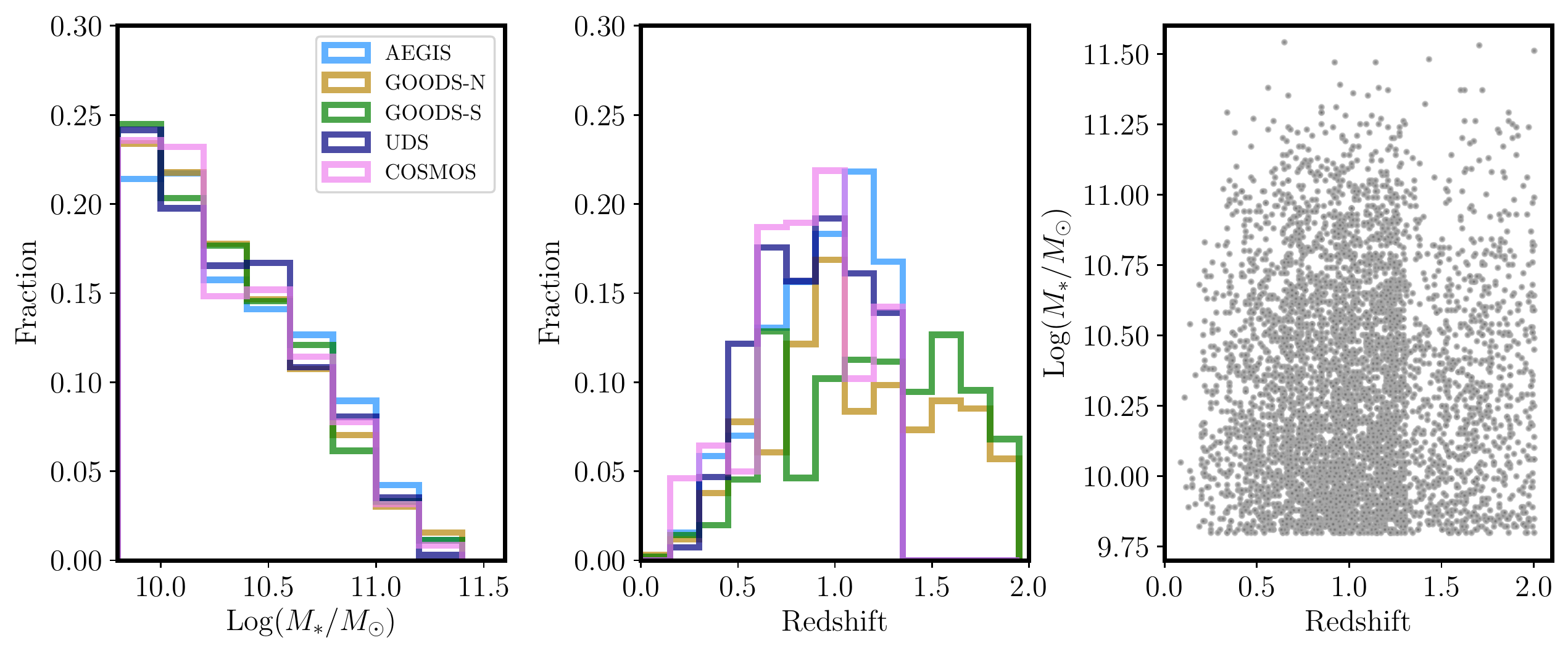}
\caption{The stellar mass (left panel) and redshift distribution (middle panel) of galaxies in our sample, color coded according to the different CANDELS fields. The stellar mass - redshift distribution of our sample is shown in the right panel. The stellar mass limit of our sample is $\log(M_{\star}/M_{\odot})>9.8$. The redshift limit is $z=1.3$ for the three fields of UDS, COSMOS, and AEGIS, and $z=2$ for GOODS-S and GOODS-N. These cuts in stellar mass and redshift are motivated by mock simulations (see Appendix \ref{sec:appendixB}): for galaxies in our sample, we are able to accurately infer the stellar mass density profiles.} 
\label{fig1}
\end{figure*}

For better understanding and constraining the physical processes involving the total mass growth and mass (re-)distribution within galaxies, studying their stellar mass profiles is essential. Normally, difficulties on estimating the mass profiles lead to use of light profiles in most studies. However, it is known that the age, dust and metallicity of the underlying stellar populations and the star-formation history (SFH) within galaxies varies from the central regions to the outskirts, and hence, causing mass-to-light ratio ($M/L$) variations (or gradients) as a function of radius \citep[e.g.,][]{franx1989, peletier1990, labarbera2005}. The consequence of this is the wavelength dependence of the structural parameters \citep{kennedy2015} and introducing differences between morphological parameters obtained from the light and stellar mass distributions \citep{szomoru2013, fang2013, tacchella2015b}. Comparing observed mass profiles with simulations is also more straightforward than the light profiles. Hence, using mass profiles are more desired and robust for imposing constrains on the physical processes and comparisons. 

How do we derive the stellar mass distribution within galaxies? Deriving two-dimensional (2D) stellar mass maps from integral field unit (IFU) spectroscopy is favorable \citep{bacon2001}. However, IFU data for a large sample of galaxies (in particular at high redshifts) is still expensive \citep{bacon2017}. The sensitivity of this technique to the bright regions also limits observation to the central regions within galaxies, even if samples would be large \citep{sanchez2012, croom2012, bundy2015}. A simple solution to this is to make use of photometric, multi-wavelength observations. 

Different approaches have been adopted by authors for converting multi-wavelength observations to 2D stellar mass maps and 1D stellar mass profiles. In majority of these studies, the stellar mass profiles are obtained from 1D light profiles (observed or point spread function corrected). These light-based profiles are then converted to mass profiles either with a constant or a radially varying $M/L$ correction, where the latter usually assumes a simple color-$M/L$ relation or is based spectra energy distribution (SED) fitting \citep{dokkum2010, patel2013a, szomoru2013, fang2013, morishita2015, tacchella2015a, Barro2017, mosleh2017, mosleh2018, suess2019a}. The second approach for deriving stellar mass profiles is to build 2D stellar mass maps by means of pixel-by-pixel SED fitting technique. This technique was first introduced by \cite{Abraham1999} and \cite{conti2003} and has been utilized for several purposes including star-formation rate profiles, color gradients and testing total stellar masses, galaxy mergers etc. \citep{Lanyon2007, zibetti2009, Lanyon2012, hemmati2014, cibinel2015, sorba2015, Abdurrouf 2018,  cibinel2019}. The structural analysis (parametric and non-parametric) based on the 2D stellar mass maps are studied by \cite{wuyts2012, lang2014, cibinel2015, chan2016, morselli2019}.

Beside these two different approaches, technical details within these approaches are diverse in the literature as well. In particular, the reliability of these methods over the range of stellar masses and redshifts are not fully examined. Beside these issues, the number of published catalogs on the structural parameters based on the stellar mass profiles are sparse. Recently, \cite{morselli2019} published a catalog for a limited ($\sim 700$) sample of sources within $0.2 < z < 1.2$ and \citet{suess2019a} presented measurements of \ser parameters (half-mass radii and $n$) for a sample of $\sim 7000$ galaxies at $1.0 \leqslant z \leqslant 2.5$ with stellar masses of $ 9.0 \leqslant\log(\mstar/\msun) \leqslant 11.5$ for three high redshift CANDELS fields. Hence, consistent measurement of the stellar mass based sizes at $0.3\leqslant z \leqslant 2.0$ for all CANDELS fields are required. Moreover, adopting a simple methodology to avoid many prior assumptions about the shape of light or $M/L$ profiles \citep[][]{mosleh2017, suess2019a} for estimating the stellar mass density profiles can help to understand uncertainties in the final results.

Therefore, in this work, we create stellar mass maps for a large sample ($\sim 5500$) of galaxies with $\log(\mstar/\msun) \geqslant 9.8$ at $z \leqslant 2$ to cover a wider range in redshift and stellar mass, as described in Section 3. The structural parameters (sizes containing $20\%$, $50\%$, and $80\%$ of the stellar mass -- $r_{20}, r_{50}$, and $r_{80}$) are then estimated based on the 1D and 2D stellar mass distributions (Section 4 and Appendix \ref{sec:appendixA}). We extensively test our methodology by using a large sample of mock galaxies (see Appendix \ref{sec:appendixB}). We present the size-mass relations ($r_{20}, r_{50}, r_{80}$ - stellar mass) for star-forming and quiescent galaxies in Section 5. The final results and their interpretations are discussed in Section 6. For a consistency with recent works, we choose the following cosmological parameters throughout this paper: $\Omega_{m}=0.3$, $\Omega_{\Lambda}=0.7$ and $H_{0}=70~\mathrm{km}\/\mathrm{s}^{-1}\/\mathrm{Mpc}^{-1}$.

\begin{figure}
\centering
\includegraphics[width=0.48\textwidth]{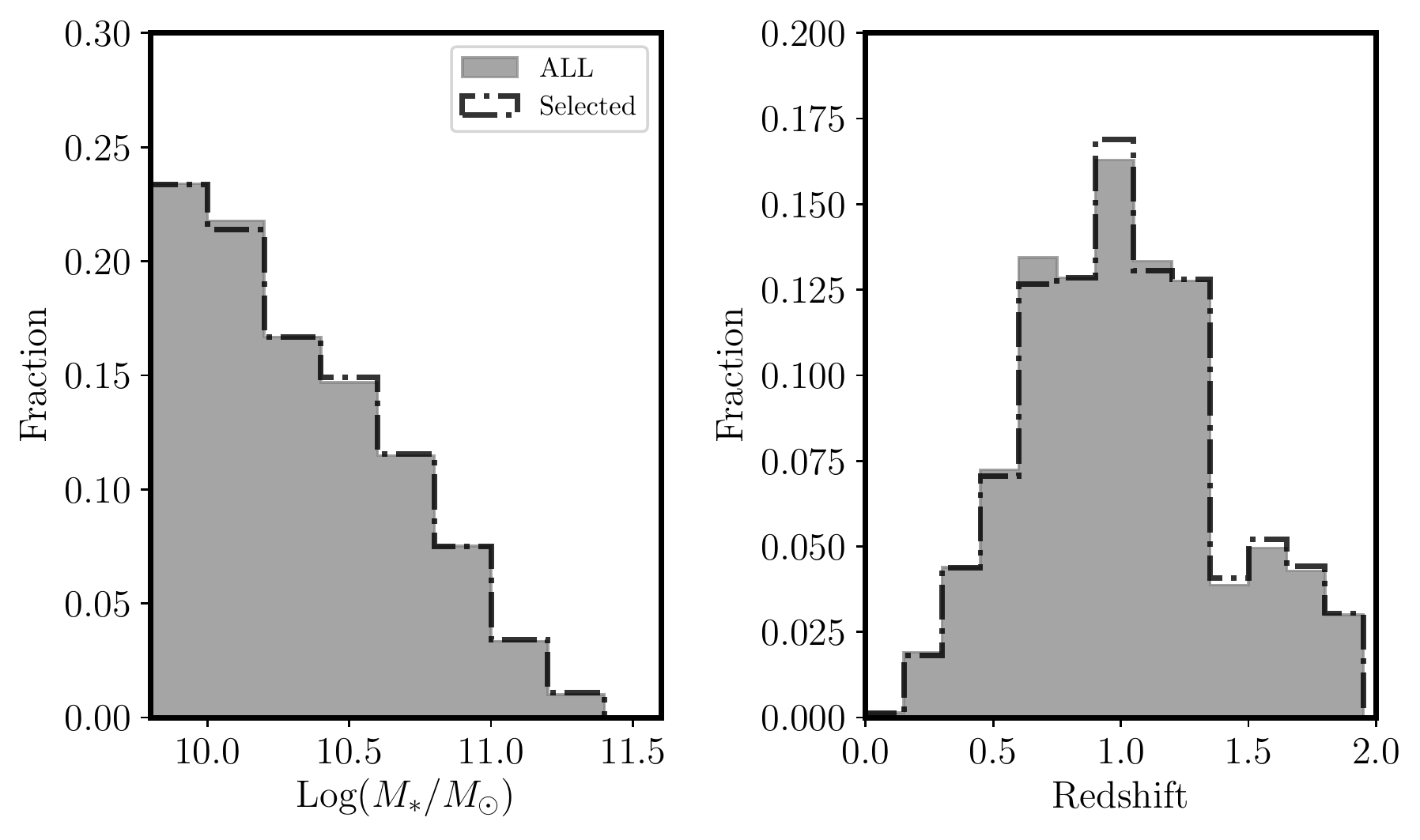}
\caption{Comparing the histograms of the stellar mass and redshift of the original sample (shaded gray regions) to the ones after excluding of those with less than 3 filter coverage (dashed-dotted lines). The sample distribution remains almost intact after applying this criteria, which removes only 10\% of the original sample (see text in Section 2 for more details).} 
\label{fig2}
\end{figure}

\section{DATA \& SAMPLE}
\label{data}

The sample of galaxies used for this study is based on the publicly available catalogs and imaging data of the 3D-HST Treasury Program \citep{brammer2012, skelton2014} and the Cosmic Assembly Near-IR Deep Extragalactic Legacy Survey \citep[CANDELS;][]{grogin2011, koekemoer2011}. We make us of all five fields (GOODS-South, GOODS-North, COSMOS, UDS, and AEGIS). The total area of these fields is about 900 arcmin$^2$. Using different fields helps to mitigate cosmic variance effects. The photometry of the sources from all 3D-HST observations and all publicly available data over a wide range of wavelengths (0.3-8 $\mu$m) are provided in the catalogs. Using these ancillary data, the stellar masses and photometric redshifts (if no spectroscopic or grism redshift is available) are determined with the $EAZY$ \citep{brammer2008} and FAST \citep{kriek2009} codes, respectively. 

For this study, the point spread function (PSF)-matched mosaic images of these fields, available on the 3D-HST website\footnote[1]{\url{http://3dhst.research.yale.edu/Home.html}}, are used. These mosaics are available in seven filters ($B_{435}$,$V_{606}$, $i_{775}$, $z_{850}$, $J_{125}$, $JH_{140}$, $H_{160}$) for GOODS-South and GOODS-North and in five filters ($V_{606}$, $z_{814}$, $J_{125}$, $JH_{140}$, $H_{160}$) for COSMOS, UDS, and AEGIS. 

For our analysis, we select galaxies with the following criteria:
\begin{itemize}
\item \texttt{use\_phot} $=1$ and \texttt{flags} $\leqslant 2$ (photometric quality and not-blended flags in the 3D-HST catalog);
\item $ \log(\mstar/\msun) \geqslant 9.8$;
\item $ z \leqslant 2$ for sources in GOODS-South \& GOODS-North;
\item $ z \leqslant 1.3$ for COSMOS, UDS \& AEGIS.
\end{itemize}

The reasons for the chosen stellar mass and redshift ranges are justified in the following sections. Briefly, these cuts are based on our simulations (Appendix \ref{sec:appendixB}) that the results (i.e. the half-mass size measurements) presented in this work are less reliable at $z\geqslant 1.3$ ($z\geqslant 2.0$) and $\log(\mstar/\msun) \leqslant 9.8$ for COSMOS, UDGS and AEGIS (GOODS-South and GOODS-North). With the above criteria, the total number of all selected objects in all CANDELS fields is 5557. However, the stellar masses are more reliable for those that have sufficient wavelength coverage; hence, we imposed an additional constraint for being detected in at least 3 HST filters. The cost of these constraints is to exclude about 10\% of the sources, which mainly lie in COSMOS and UDS. In addition, we fail to determine the stellar mass maps in $\sim 2$\% of the sample, usually due to their proximity to the edges of the mosaic images or contamination of very bright stars. These leave us with a final sample of 4887 galaxies.  The histograms in Figure \ref{fig1} show the distributions of the stellar masses and redshifts (left and middle panels, respectively) for objects in this study. As shown in the histograms of Figure \ref{fig2}, excluding sources with less sufficient number of filters does not affect the general stellar mass and redshift distributions of the sample. The right panel of Figure \ref{fig1} also illustrates the distribution of the stellar mass of the sample as a function of redshift. This shows that the sample is complete down to the redshift of $\sim 0.3$. Hence, for the results of this paper, we set this redshift as our lower limit.  

The star-forming and quiescent galaxies in our sample are selected based on their location in the $UVJ$ color-color diagram \citep{williams2009}. We use the same criteria as given by Equations 1 to 3 in \cite{mosleh2017} to separate galaxies on the $U-V$ versus $V-J$ rest-frame color diagrams. Our sample consists of 3524 and 1363 star-forming and quiescent galaxies, respectively.

\begin{figure*}
\centering
\includegraphics[width=\textwidth]{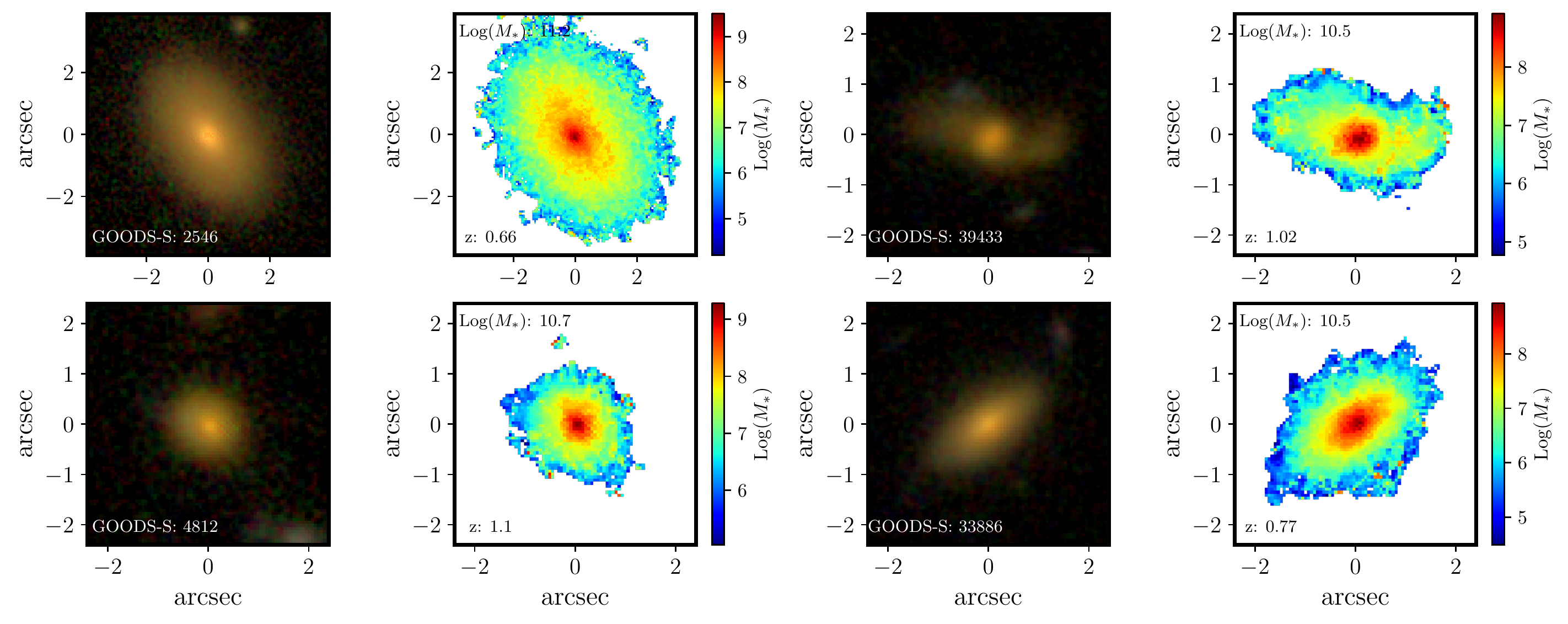}
\centering
\includegraphics[width=\textwidth]{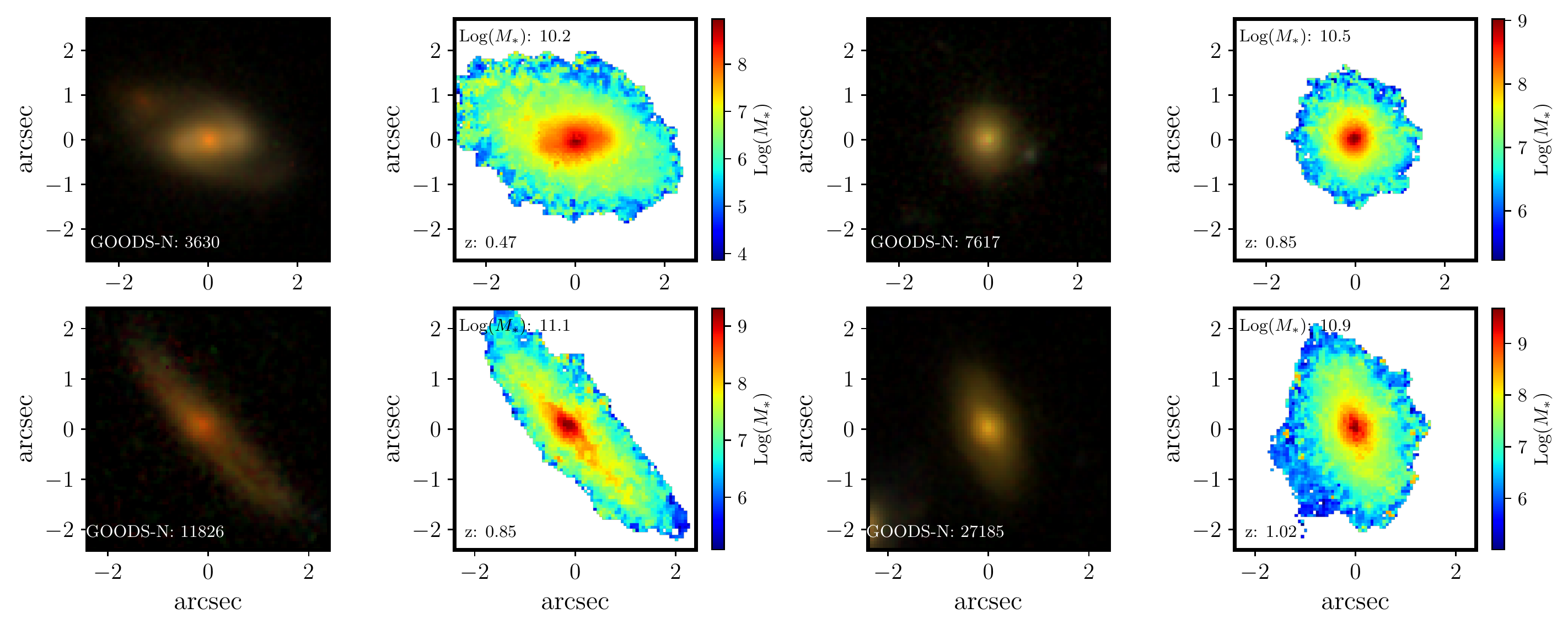}
\centering
\includegraphics[width=\textwidth]{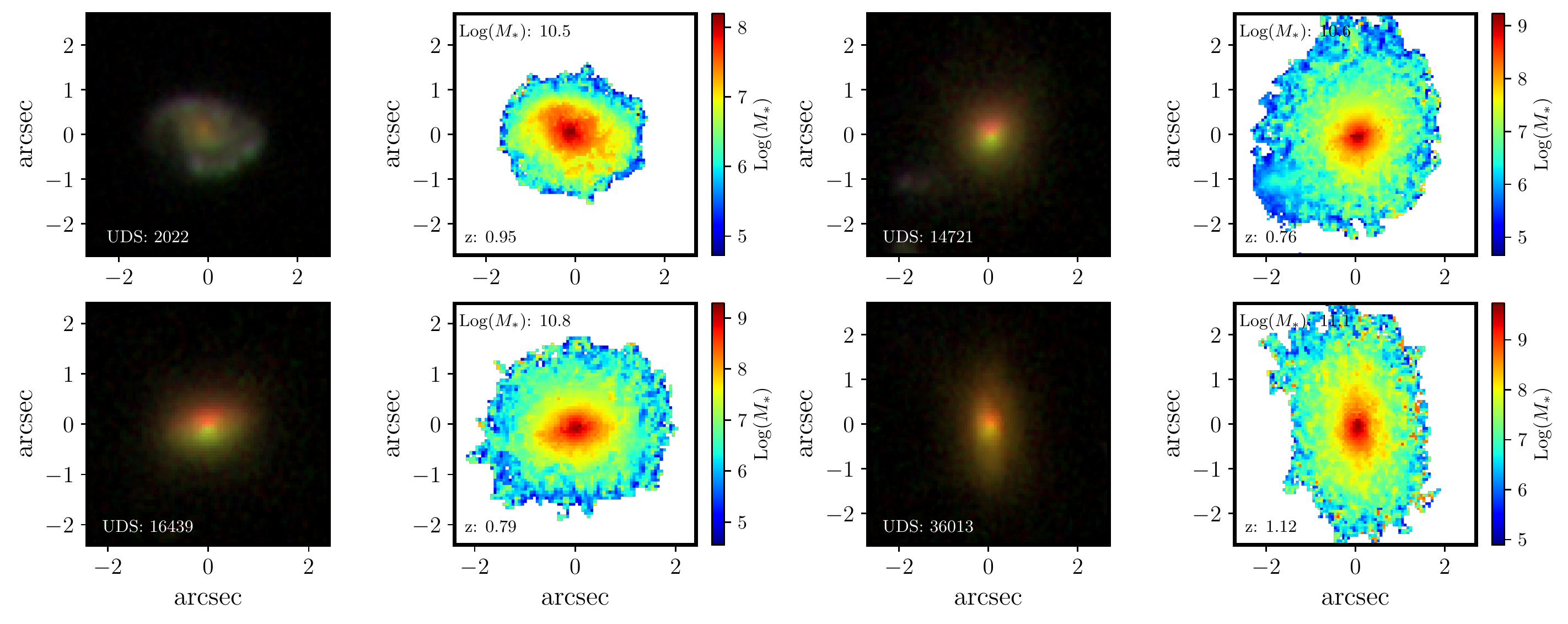}
\caption{The color ($IJH$) images and the estimated stellar mass maps of a few example galaxies from different fields are shown. The stellar mass maps are derived using the pixel-by-pixel SED fitting method (Section 3). The stellar mass maps are relatively smooth for different types of galaxies. The ID, redshift ($z$) and stellar mass of these galaxies are from the 3D-HST catalog.} 
\label{fig3}
\end{figure*}

\section{Creating Resolved Mass Maps}
\subsection{2D Stellar Mass Maps}

We create spatially resolved stellar mass maps by SED fitting each individual pixel (pixel-by-pixel method). We use the PSF-matched mosaic images to identify the same physical region of each galaxy in the different filters. For each galaxy, we create postage stamps of $48''\times 48''$. The segmentation (mask) maps of galaxies, provided by the 3D-HST team, are used to select pixels belong to each object. Fluxes of each pixel in all available filters are extracted separately and their associated flux errors are determined using empty the aperture method \citep{skelton2014} of the noise-equalized regions around each object.

The resolved stellar mass maps are derived by finding the best-fit SED model for each pixel. We use iSEDfit, a Bayesian code \citep{moustakas2013}, to perform the SED fitting. The full grid of 100,000 models are created based on the \citet{BC2003} stellar population evolution models with ages between 0.1 and 13.5 Gyr. The star-fromation history for these models are assumed to be exponentially declining ($\mathrm{SFR} \propto \exp(-t/\tau)$, with the $e$-folding timescale $\tau$ between $0.01-1.0$ Gyr) and the \cite{chabrier2003} initial mass function (IMF) is adopted. We set the metallicity range to 0.004-0.03 and assume the \citet{calzetti2000} dust attenuation law. For each object, the redshift of all pixels are set to the redshift of the galaxies from the 3D-HST catalog. In Figure \ref{fig3}, color images and the corresponding stellar mass maps of a few example galaxies are shown. The stellar mass maps are relatively smooth compared to the observed images due to the contribution of the old stellar population to the total stellar mass, consistent with earlier findings \citep{wuyts2012, lang2014, tacchella2015b, sorba2015}. We should note that to reduce the effects of mass loss\footnote{There are some scientific questions, in particular related to the evolution of quiescent galaxies, where it is more useful to adopt the integral of the star-formation history as the stellar mass, since with this definition the stellar mass of a quiescent galaxy remains constant with time; see \cite{carollo2013a, fagioli2016, tacchella2017}.}, the stellar-mass maps in this study are based on the total masses which is derived from the integrated star-formation history.

\subsection{Robustness of the Stellar Masses}

\begin{figure*}
\centering
\includegraphics[width=0.48\textwidth]{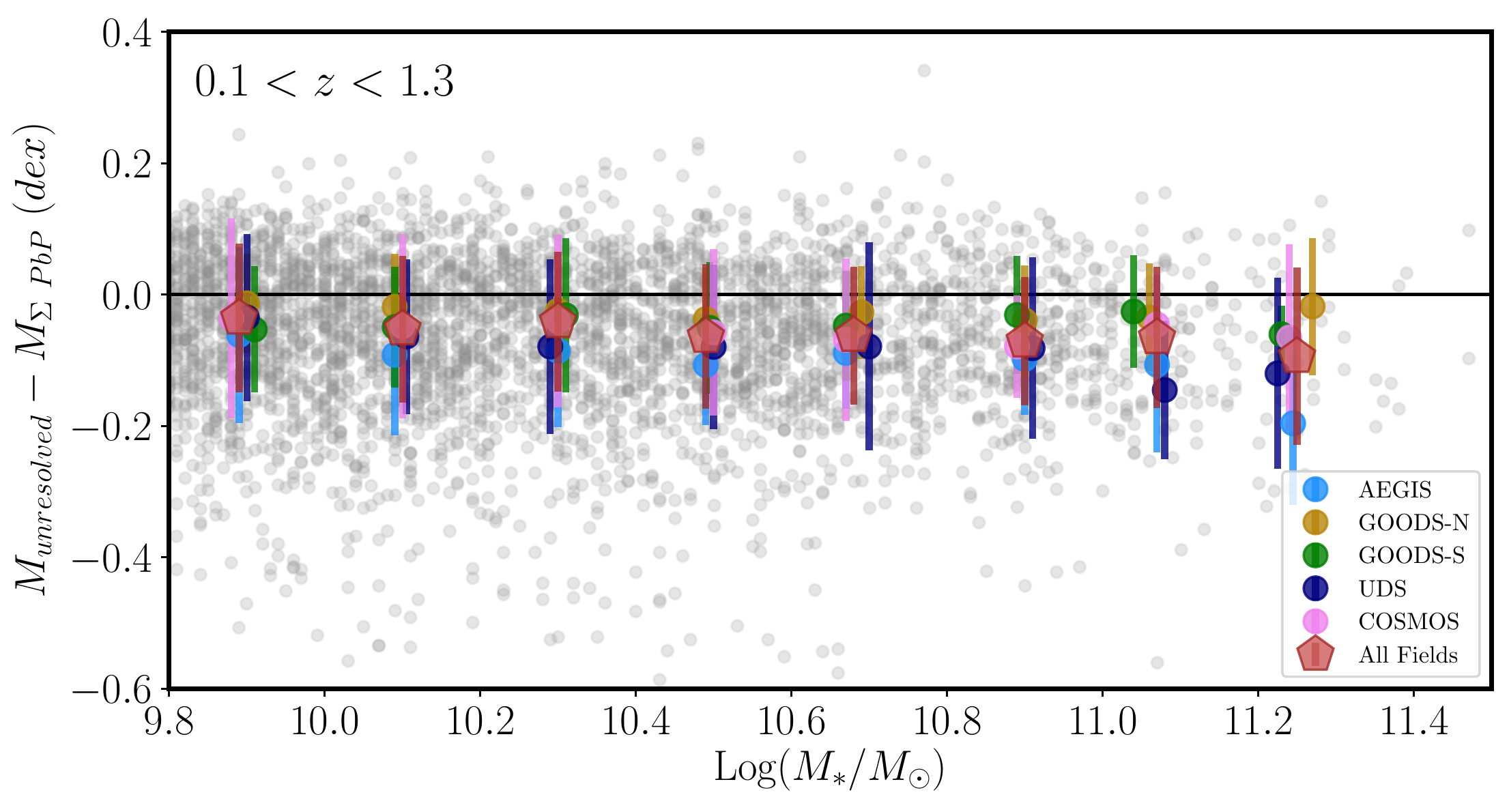}
\hspace{2. mm}
\centering
\includegraphics[width=0.48\textwidth]{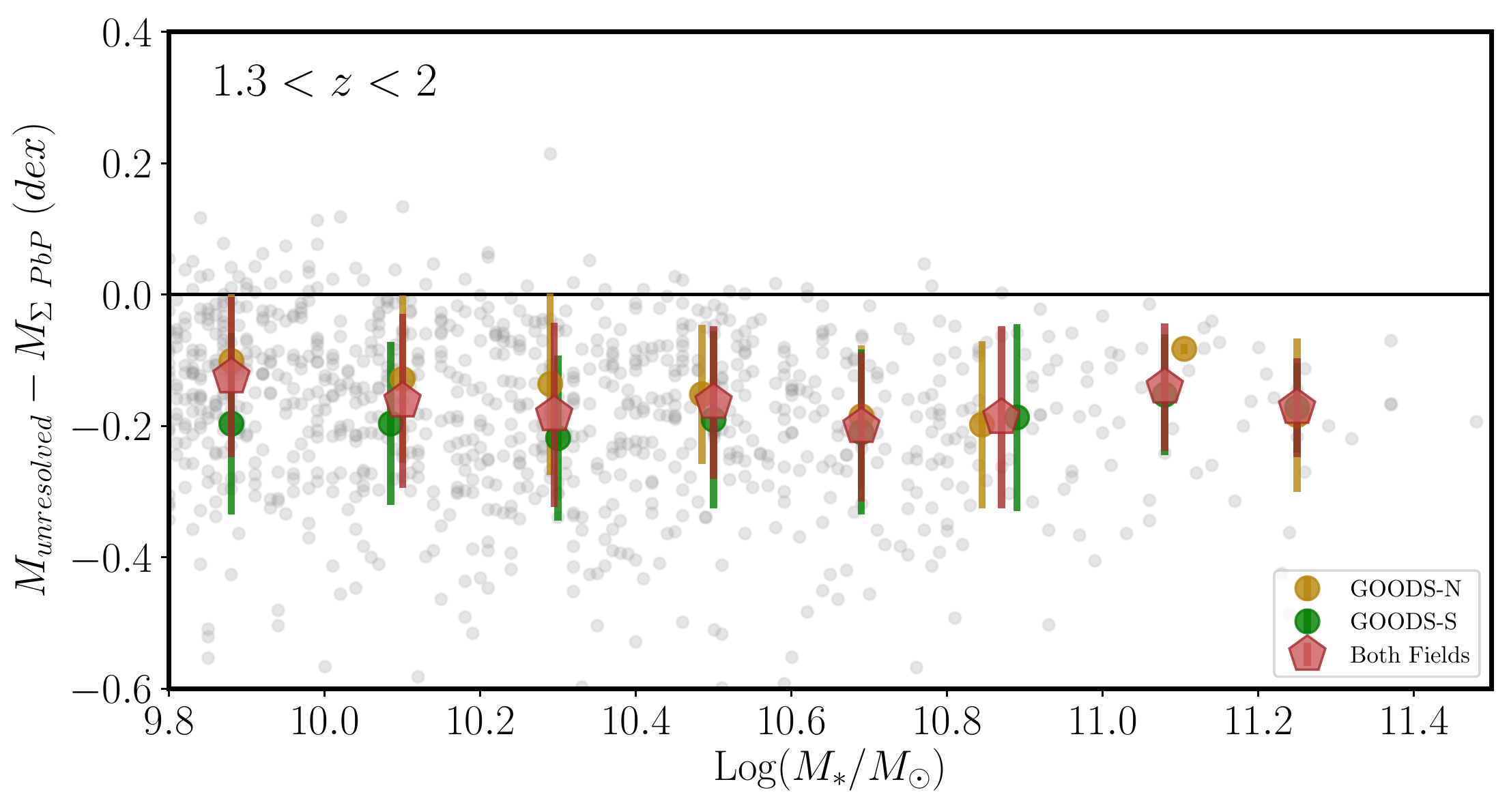}
\caption{The stellar mass differences between resolved and unresolved stellar mass for objects within $0.3<z<1.3$ and $1.3<z<2.$ (left and right panels, respectively). There is a weak systematic difference ($\sim 0.06$ dex) between resolved and unresolved for the lowest redshift range. This difference increases ($\sim 0.17$ dex) for the higher redshift range. The unresolved stellar masses normally smaller compared to the resolved ones. This is independent of field and the number of filters used.}
\label{fig4}
\end{figure*}

It is important to verify the robustness the stellar mass maps. For this purpose, we first quantify the differences of the total stellar masses from the pixel-by-pixel method and SED fitting of the total fluxes of all pixels. We compare the stellar mass obtained from the integrated photometry (unresolved) and the total stellar mass by summing individual pixels from the mass map (resolved) in Figure \ref{fig4}. For the redshift range of $z < 1.3$ (left panel), the median differences between unresolved and resolved stellar masses for all fields is $\sim 0.06$ dex. The discrepancy increases to $\sim 0.17$ dex for the redshift range of $1.3 < z < 2$ (right panel). This means that the total stellar masses obtained from the integrated photometry of the galaxies (unresolved) are on average less than the resolved total stellar masses from the pixel-by-pixel analysis. This does not depend on the field or the number of filters used. The difference at $z<1.3$ might be negligible, though at higher redshifts, this difference is more significant. This discrepancy is within the order of the uncertainties of the stellar mass estimates due to the stellar population modeling \citep{conroy2009}, though, the systematic needs to be better understood.  

\begin{figure*}	
\centering
\includegraphics[width=0.485\textwidth]{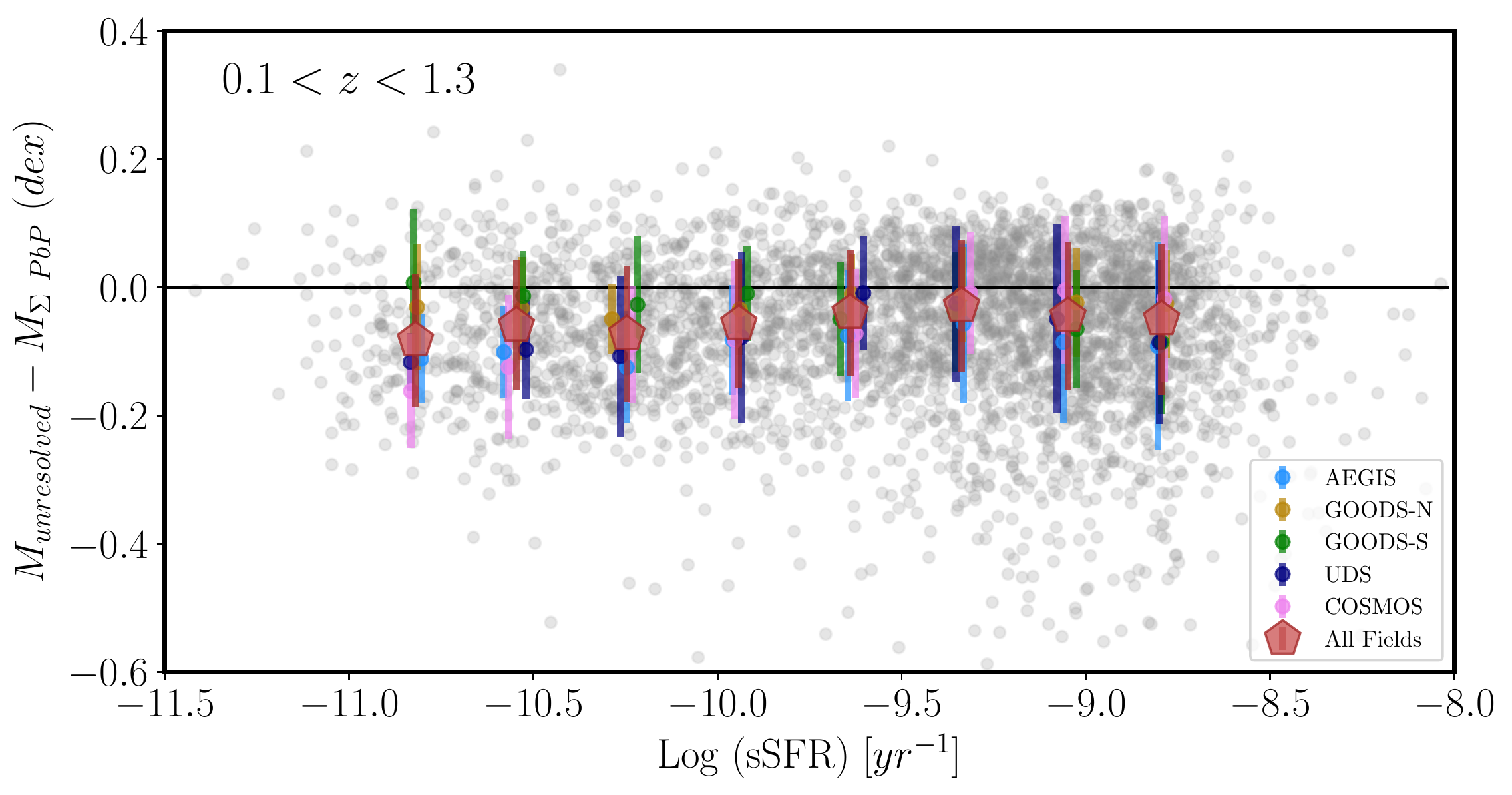}
\centering
\includegraphics[width=0.485\textwidth]{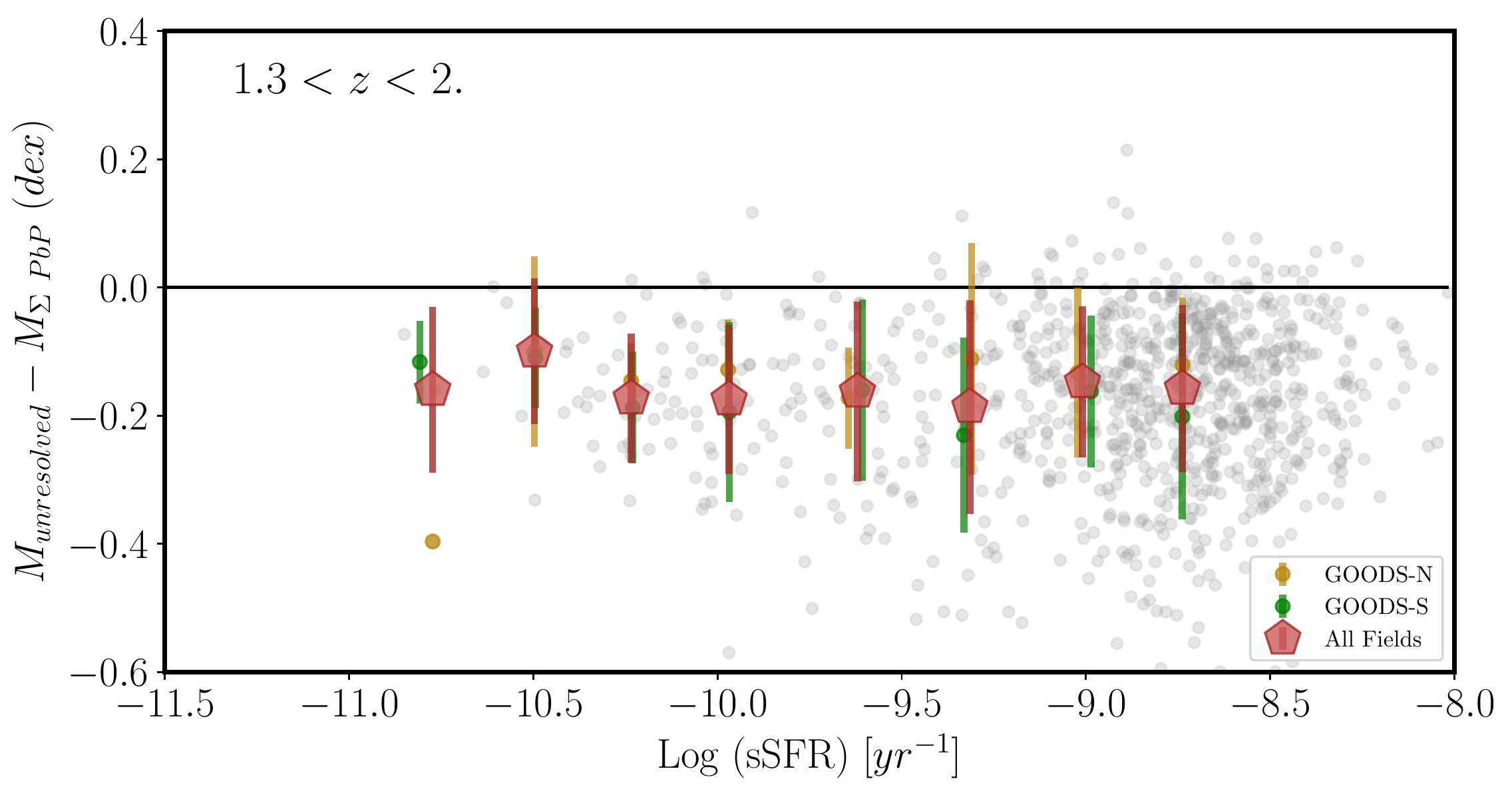}
\caption{The differences between resolved and unresolved total stellar masses as a function of specific star formation rate (sSFR) for two redshift bins as in Figure \ref{fig4}. There is no significant trend with sSFR, at least for the stellar mass range of this study ($\log(\mstar/\msun) \geqslant 9.8$).}
\label{fig5}
\end{figure*}

The origin of this difference is still not clear \citep[see, e.g.,][]{zibetti2009, martinez2017, sorba2015} and not reported in some studies \citep[e.g.,][]{wuyts2012}. In a recent study, \citet{sorba2018} argued that this effect depends on the specific star-formation rate (sSFR) of galaxies and caused by the outshining effect on the SED of galaxies, i.e., contribution of the young massive stars with lower mass-to-light ratio ($M/L$) to the total flux of galaxies \citep{papovich2001, maraston2010}. We used the combined UV+IR SFR from the analysis by \cite{whitaker2014} to test this in Figure \ref{fig5} for the two redshift ranges and could not see any considerable trend with the sSFR for any of these redshift ranges. An almost constant systematic offset exists at all sSFRs in the redshift bin of $1.3 < z < 2$. We should also note that the available HST filters do not cover the rest-frame near-infrared (NIR) wavelength ranges ($\gtrapprox 7000 $ \AA) of the sources beyond redshift of $\approx 1.3$. Many studies emphasized that the NIR SED of galaxies is crucial for determining their robust stellar masses \citep[see e.g., ][]{maraston2006, ilbert2010}, though, as shown in the Appendix \ref{sec:appendixB}, this issue mainly increases the uncertainties but does not introduce a significant systematics in this redshift range.

It is worth noting that the assumption of the star-formation history model might also contribute to this effect \citep[see recent work by][]{lower2020}. In this work, the best-fit time-scale ($ \tau$) varies for each pixel, and hence differs from a single value for the entire galaxy, which can mimic more complex SFHs. As shown in \citet{lower2020}, simplistic (delayed) $\tau$ models underpredict the total stellar mass. Therefore this might contribute to this systematic offset. However, testing this is beyond the scope of this paper.

\begin{figure*}
\centering
\includegraphics[width=\textwidth]{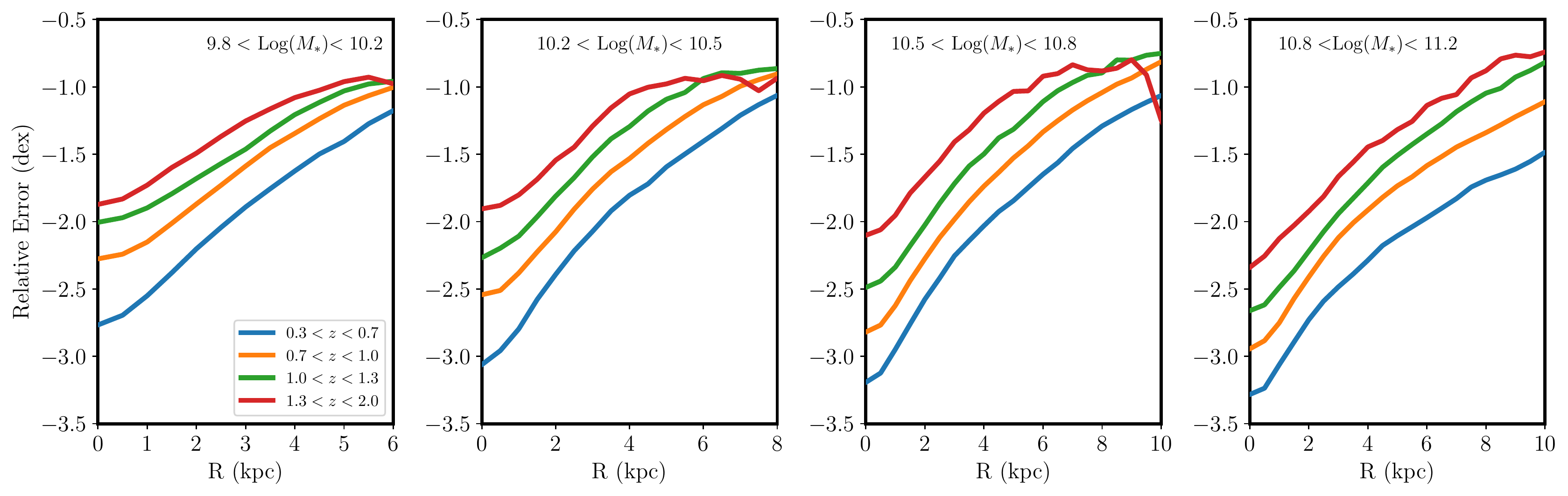}
\caption{Median stacked relative errors of the stellar mass density profiles as a function of radius for different stellar masses and redshift ranges. At fixed mass, the uncertainties of the stellar mass profiles increase towards the outskirts and towards higher redshifts. Typical relative errors are of the order of a 1\% in the central region and roughly 10\% in the outskirts.}
\label{fig6}
\end{figure*}

At fixed stellar mass, comparing the stellar mass density error profiles for different redshift bins shows that uncertainties in the stellar mass estimates increase towards higher redshifts and towards larger radius (see Figure \ref{fig6}). In general, the stellar mass maps for sources beyond $z>1.3$ might suffer from some uncertainties due to a combination of different effects such as less coverage of their SEDs, outshining effects, variation of the SFH across the galaxy or less signal to noise (S/N) per pixel. Furthermore, as discussed later in Section 4 and Appendix \ref{sec:appendixB}, the fraction of objects without reliable size estimate increases at this high redshift range, mainly due to the uncertainties of their mass maps. Therefore, we should treat this redshift range with caution.


\section{Stellar Mass Structural Parameters}

For this study, we use two methods for deriving the stellar mass weighted structural parameters by finding their best-fit \ser models \citep{sersic1963}. These methods are based on one and two dimensional (1D \& 2D) profile fitting approaches. As discussed in the Appendix \ref{sec:appendixA}, both methods are giving consistent results. However, since our 1D method is relatively more robust compared to the 2D method, we choose this as our fiducial method for the rest of this paper. We describe this 1D method in detail in this section. The second method and the reliability of these two methods are fully described in the Appendix \ref{sec:appendixA} and \ref{sec:appendixB}. The key results adopting the second method are presented in Appendix \ref{sec:appendixC}. The structural parameters from both methods are presented in Table \ref{tableA1} in the Appendix \ref{sec:appendixC}.


\subsection{1D Profile Fitting Method ($M_{I}$)}

In this procedure, we find the best-fit \ser models from their one-dimensional (1D) stellar mass profiles. The main motivation for using this method is to reduce the effect of uncertainties in the stellar mass maps in the outskirts. Smoothing the stellar mass maps and increasing their S/N ratio by means of smoothing methods such as ADAPTSMOOTH \citep[]{zibetti2009} or Voronoi binning method \citep{Cappellari2003} prior to the SED fitting can introduce biases on the \ser parameters and can create large fluctuations (steps) in their mass maps. Converting to the 1D profiles has the advantage to reduce the uncertainties introduced in the 2D method, while preserving the general shape of profiles. In addition, as described below, the procedure of finding the best-fit models for 1D profiles benefits from the evaluation of all models over a wide range of the parameter space. Therefore, this will give us a better constraint on the model parameters. In summary, this method allows us to estimate the sizes robustly.

\begin{figure*}
\centering
\includegraphics[width=0.9\textwidth]{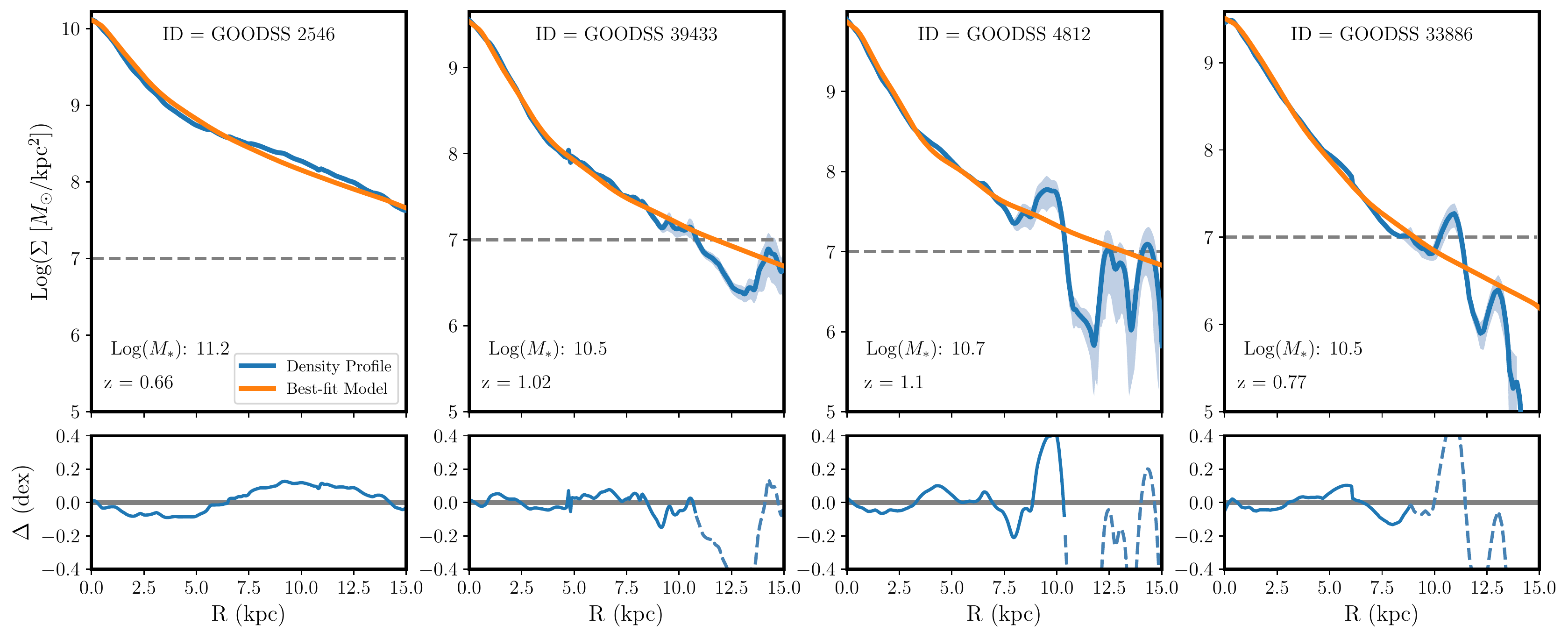}
\vspace{1. mm}
\centering
\includegraphics[width=0.9\textwidth]{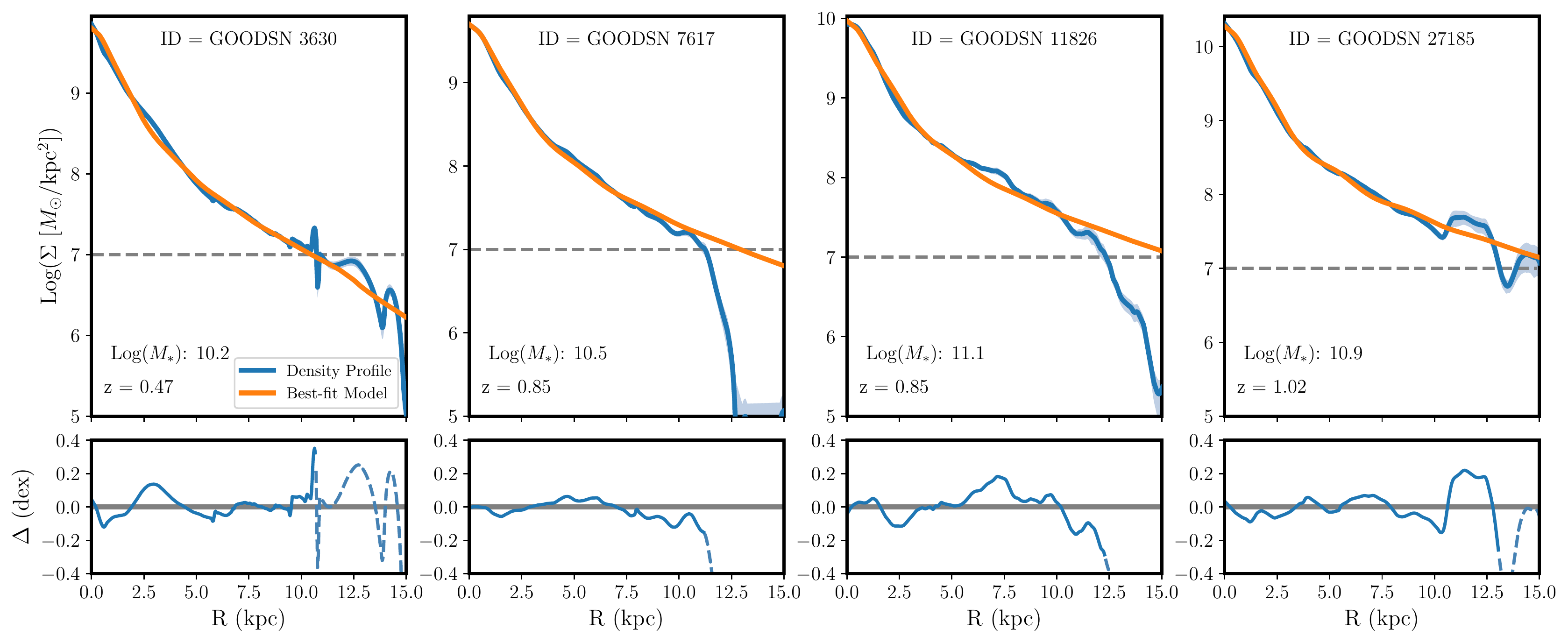}
\vspace{1. mm}
\centering
\includegraphics[width=0.9\textwidth]{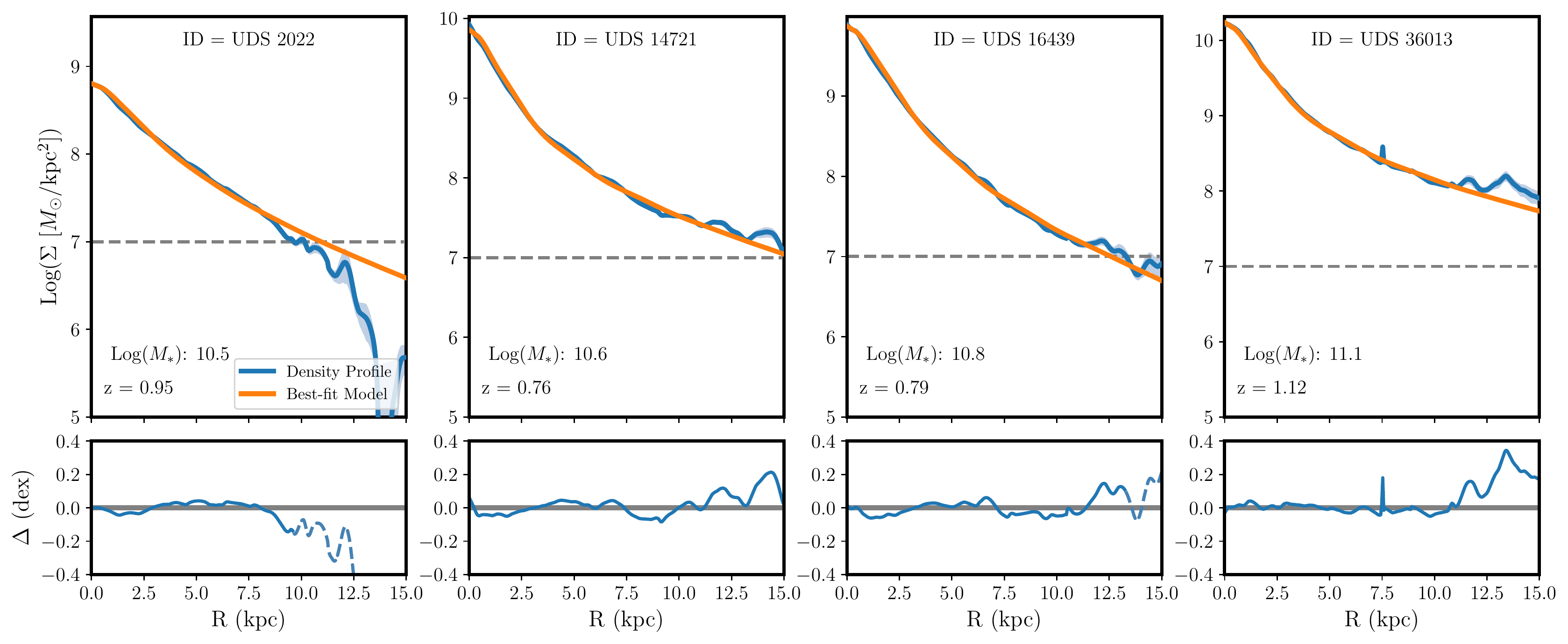}
\caption{Results from the first method for finding the best-fit 1D \ser models (orange lines) from the 1D stellar mass density profiles (blue lines) for the objects in Figure \ref{fig3}. The residuals are shown in bottom of each panel. For this technique, a surface brightness limit of $\Sigma = 10^7 \msun/\mathrm{kpc}^2$ is assumed during the fit. This approach helps to reduce the effect of uncertainties of the stellar mass estimates from the pixel-by-pixel SED method, in particular in the outer parts, and to evaluate all models in a wide range of the $n - r_{50}$ parameter space.} 
\label{fig7}
\end{figure*}

For this purpose, we follow the methodology used by \cite{maltby2018}. First, the stellar mass surface density profiles are measured using elliptical isophot (isomass for our case) fitting on the 2D stellar mass maps by means of IRAF task ELLIPSE \citep{jedrzejewski1987}. This provides 1D stellar mass surface density profiles as a function of radius. The fitting performed for each galaxy by fixing the same center for all ellipses and assuming ellipticity ($e$) and position angle (PA) as free parameters. The profiles are then circularized at each radius using $a \sqrt{1-e}$, in which $a$ is the semi-major axis. 

The next step is to build a library of \ser models to find the best-fitting ones. To this end, we first generate a table of 29,400 \ser models over the size ($r_{50}$) - \ser ($n$) parameter space ($0.006'' \leqslant r_{50} \leqslant1.8''$ and $0.3 \leqslant n \leqslant 10$), using steps of 0.1 for each parameter. The $e$ and PA are assumed to be zero for the models. We then use \texttt{GALFIT} to create 2D model images convolved with the $H_{160}$-band PSFs. To ensure that any field-to-field PSF variation does not affect the results, the models are created for each CANDELS field separately. The 1D density profiles of the models are then determined by fitting ellipses to these 2D galaxy models (fixing the center, PA and $e$ parameters). This grid of profile models are afterwards used for finding the best-fit 1D model. This is done by comparing the normalized 1D mass density profiles of galaxies with all normalized model profiles in the library and finding the minimum $\chi^2$ value. To reduce the uncertainties from the outer regions and the background, the fitting is performed down to a stellar mass surface density of $\Sigma = 10^7 \msun/\mathrm{kpc}^2$. This also ensures that the models catch the main part of the profiles. Our tests show that changing this surface mass density limit does not affect our results. The errors of the parameters are estimated by perturbing the galaxies' mass profiles within their uncertainties for 100 realizations, and finding their best-fit models. The 1$\sigma$ scatter of the values are then used to estimate their errors.

The blue lines in Figure \ref{fig7} are the surface stellar mass density profile determined from the mass maps as described above for the same sample of galaxies as shown in Figure \ref{fig2}. The best-fit models for these galaxies are illustrated as orange lines. The gray dashed lines depict the surface density limit of $10^7 \msun/\mathrm{kpc}^2$. The residuals are also shown as blue lines at the bottom of each panel. Despite some noisy features in the outer regions of the stellar mass profiles, this method is able recover the true shape of the mass profiles of these galaxies. This approach and its reliability is tested with simulated objects in Appendix \ref{sec:appendixB}. The average recovery rates of reliable size measurements are 96.4\% and 91.1\% for the low ($z<1.3$) and high redshift ($z>1.3$) bin, respectively. The majority of this loss is due to galaxies with stellar masses below $\log(\mstar/\msun) \sim 10.5$. This mainly affect the size-mass relation of low-mass galaxies at $z>1$.

\subsection{Measuring sizes: $r_{20}$, $r_{50}$ and $r_{80}$}

We use the half-mass radii ($r_{50}$) and \ser ($n$) values obtained from the best-fit models to find the radius of galaxies containing 20 and 80 percent of their total stellar mass ($r_{20}$ and $r_{80}$, respectively). We use Equation 3 of \cite{miller2019} to convert $r_{50}$ and $n$ to $r_{20}$ and $r_{80}$ as:

\begin{equation}
\frac{r_{20}}{r_{50}} (n) = -0.0008n^3 + 0.0178n^2 - 0.1471n + 0.6294
\label{eq:r20}
\end{equation}
\begin{equation}
\frac{r_{80}}{r_{50}} (n) = 0.0012n^3 - 0.0123n^2 + 0.5092n + 1.2646.
\label{eq:r80}
\end{equation}

Finally, we note that based on the results from our simulations (Appendix \ref{sec:appendixB}, the \ser indices from the 1D method are underestimated for the average values of $12\%$ and $16\%$ at low and high redshift bins ($z<1.3$ and $z>1.3$), respectively. Therefore, the \ser values are corrected for estimating the $r_{20}$ and $r_{80}$. 

\section{Results}
\label{results}

As mentioned in the Introduction, the stellar mass assembly history of galaxies can be traced by the morphological properties such as size and concentration. In this section, we use the measured mass-weighted sizes and concentrations from the stellar mass surface density profiles to explore these quantities for different types of galaxies and their dependence on redshift. First, we quantify and compare the size-mass relation of star-forming and quiescent galaxies. Following this, we measure the size evolution as a function of cosmic time at fixed stellar mass. Finally, we show results on the scatter of the size-mass relation and the \ser index.

\begin{figure*}
\centering
\includegraphics[width=\textwidth]{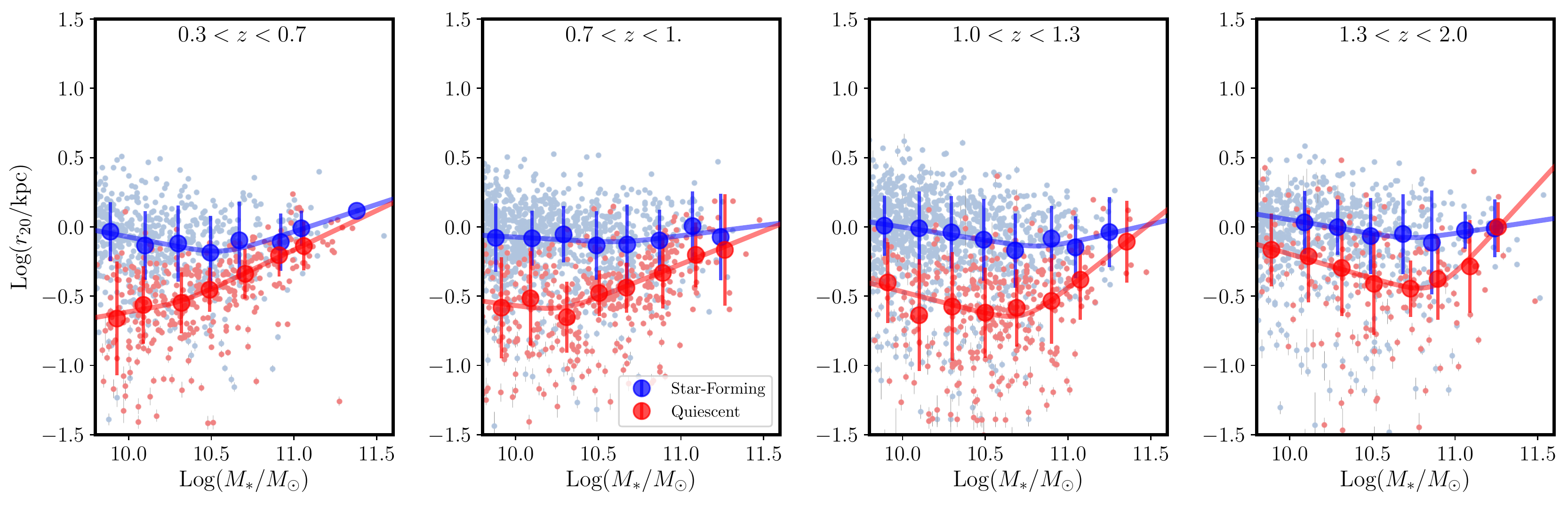}
\vspace{3. mm}
\centering
\includegraphics[width=\textwidth]{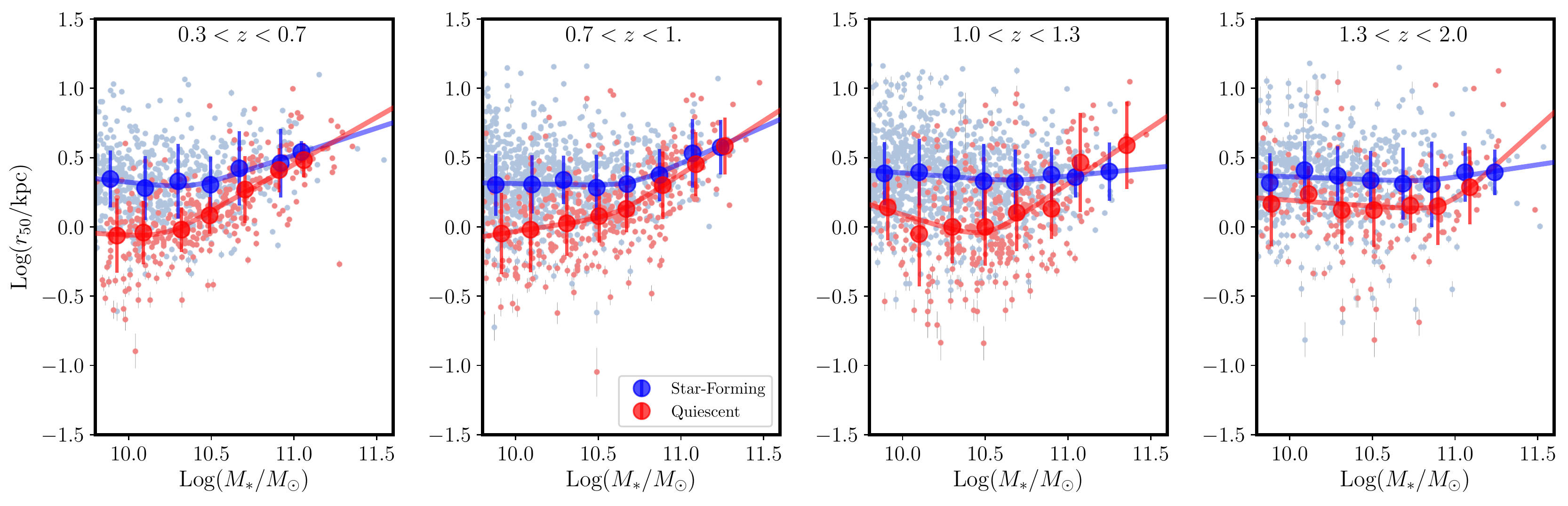}
\vspace{3. mm}
\centering
\includegraphics[width=\textwidth]{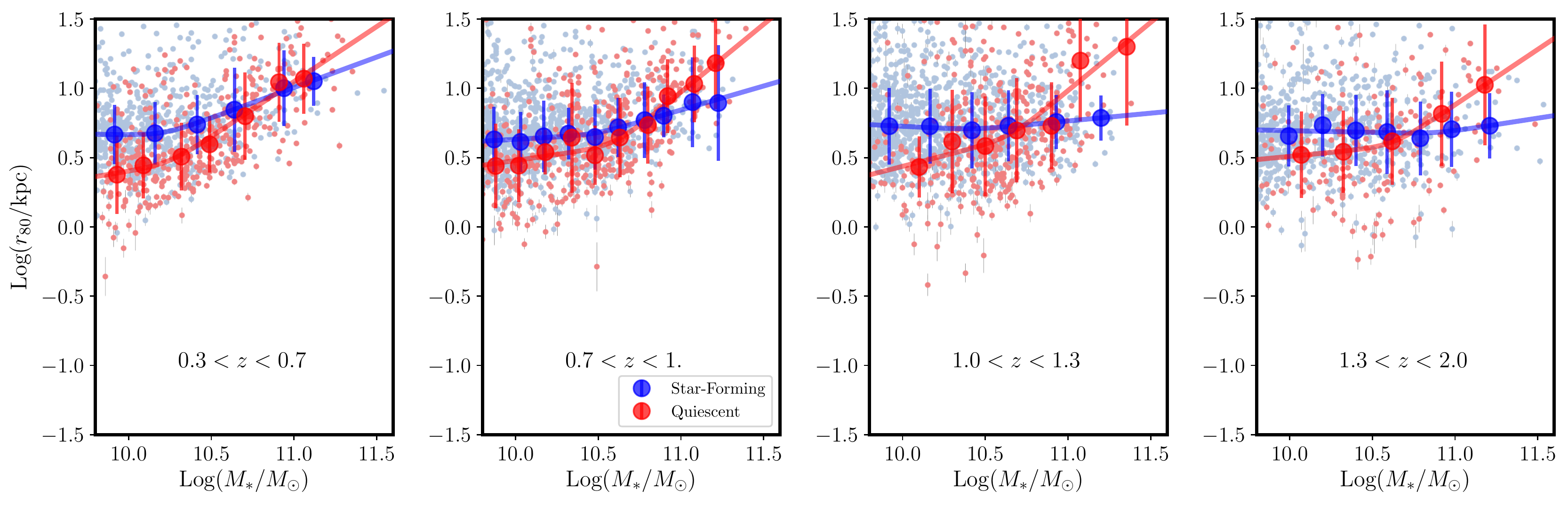}
\caption{The stellar mass-size plane for star-forming (blue) and quiescent (red) galaxies. From top to bottom, we show the size-mass relation for the $r_{20}$, $r_{50}$ and $r_{80}$ sizes. The panels from left to right show increasing redshift bins. The large symbols illustrate the median of sizes in each stellar mass bin. The solid line shows the best fit relation (see Table 1 \& 2 for the best-fit parameters). Overall, the size-mass relation is relatively flat for star-forming galaxies: sizes only depend weakly on stellar mass. On the other hand, the size-mass relation for quiescent galaxies is steeper, in particular above the pivot mass of $\log(\mstar/\msun) \sim 10.5$, which only depends weakly on cosmic epoch (see also Fig. \ref{fig10}).} 
\label{fig8}
\end{figure*}


\subsection{Size-Mass Relation}

We present the size-mass relation at different cosmic epochs for star-forming and quiescent galaxies in Figure \ref{fig8}. This relation has been explored extensively using light profiles up to high redshifts in many studies \citep[e.g.,][]{mosleh2012, vanderwel2014, Holwerda2015, Allen2017, miller2019, damjanov2019, Whitney2019}. \cite{mosleh2017} studied the half-mass size evolution of galaxies at fixed mass based on the 1D stellar mass profiles for the GOODS-North and GOODS-South fields, and \cite{suess2019a} presented the half-mass size-mass relation for galaxies at $1.0<z<2.5$. In this work, we expand this to the different measures of mass-based sizes and look at their evolution from $z=0.3$ to 2.0. This helps to explore how the stellar mass has been assembled in the inner and outer parts of the star-forming and quiescent galaxies over the last 10 Gyrs. 

In Figure \ref{fig8} we examine the size-mass relation for the mass-based $r_{20}$ (top panels), $r_{50}$ (middle panels), and $r_{80}$ (bottom panels) sizes, which include 20\%, 50\%, and 80\% of the total stellar mass, respectively. Panels from left to right show bins of increasing redshift. The individual galaxies are shown as blue (star-forming galaxies) and red (quiescent galaxies) points. The large blue and red circles present the median values of sizes as a function of stellar mass for star-forming and quiescent galaxies, respectively. The error bars are the $1\sigma$ scatter in each bin. We use a broken power-law relation from Equation 2 of the \cite{mowla2019a} to quantify the size-mass relation:
\begin{equation}
r = r_p \left(\frac{\mstar}{M_p}\right)^{\alpha} \left[\frac{1}{2}\left\{ 1+\left(\frac{\mstar}{M_p}\right)^{\delta}\right\} \right]^{(\beta -\alpha)/\delta}
\end{equation}
where $\alpha$ and $\beta$ are the slopes at the low- and high-mass end, respectively. The variables $r_p$ and $M_p$ are the pivot radius and stellar mass at which the slope transitions from $\alpha$ to $\beta$. We fit the median values with this relation while setting the smoothing factor ($\delta$) to 6. The results of the best fit values are presented in Tables \ref{table1} and \ref{table2}.

\begin{figure}
\centering
\includegraphics[width=0.48\textwidth]{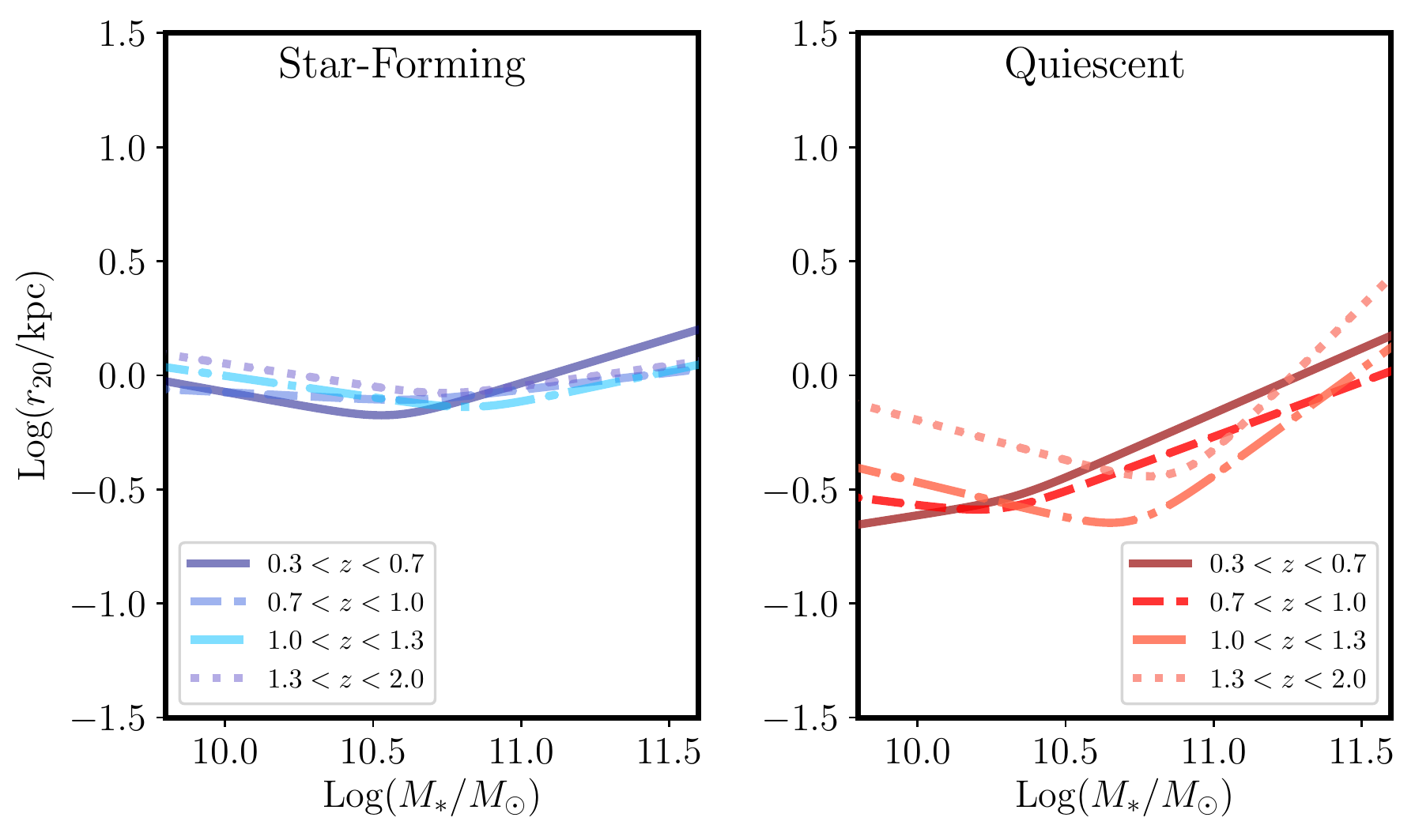}
\hspace{.2 mm}
\centering
\includegraphics[width=0.48\textwidth]{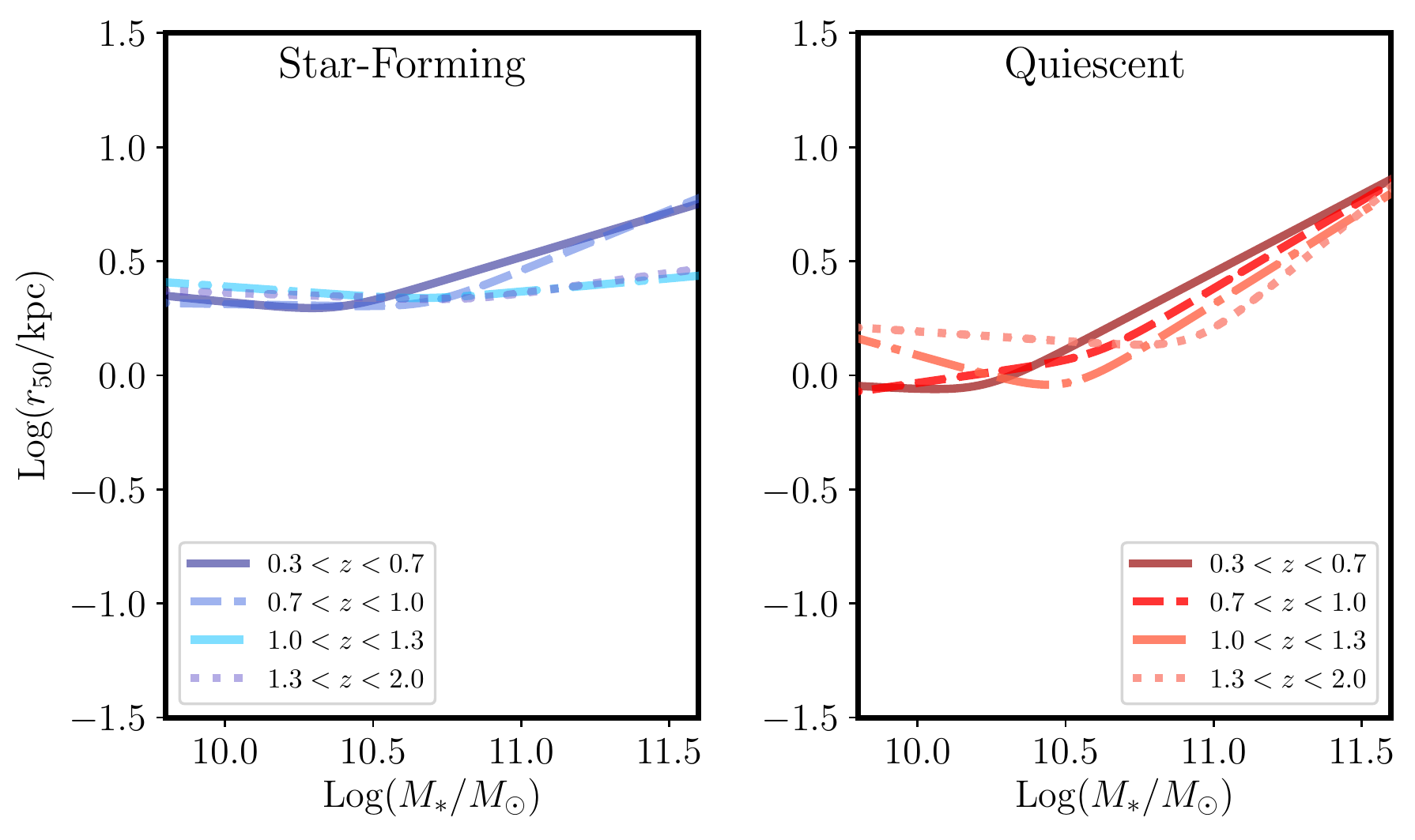}
\hspace{.2 mm}
\centering
\includegraphics[width=0.48\textwidth]{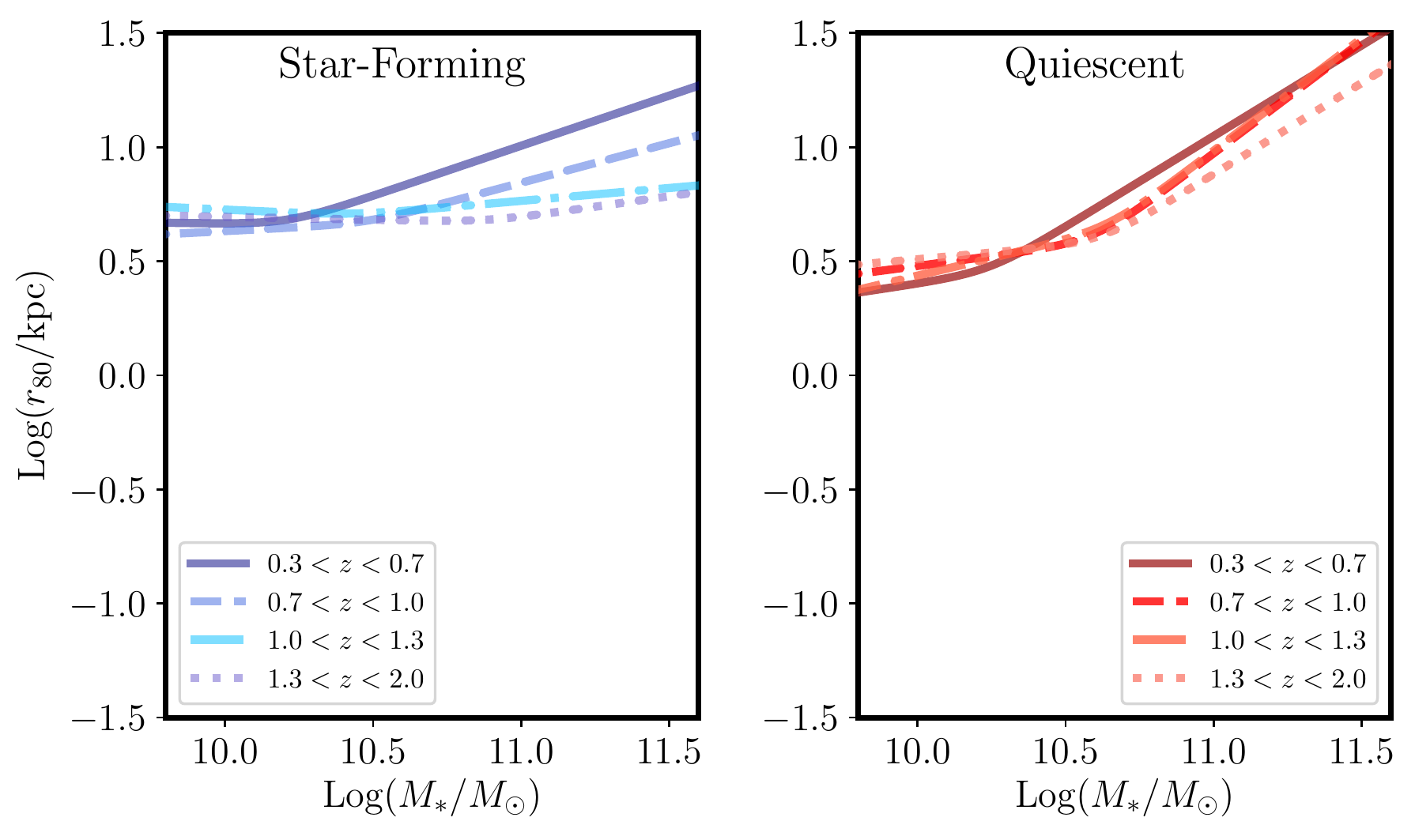}
\caption{Comparison of the size-mass relation of star-forming and quiescent galaxies at different redshifts. The relations are illustrated for the $r_{20}$, $r_{50}$ and $r_{80}$ sizes from top to bottom.}
\label{fig9}
\end{figure}

We first focus on the size-mass relation given by $r_{20}$, which is shown in the top panels Figure \ref{fig8}. For the star-forming galaxies, the size-mass relation is relatively flat in all redshift bins. This relation has only a shallow slope at the high-mass end at $z<1$. The slope at low masses is almost zero, but there is a hint that in the lowest redshift bin that this slope ($\alpha$) tends to be slightly negative. This indicates that the median $r_{20}$ sizes are almost independent of redshift and stellar mass. The median size of $r_{20}$ is about 1 kpc, which corresponds to the assumed bulge size of the star-forming galaxies in some recent studies \citep[e.g.,][]{cheung2012, fang2013}.

Quiescent galaxies have smaller $r_{20}$ sizes compared to the star-forming ones at all stellar masses, reflecting the higher concentration of these galaxies (see also Section 5.3). For quiescent galaxies, the relation is steeper at high masses above the pivot mass ($M_p$). The existence of the $r_{20}$-mass relation with a larger slope for the massive quiescent galaxies suggests that as the total stellar mass increases, these objects have larger cores and higher stellar mass concentrations. Below the pivot stellar mass, the relation tends to be flattened, at least for the redshift bins of $z<1$. For the higher redshift bins ($z>1$), the trends seem to be reversed, though this might be affected by the incompleteness in our sample (see Section 4 and also Section 6.2 for discussion whether this trend is reliable). A larger sample of low mass galaxies with robust stellar mass size measurements at these high redshift ranges is required for better understanding this behavior.

\begin{figure*}
\centering
\includegraphics[width=0.47\textwidth]{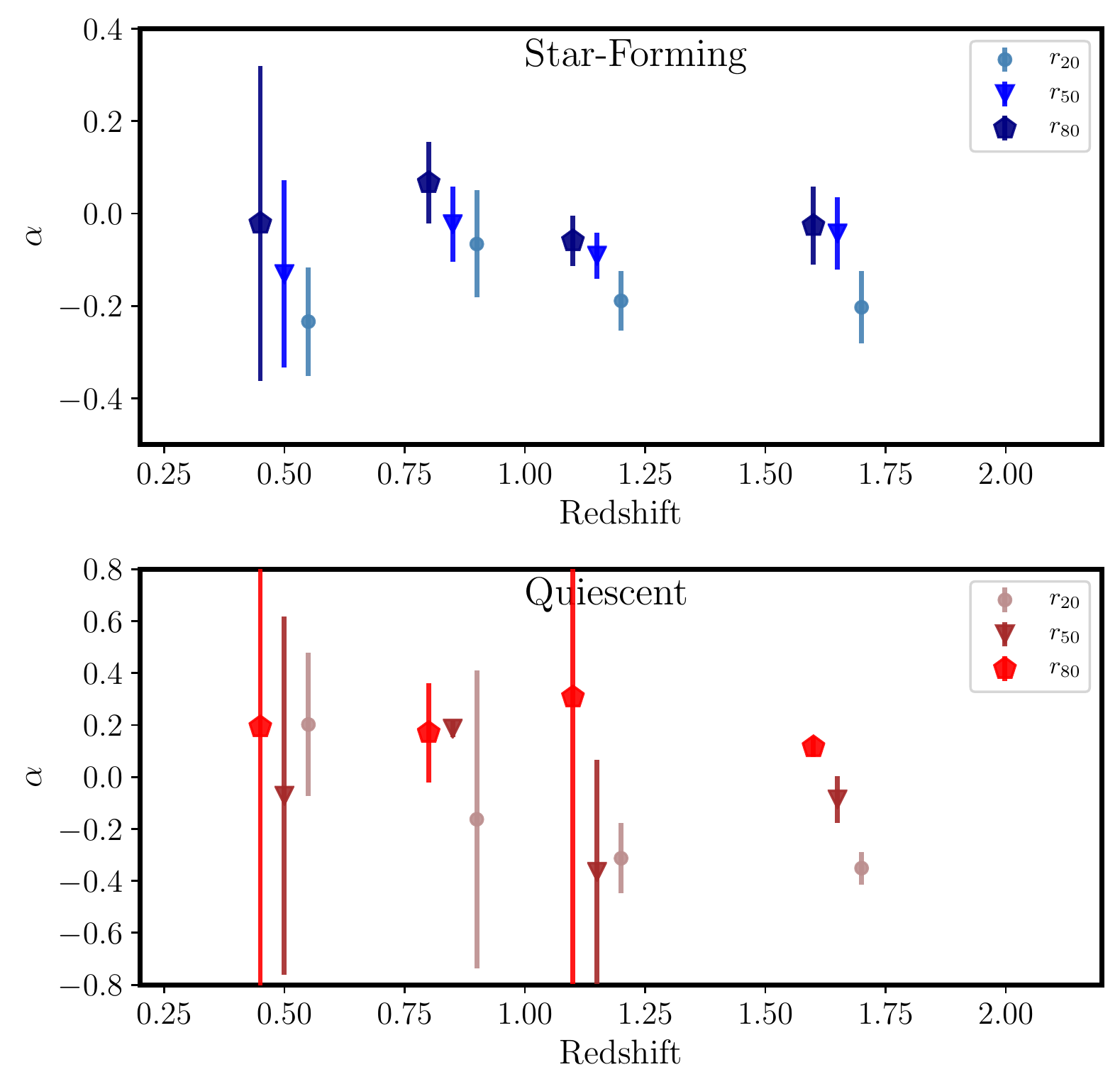}
\hspace{1. mm}
\centering
\includegraphics[width=0.47\textwidth]{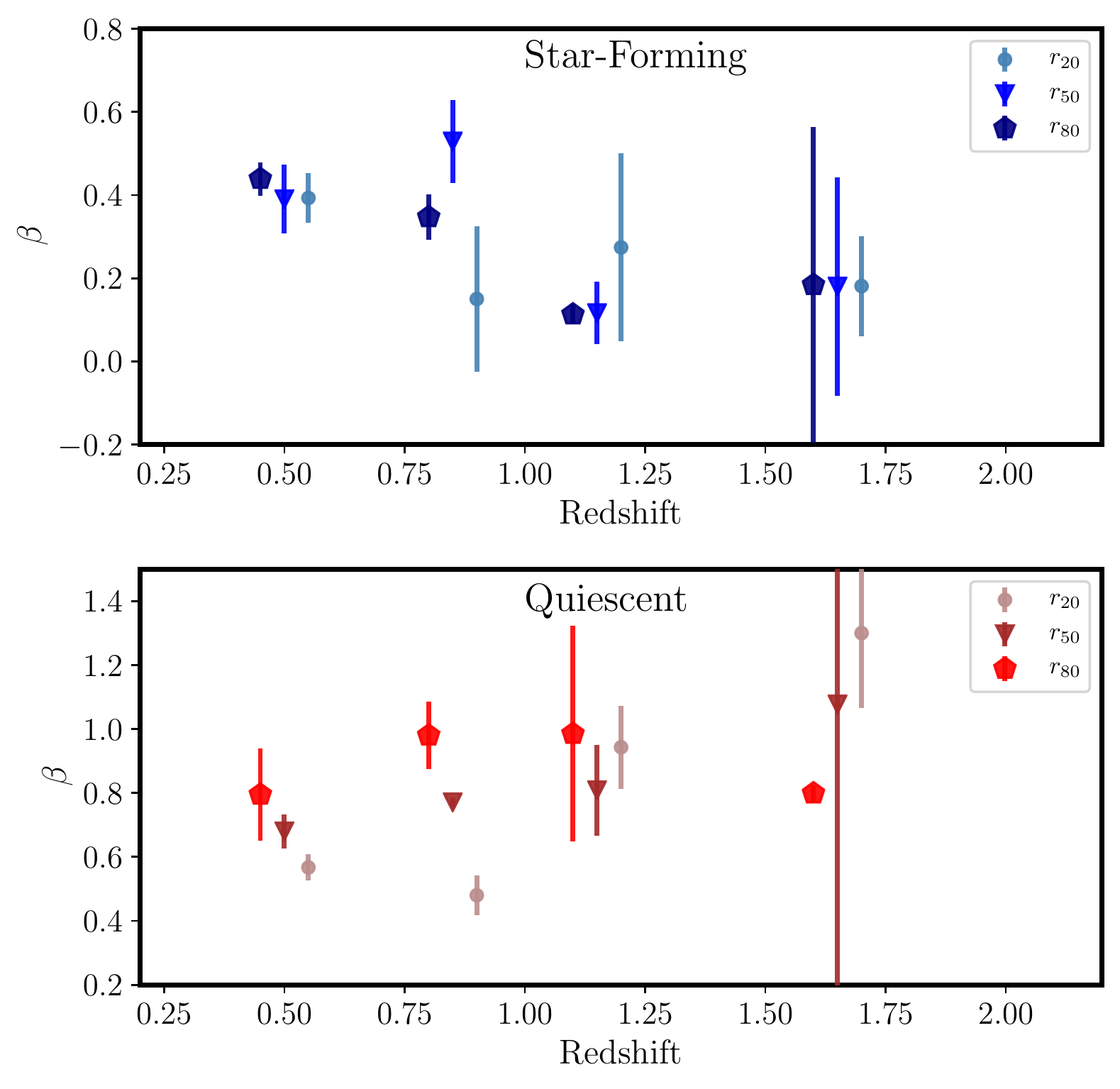}
\caption{\textit{Left panels:} The evolution of the low-mass end slope of the size-mass relation ($\alpha$) with redshift for star-forming and quiescent galaxies (top and bottom panels, respectively). \textit{Right panels:} The same as left ones, but for the high-mass end slope ($\beta$). The exact values are given in Table 1 and 2. Low- and high-mass end slopes only depend weakly on redshift for both star-forming and quiescent galaxies. The strongest trend with redshift can be found for the high-mass end slope of star-forming galaxies: the size-mass relation steepens toward low redshifts for star-forming galaxies.}
\label{fig10}
\end{figure*}

\begin{figure*}
\centering
\includegraphics[width=0.47\textwidth, height=8. cm]{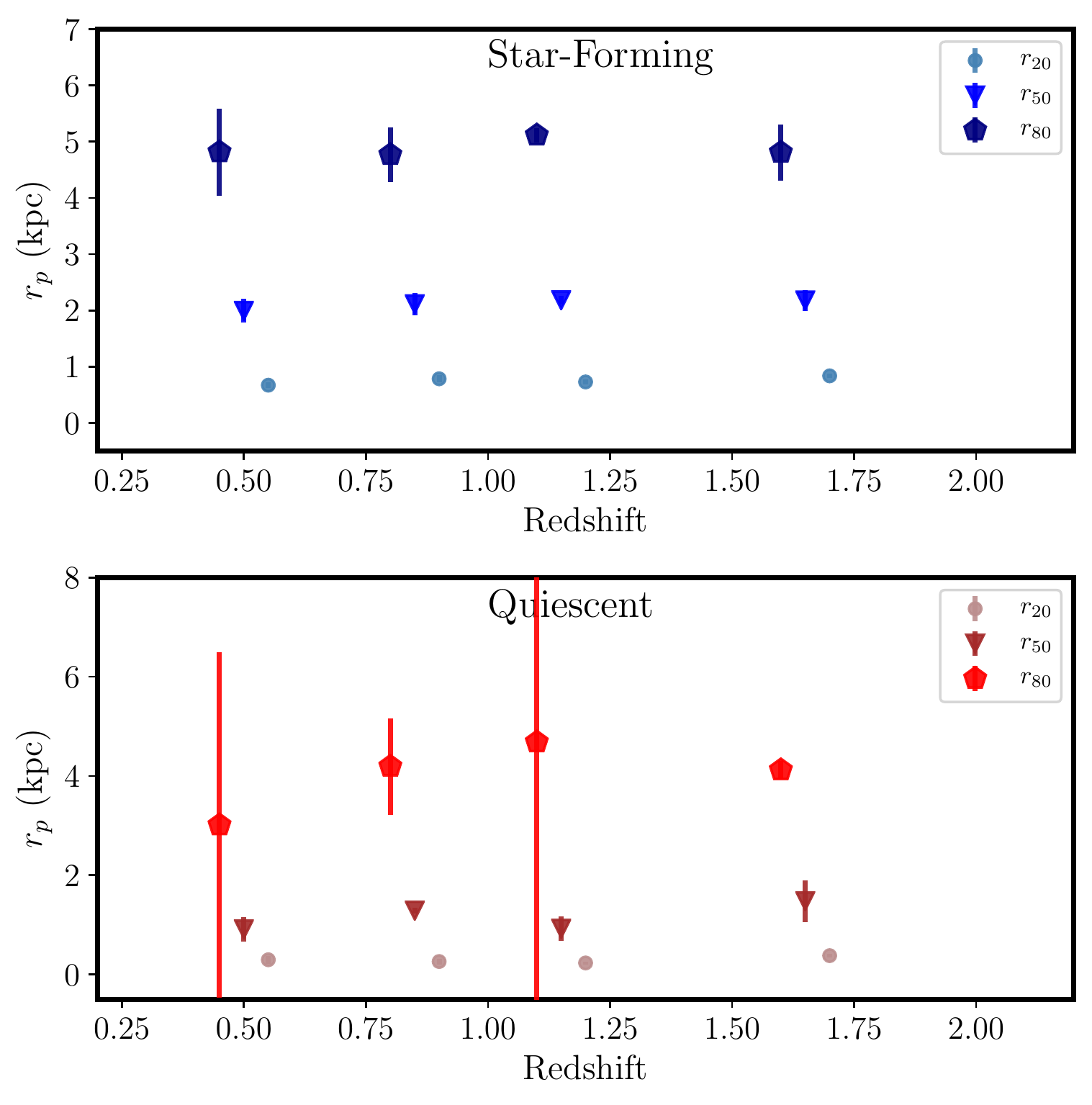}
\centering
\includegraphics[width=0.47\textwidth, height=8 cm]{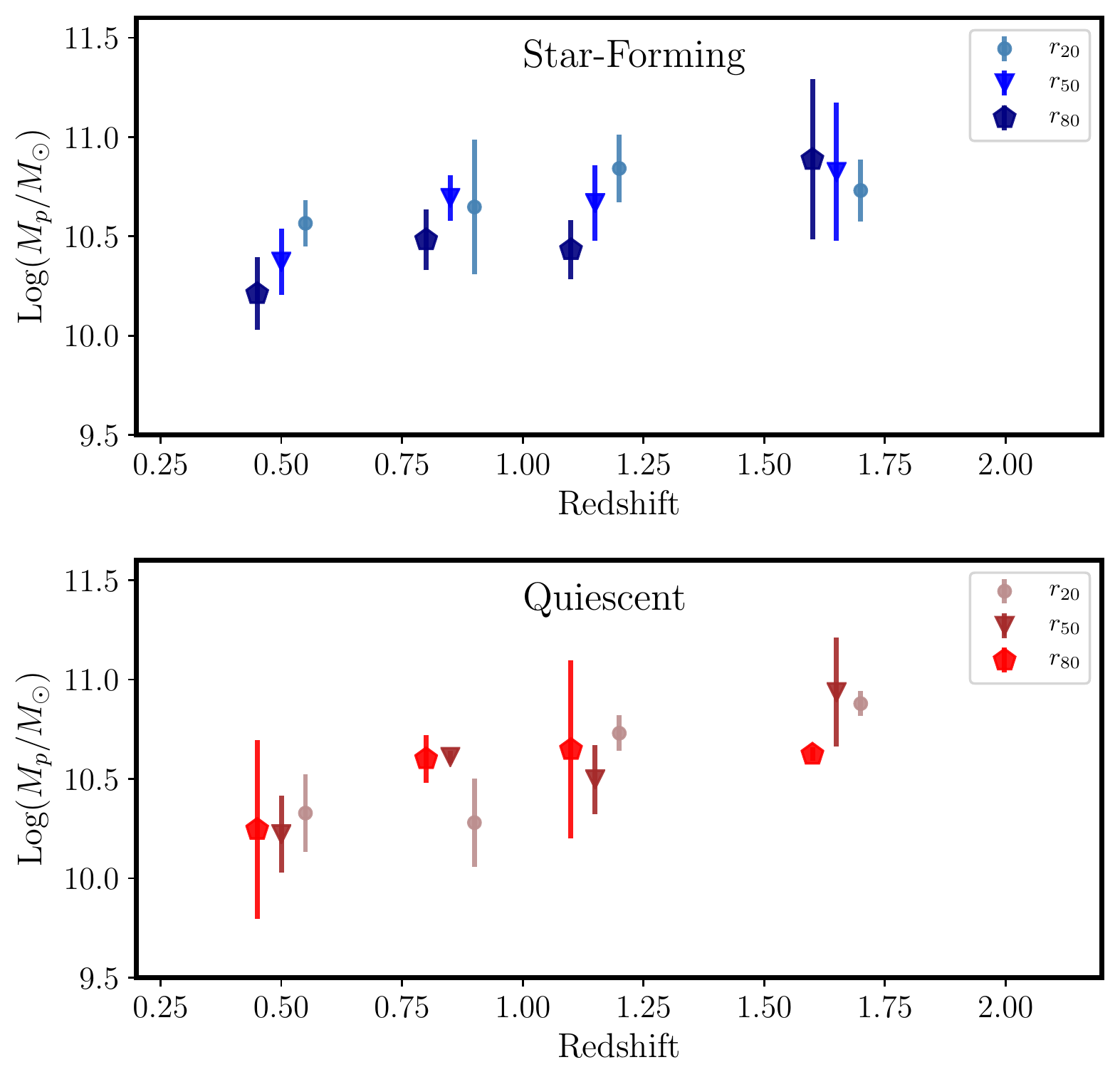}
\caption{Evolution of the pivot radius and pivot stellar mass (left and right panels, respectively) of the size-mass relation as a function redshift. The top and bottom panel are for star-forming and quiescent galaxies, respectively. The exact values are given in Table 1 \& 2. We find that the pivot radius does not evolve with redshift, while the pivot stellar mass increases with redhsift. Interestingly, this holds for all three sizes ($r_{20}$, $r_{50}$ and $r_{80}$).}
\label{fig11}
\end{figure*}

The half-mass size $r_{50}$-mass relations are shown in the middle panels of the Figure \ref{fig8}. The half-mass radii of star-forming galaxies are larger than the ones of quiescent galaxies at fixed stellar mass, consistent with the results based on half-light radii \citep[e.g.,][]{trujillo2006a, williams2010, vanderwel2014}. In the high-$z$ bins ($z>1$), the size-mass relation is relatively flat for star-forming galaxies, while at later epochs, the relation gets steeper above the pivot mass ($\beta$ increases by about a factor of $\sim 2$). Furthermore, the slope $\beta$ is steeper for quiescent galaxies than for star-forming ones. Moreover, the most massive quiescent galaxies ($\log(\mstar/\msun) > 11$) have comparable $r_{50}$ sizes to their star-forming counter parts \citep{faisst2017}, although the statistic is low at the high-mass end. 

We find a similar behavior for star-forming and quiescent galaxies regarding the $r_{80}$-mass plane (bottom panels of Figure \ref{fig8}). In this plane, the scatter of the size distribution decreases slightly, and the average size difference between the star-forming and quiescent ones is reduced compared to the $r_{20}$ and $r_{50}$ sizes (see discussion in Section 6). The $r_{80}$ sizes are representative of the radii where the bulk of (80\%) of the total stellar mass content of the galaxies are located. Selecting such a radius can reduce differences of the measured sizes (less sensitive to the \ser index), and hence decreases the scatter of the size-mass relation (see also discussion by \citealt{sanchez2020} and recent work by \citealt{trujillo2020}). There is a hint that the distances between the median points for different populations (red and blue points) are less at low redshift bins compared to the high redshift ones, though, further examination is required. 

In Figure \ref{fig9}, the size-mass relations as a function of redshift are plotted for star-forming (left panels) and quiescent (right panels) galaxies. Overall, there is surprisingly little evolution in the size-mass relations for all three size definitions and galaxy types. This can also be seen in Figure \ref{fig10}, where the evolution of the low- and high-mass end slopes ($\alpha$ and $\beta$, respectively) as a function of redshift is shown. The low-mass end slopes are always shallower than the high-mass end slopes. The high-mass slopes are also steeper for quiescent than star-forming galaxies. Variation of the size-mass relation slopes at high and low mass ends, has already been reported for different types of galaxies and suggested to be related to different mechanisms for the growth of galaxies \citep[e.g.,][]{shen2003, janz2008, mosleh2013, mowla2019a}. The behavior of this relation for different size definition and their corresponding pivot mass for each type of galaxies, can also be informative to find possible scenarios for the transition between galaxy populations. 

For star-forming galaxies, the size-mass relation does not evolve significantly below the pivot mass with cosmic time for all three size definitions, i.e. the normalization as well as the low-mass slope $\alpha$ remain constant. Even above the pivot mass scale, the $r_{20}$-mass relation remains constant. On the other hand, the $r_{50}$-mass and $r_{80}$-mass relations both steepen above the pivot mass scale with cosmic time (the high-mass slope $\beta$ increases). The size-mass relation of quiescent galaxies shows little evolution at low masses as well. In particular, the $r_{50}$-mass and $r_{80}$-mass relations stay roughly constant with cosmic time below the pivot mass scale. Above the pivot mass, sizes are typically larger at later epochs. The $r_{20}$-mass relation shows an overall more complex behavior. We discuss the size evolution at fixed stellar mass further in the next section.

We should note that the low and high-mass end slopes (of the $r_{50}$ size-mass relation) for both star-forming and quiescent galaxies in the lowest redshift bin are consistent (within the error bars) with the results for the galaxies in the local universe  \citep[e.g.,][]{mosleh2013, lang2015}.

Finally, the pivot stellar masses for both samples regardless of the size definition increase with redshift, as also depicted in the right panels of the Figure \ref{fig11}. For both star-forming and quiescent galaxies, the pivot radii are almost constant at different redshifts. As we discuss in Section 6, this may be an indication that the mechanisms that drive galaxy structure start to act at lower stellar masses at the later cosmic times.

\begin{deluxetable*}{cccccc}
\tablecolumns{6}
\centering
\tablewidth{\textwidth}
\tabletypesize \footnotesize
\tablecaption{The fitting results for the size-mass relation of Equation 3 for the star-forming galaxies.}\label{table1}
\tablehead{
\colhead{Redshift} &
\colhead{Size} &
\colhead{$\alpha$} &
\colhead{$\beta$} & 
\colhead{$r_{p}$ (kpc)}&
\colhead{Log($M_{p}/\mstar$)}
}
\startdata
$0.3<z<0.7$ & $r_{20}$ & $-0.234\pm0.118$ & $0.393\pm0.059$ & $0.670\pm0.050$ & $10.566\pm0.115$\\
$0.7<z<1.0$  &   & $-0.066\pm0.116$ & $0.150\pm0.175$ & $0.784\pm0.065$ & $10.648\pm0.338$\\
$1.0<z<1.3$  &   & $-0.189\pm0.064$ & $0.274\pm0.226$ & $0.728\pm0.043$ & $10.842\pm0.170$\\
$1.3<z<2.0$  &   & $-0.203\pm0.078$ & $0.181\pm0.120$ & $0.836\pm0.039$ & $10.731\pm0.156$\\
\hline
$0.3<z<0.7$ & $r_{50}$ & $-0.131\pm0.203$ & $0.390\pm0.083$ & $1.993\pm0.212$ & $10.373\pm0.166$\\
$0.7<z<1.0$  &  & $-0.023\pm0.082$ & $0.529\pm0.100$ & $2.110\pm0.196$ & $10.694\pm0.115$\\
$1.0<z<1.3$  &  & $-0.091\pm0.050$ & $0.116\pm0.075$ & $2.182\pm0.071$ & $10.668\pm0.189$\\
$1.3<z<2.0$  &  & $-0.043\pm0.078$ & $0.180\pm0.263$ & $2.177\pm0.181$ & $10.825\pm0.347$\\
\hline
$0.3<z<0.7$ & $r_{80}$ & $-0.021\pm0.341$ & $0.439\pm0.040$ & $4.818\pm0.775$ & $10.212\pm0.183$\\
$0.7<z<1.0$  &  & $0.067\pm0.088$ & $0.347\pm0.054$ & $4.770\pm0.488$ & $10.483\pm0.153$ \\
$1.0<z<1.3$  &   & $-0.059\pm0.055$ & $0.113\pm0.024$ & $5.118\pm0.128$ & $10.433\pm0.149$ \\
$1.3<z<2.0$  &   & $-0.026\pm0.084$ & $0.184\pm0.379$ & $4.811\pm0.500$ & $10.887\pm0.403$\\
\enddata
\end{deluxetable*}

\begin{deluxetable*}{cccccc}
\tablecolumns{6}
\centering
\tablewidth{\textwidth}
\tabletypesize \footnotesize
\tablecaption{The fitting results for the size-mass relation of Equation 3 for the quiescent galaxies.}\label{table2}
\tablehead{
\colhead{Redshift} &
\colhead{Size} &
\colhead{$\alpha$} &
\colhead{$\beta$} & 
\colhead{$r_{p}$ (kpc)}&
\colhead{Log($M_{p}/\mstar$)}
}
\startdata
$0.3<z<0.7$ & $r_{20}$ & $0.202\pm0.276$ & $0.567\pm0.041 $ & $0.296\pm0.077$  & $10.328\pm0.195$\\
$0.7<z<1.0$  &   & $-0.163\pm0.574$  & $0.480\pm0.062$  & $0.261\pm0.058$  & $10.280\pm0.223$\\
$1.0<z<1.3$  &   & $-0.313\pm0.135$ & $0.943\pm0.130$ & $0.232\pm0.030$ & $10.730\pm0.090$\\
$1.3<z<2.0$  &   & $-0.351\pm0.063$ & $1.300\pm0.235$ & $0.379\pm0.030$ & $10.879\pm0.063$\\
\hline
$0.3<z<0.7$ & $r_{50}$ & $-0.072\pm0.690$ & $0.680\pm0.053$ & $0.913\pm0.249$ & $10.222\pm0.195$\\
$0.7<z<1.0$  &  & $0.186\pm0.033$ & $0.771\pm0.016$ & $1.288\pm0.054$ & $10.611\pm0.031$\\
$1.0<z<1.3$  &  & $-0.365\pm0.431$ & $0.809\pm0.142$ & $0.922\pm0.252$ & $10.498\pm0.174$\\
$1.3<z<2.0$  &  & $-0.086\pm0.090$ & $1.077\pm2.235$ & $1.479\pm0.419$ & $10.937\pm0.273$\\
\hline
$0.3<z<0.7$ & $r_{80}$ &$0.193\pm1.710$ & $0.795\pm0.144$ & $3.013\pm3.479$ & $10.246\pm0.450$ \\
$0.7<z<1.0$  &   &$0.171\pm0.191 $ & $0.980\pm0.105$ & $4.194\pm0.972$& $10.601\pm0.121$  \\
$1.0<z<1.3$  &   & $0.308\pm1.105$ & $0.985\pm0.337$ & $4.684\pm6.834$ & $10.648\pm0.449$\\
$1.3<z<2.0$  &   &$0.116\pm0.039$ & $0.799\pm0.025$ & $4.123\pm0.178$ & $10.624\pm0.034$\\
\enddata
\end{deluxetable*}


\begin{figure*}
\centering
\includegraphics[width=0.32\textwidth]{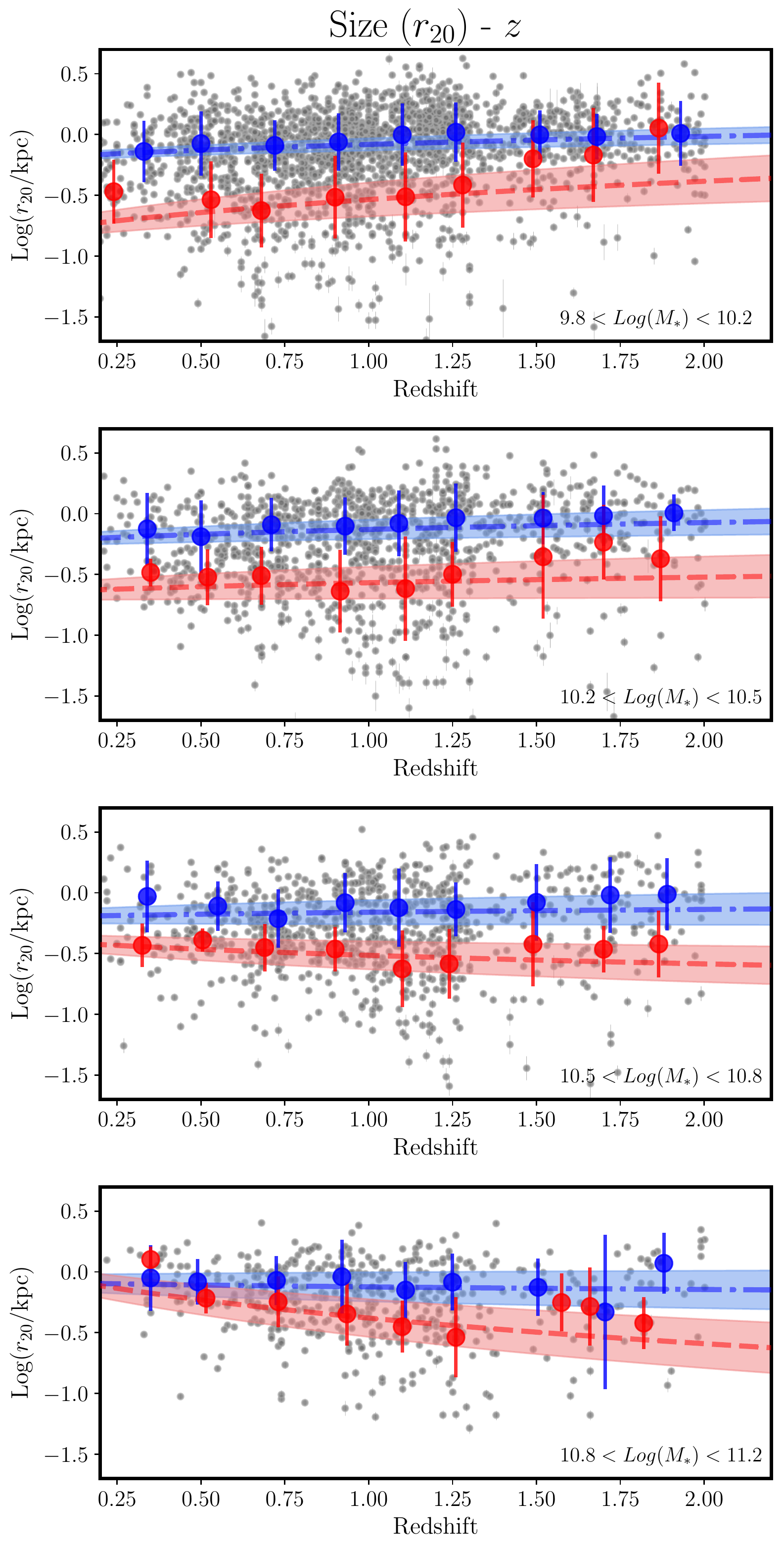}
\vspace{.1 mm}
\centering
\includegraphics[width=0.32\textwidth]{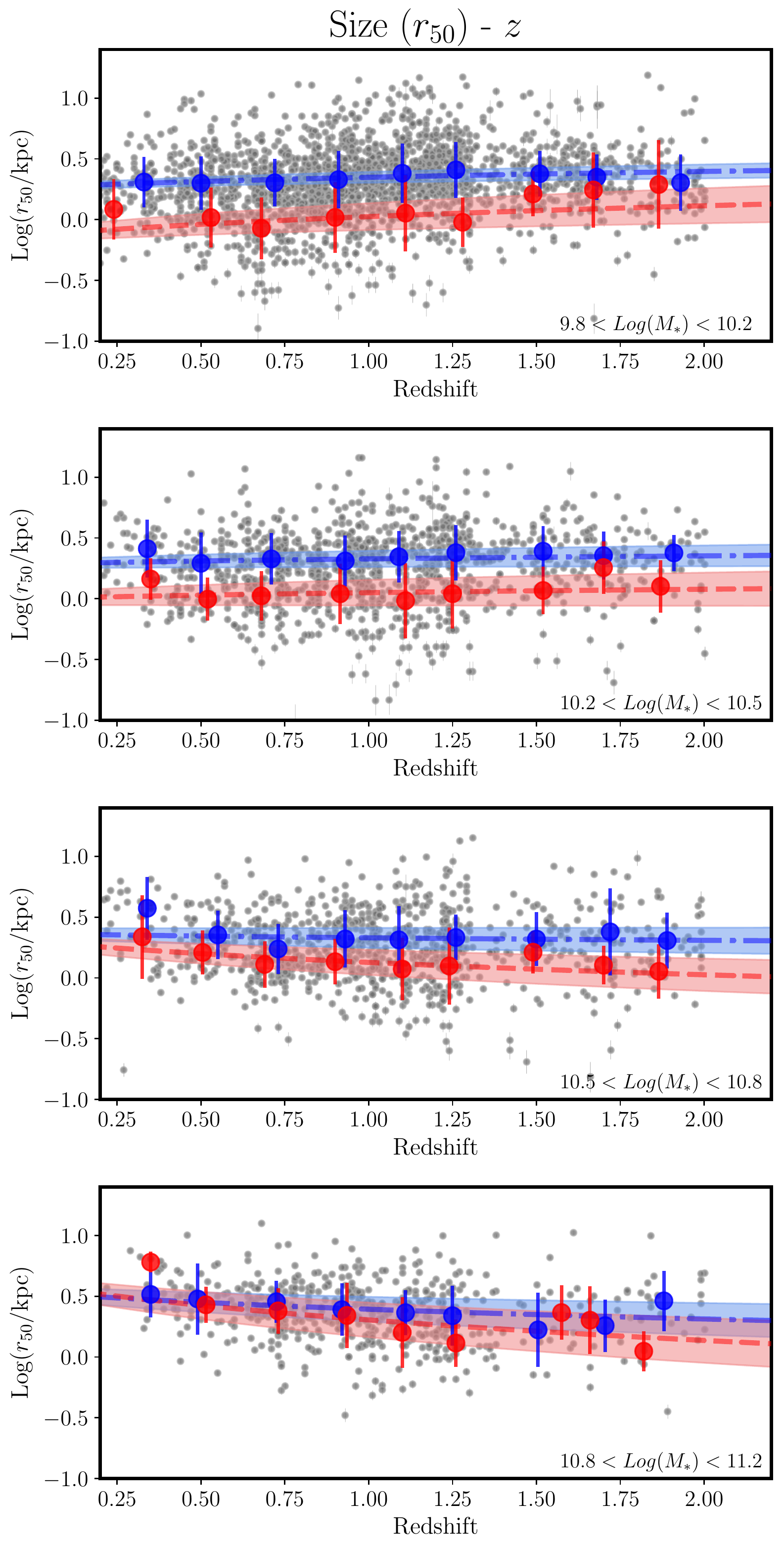}
\vspace{.1 mm}
\centering
\includegraphics[width=0.32\textwidth]{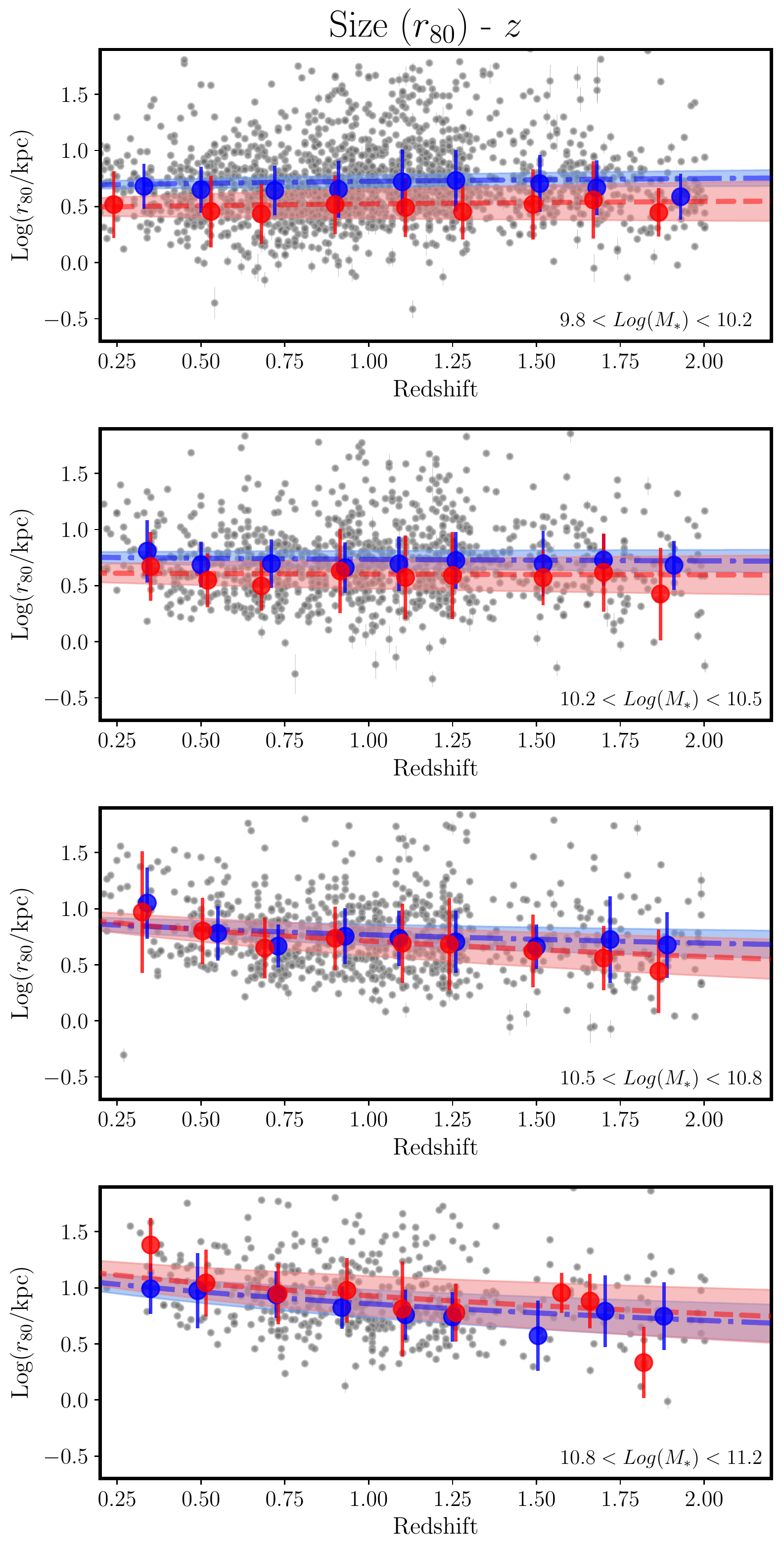}
\caption{\textit{Left Panel}: Size ($r_{20}$) evolution of galaxies with redshift for different stellar masses, increasing from top to bottom panels. The $r_{20}$ sizes of the star-forming galaxies do not change with redshift, however, the size of $r_{20}$ increases for massive quiescent galaxies, and become closer to the $r_{20}$ sizes of the star-forming galaxies at low redshifts. \textit{Middle Panel}: Same as left panel but for $r_{50}$ size evolution of galaxies with at different stellar masses. This evolution is more significant for the quiescent galaxies, in particular massive ones. \textit{Right Panel}: $r_{80}$ size evolution of galaxies at different stellar masses. Massive quiescent and star-forming galaxies show considerable evolution with redshift}
\label{fig12}
\end{figure*}


\subsection{Size Evolution at Fixed Stellar Mass}

In this section, we explore the evolution of the mass-based sizes of galaxies at fixed stellar mass over a redshift range of $0.3<z<2.0$. We emphasize that this does not trace the size evolution of individual galaxies, rather, this compares the sizes of similar-mass galaxies at different cosmic epochs. We split the samples into four stellar mass bins as shown in Figure \ref{fig12} (mass bins increase from top to bottom). Similar to the size-mass relation figures, the median values of the star-forming and quiescent galaxies in each redshift bin are illustrated with the blue and red points, respectively. The evolution of sizes are quantified as $r_{m}\propto (1+z)^{\gamma}$. The best-fit values of $\gamma$ parameter for different sizes are reported in Tables \ref{table3}, \ref{table4} \& \ref{table5}.

The evolution of the $r_{20}$ sizes are shown in left panels of Figure \ref{fig12}. There is no sign for evolution of the $r_{20}$ sizes for the star-forming galaxies in all stellar mass bins, consistent with the size-mass relation shown in Figures \ref{fig8} and \ref{fig9}. There is only a hint that below $10^{10.5} \msun$ the sizes are smaller at lower redshift, perhaps a sign for building a central concentration, however, this trend is weak (i.e. $\gamma \sim 0.3\pm0.1$). For quiescent galaxies with stellar mass of $9.8 < \log (\mstar/\msun) < 10.2$, the evolution of $r_{20}$ sizes at fixed mass is significant but positive ($\gamma = 0.85\pm0.24$), which can be a sign that the low-mass quiescent galaxies at later epochs have more prominent central densities compared to their counterparts at high redshifts. The trend can be insignificant if only $z<1.3$ would be assumed (upper left panel of Figure \ref{fig12}). This needs to be confirmed by a large sample of these galaxies with robust size measurement throughout the whole redshift range. The evolution is negligible for galaxies within $10.2 < \log (\mstar/\msun) <10.5$ with $\gamma = 0.26\pm0.22$. However, for massive quiescent galaxies with stellar masses of $10.8<\log \mstar < 11.2$, the $r_{20}$ sizes decrease with redshift (i.e., $r_{20} \propto (1+z)^{-1.20\pm0.26}$).

The evolution of the half-mass ($r_{50}$) sizes are not significant for massive star-forming galaxies: star-forming galaxies above $10^{10.5} \msun$ evolve in size ($r_{50}$) with redshift ($\gamma=-0.45\pm0.16$). The results are consistent with recent studies based on the mass-weighted radii by \cite{mosleh2017} and \cite{suess2019b}. The evolution reported for the similar mass range by \cite{mosleh2017} is $\alpha=-0.46\pm0.11$. On the other hand, low mass star-forming galaxies (i.e., $\log(\mstar/\msun)<10.5$), have similar half-mass radii at all redshift bins, indicating no size evolution. The slow evolution of $r_{50}$ sizes at low redshifts ($z\leqslant 1$) has been reported previously based on the half-light radii \citep{lilly1998, simard1999, ravindranath2004, barden2005, dutton2011, vanderwel2014, straatman2015}. However, the slope of the $r_{50}$ evolution obtained in this study is smaller than most of these works, most probably because other works focused on light-based sizes. 

Quiescent galaxies (particularly massive ones with $ 10.8 < \log(\mstar/\msun) < 11.2$) show a strong half-mass radii evolution with redshift with $\gamma=-0.96\pm0.24$, again consistent with previous studies \citep{Buitrago2008, williams2010, newman2012, vanderwel2014, Mowla2019b}. This size evolution is weaker at intermediate masses ($\gamma\sim-0.58$ for $10.5<\log \mstar < 10.8$) and, hence, changes depend on the stellar mass. Therefore, the origin of physical processes contributing to these $r_{50}$ evolution for quiescent galaxies might vary for different stellar masses.

The $r_{80}$ size evolution slightly differs from the other ones (see right panels of Figure \ref{fig12} and Table \ref{table5}). The massive star-forming galaxies show higher rate of evolution ($\gamma=-0.42\pm0.15$ and $-0.85\pm0.20$ for the last two massive stellar mass bins, respectively), compared to the $r_{20}$ and $r_{50}$ sizes, which is related to the pivot mass scale that is lower for the $r_{80}$-mass relation than for the $r_{50}$-mass relation. This can be an indication that stellar mass has been build up in the outskirt of the star-forming galaxies with stellar mass of $\gtrsim 10^{10.5} \msun$ at lower redshifts compared to the hight-$z$ ones. As we discuss in Section 6.1, this is consistent with accretion of gas onto the outer regions of these galaxies, leading to inside-out disk growth with cosmic time \citep[see, e.g.,][]{nelson2016}. 

The $r_{80}$ sizes of massive quiescent galaxies evolve as $\gamma=-0.90\pm0.29$, which is similar to the $r_{50}$ size evolution. Again, the $r_{80}$ size evolution of intermediate-mass quiescent galaxies ($10.5 < \log (\mstar/\msun) <10.8$) is with $\gamma=-0.79\pm0.22$ more strongly than the evolution seen by $r_{50}$. This is an indication for the assembly of the stellar masses in the outskirts via minor mergers at large radii \citep{oser2012, matharu2019}. 

\begin{figure}
\centering
\includegraphics[width=0.4\textwidth]{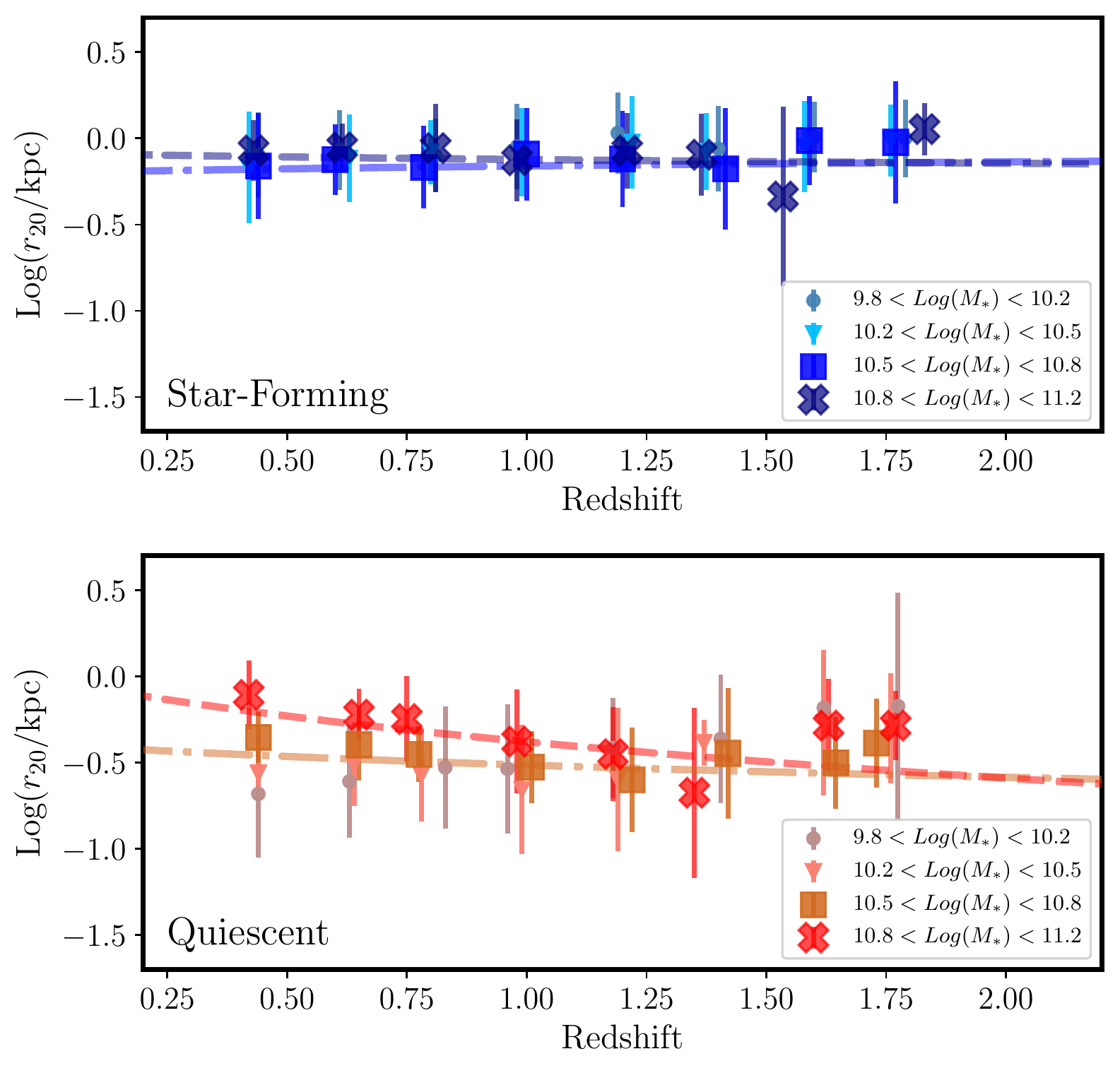}
\hspace{.2 mm}
\centering
\includegraphics[width=0.4\textwidth]{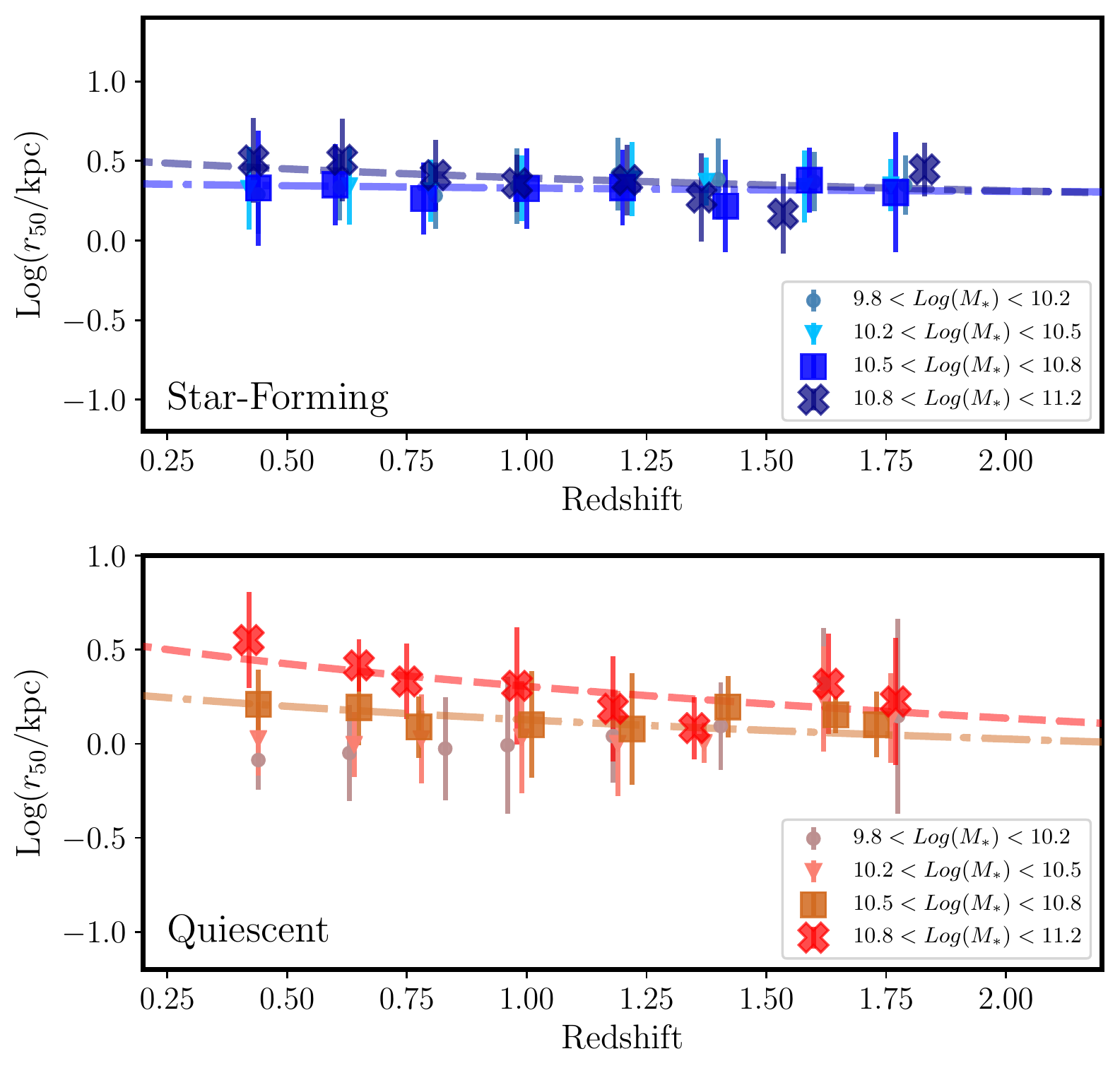}
\hspace{.2 mm}
\centering
\includegraphics[width=0.4\textwidth]{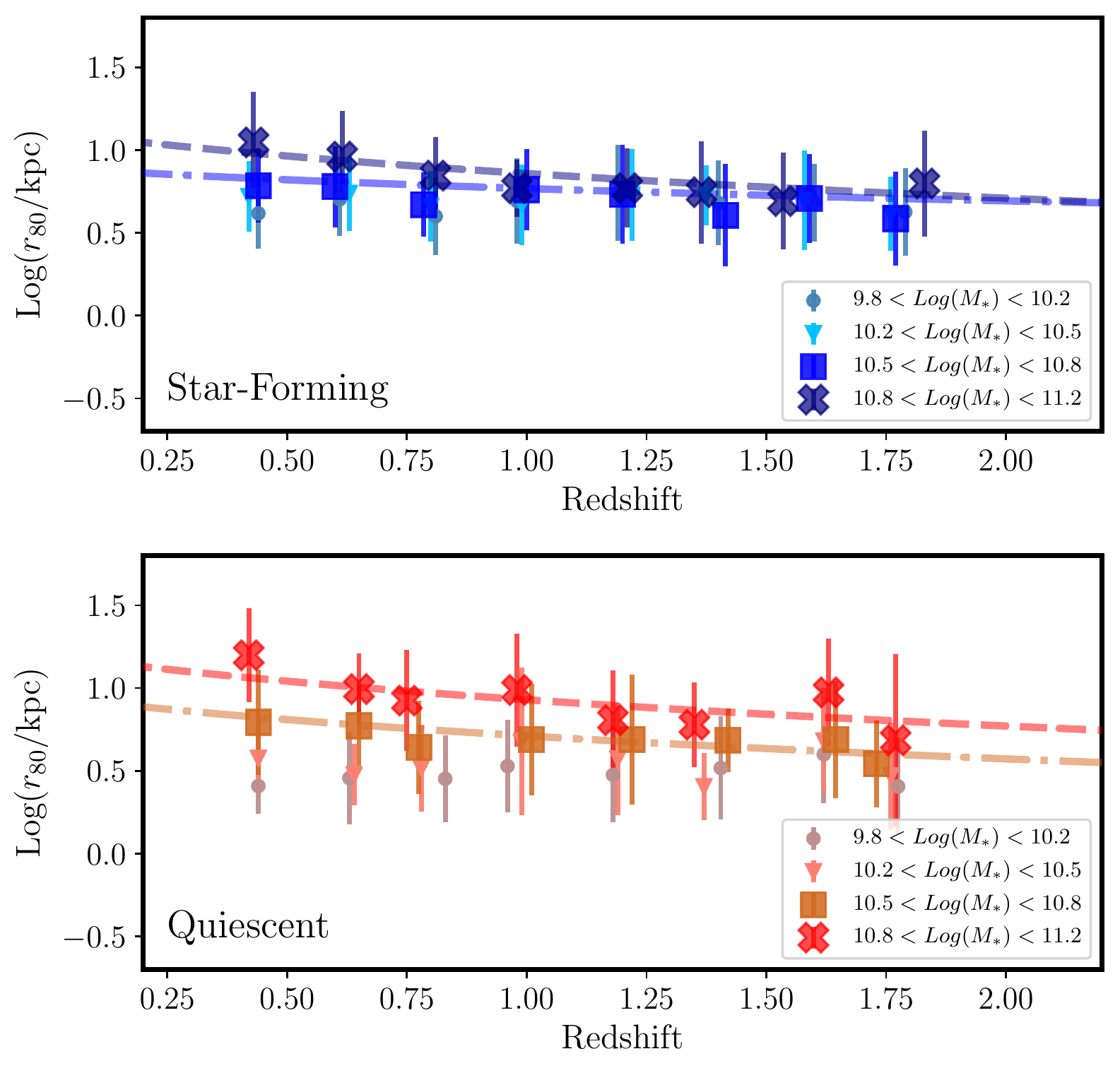}
\caption{Comparison of the size ($r_{20}$, $r_{50}$ and $r_{80}$) evolution of star-forming and quiescent galaxies with redshift at fixed stellar mass. The best-fit size evolution is shown for the two massive stellar mass bins in each panel. }
\label{fig13}
\end{figure}

We summarize these evolutionary trends in Figure \ref{fig13}, where we plot the size evolution of galaxies at fixed mass for star-forming and quiescent galaxies. In all panels, the median of sizes for each stellar mass bin are depicted with different symbols. The best-fit size evolution are also shown for the two most massive bins. This figure highlights that the median size of the star-forming galaxy population only evolves weakly with cosmic time: low-mass star-forming galaxies show little evolution, while higher mass galaxies have slightly larger $r_{50}$ and $r_{80}$ sizes at late epochs. Quiescent galaxies at low masses also have similar sizes at all epochs, while the median size massive quiescent galaxies shows significant evolution with cosmic time. 

\begin{deluxetable*}{ccccccc}
\tablecolumns{5}
\centering
\tablewidth{\textwidth}
\tabletypesize \footnotesize
\tablecaption{The fitting results for the $r_{20}$ size evolution exponent parameter ($\gamma$) in $r_{20}\propto (1+z)^{\gamma}$}\label{table3}
\tablehead{
\colhead{Sample} &
\colhead{$9.8 < \log(\mstar) < 10.2$} &
\colhead{$10.2 < \log(\mstar) < 10.5$} & 
\colhead{$10.5 < \log(\mstar) < 10.8$}&
\colhead{$10.8 < \log(\mstar) < 11.2$}
}
\startdata
 \textbf{SF ($M_{I}$)} & \boldmath{$0.375\pm0.082$} & \boldmath{$0.320\pm0.131$}	 & \boldmath{$0.132\pm0.160$} & \boldmath{$-0.123\pm0.195$} \\
 SF ($M_{II}$) & $0.236\pm0.093$ & $0.014\pm0.170$ & $0.072\pm0.200$ & $-0.645\pm0.279$ \\
 \textbf{Q ($M_{I}$)} & \boldmath{$0.854\pm0.243$} & \boldmath{$0.260\pm0.219$} & \boldmath{$-0.400\pm0.193$} & \boldmath{$-1.197\pm0.259$} \\
 Q ($M_{II}$) & $0.103\pm0.305$ & $-0.095\pm0.289$ & $-1.030\pm0.270$ & $-1.675\pm0.383$ \\
 \enddata
\tablecomments{The best-fit parameters for the stellar $r_{20}$ size relation for star-forming (SF) and quiescent (Q) galaxies, for the first and second methodology described in the text ($M_{I}$ \& $M_{II}$). Note that the results of this paper is based on out fiducial size-measurement method ($M_{I}$, bold numbers).}
\end{deluxetable*}

\begin{deluxetable*}{ccccccc}
\tablecolumns{5}
\centering
\tablewidth{\textwidth}
\tabletypesize \footnotesize
\tablecaption{The fitting results for the $r_{50}$ size evolution exponent parameter ($\gamma$) in $r_{50}\propto (1+z)^{\gamma}$}\label{table4}
\tablehead{
\colhead{Sample} &
\colhead{$9.8 < \log(\mstar) < 10.2$} &
\colhead{$10.2 < \log(\mstar) < 10.5$} & 
\colhead{$10.5 < \log(\mstar) < 10.8$}&
\colhead{$10.8 < \log(\mstar) < 11.2$}
}
\startdata
\textbf{SF ($M_{I}$)} & \boldmath{$0.276\pm0.073$} &  \boldmath{$0.140\pm0.109$} &  \boldmath{$-0.117\pm0.132$} &  \boldmath{$-0.456\pm0.166$} \\
 SF ($M_{II}$) & $0.138\pm0.073$ & $-0.036\pm0.123$ & $-0.169\pm0.159$ & $-0.826\pm0.234$ \\
 \textbf{Q ($M_{I}$)} &  \boldmath{$0.508\pm0.192$} &  \boldmath{$0.162\pm0.176$} &  \boldmath{$-0.578\pm0.172$} &  \boldmath{$-0.959\pm0.238$} \\
 Q ($M_{II}$) & $0.231\pm0.226$ & $0.322\pm0.222$ & $-0.672\pm0.222$ & $-0.844\pm0.321$ \\
 \enddata
\tablecomments{Same as Table 3, but for $r_{50}$ sizes.}
\end{deluxetable*}

\begin{deluxetable*}{ccccccc}
\tablecolumns{5}
\centering
\tablewidth{\textwidth}
\tabletypesize \footnotesize
\tablecaption{The fitting results for the $r_{80}$ size evolution exponent parameter ($\gamma$) in $r_{80}\propto (1+z)^{\gamma}$}\label{table5}
\tablehead{
\colhead{Sample} &
\colhead{$9.8 < \log(\mstar) < 10.2$} &
\colhead{$10.2 < \log(\mstar) < 10.5$} & 
\colhead{$10.5 < \log(\mstar) < 10.8$}&
\colhead{$10.8 < \log(\mstar) < 11.2$}
}
\startdata
 \textbf{SF ($M_{I}$)} & \boldmath{$0.144\pm0.088$} & \boldmath{$-0.078\pm0.124$} & \boldmath{$-0.423\pm0.149$} & \boldmath{$-0.847\pm0.203$} \\
 SF ($M_{II}$) & $0.030\pm0.086$ & $-0.133\pm0.142$ & $-0.389\pm0.186$ & $-1.107\pm0.288$ \\
 \textbf{Q ($M_{I}$)} & \boldmath{$0.123\pm0.227$} & \boldmath{$-0.037\pm0.218$} & \boldmath{$-0.788\pm0.220$} & \boldmath{$-0.900\pm0.294$} \\
 Q ($M_{II}$) & $0.302\pm0.277$ & $0.468\pm0.271$ & $-0.575\pm0.271$ & $-0.436\pm0.374$ \\
 \enddata
\tablecomments{Same as Table 3, but for $r_{80}$ sizes.}
\end{deluxetable*}

\begin{figure}
\centering
\includegraphics[width=0.48\textwidth]{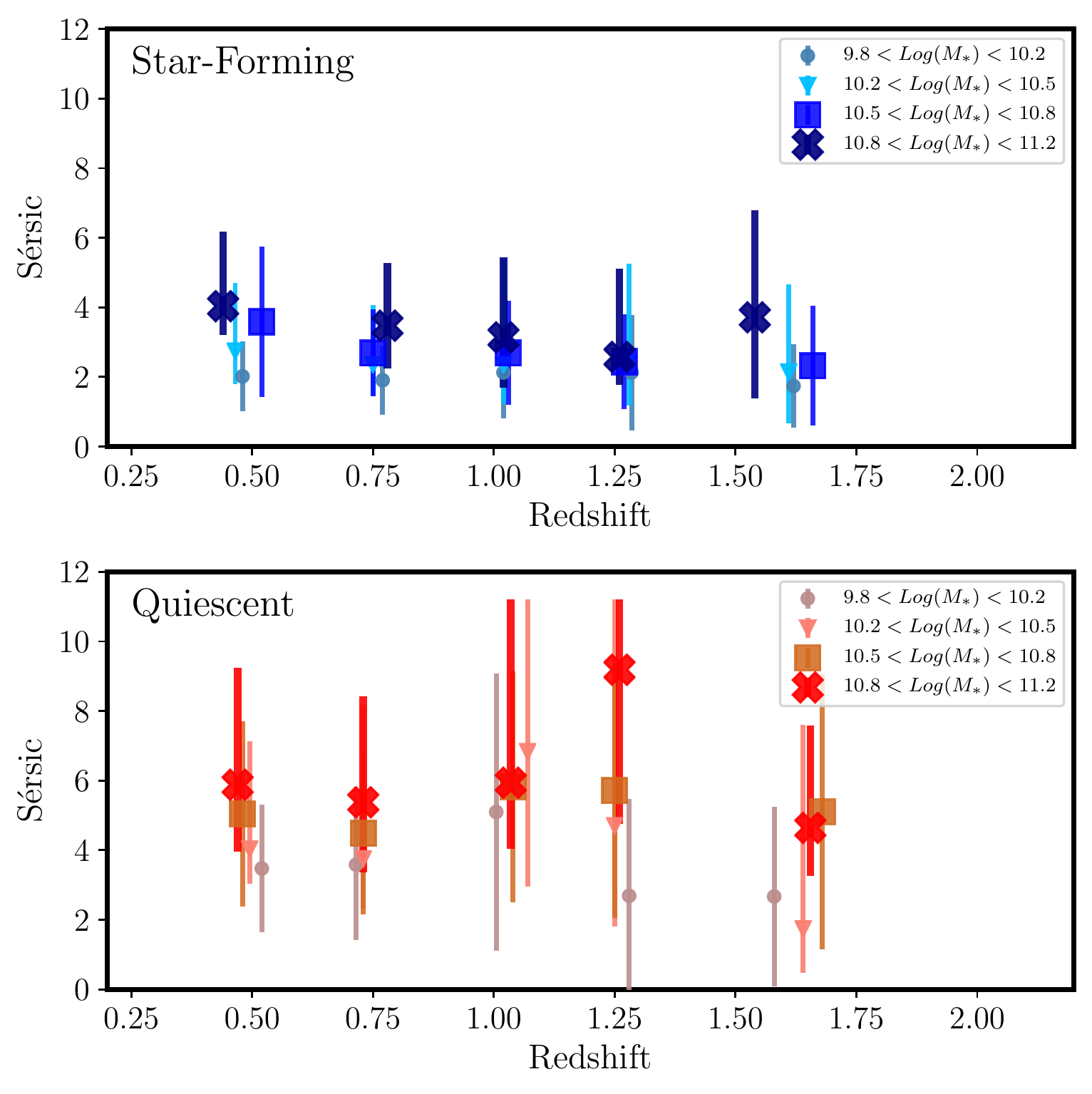}
\caption{Median \ser evolution with redshift for star-forming and quiescent galaxies (top and bottom panels, respectively). Each symbol represents a different stellar mass bin.}
\label{fig14}
\end{figure}

\begin{figure}
\centering
\includegraphics[width=0.48\textwidth]{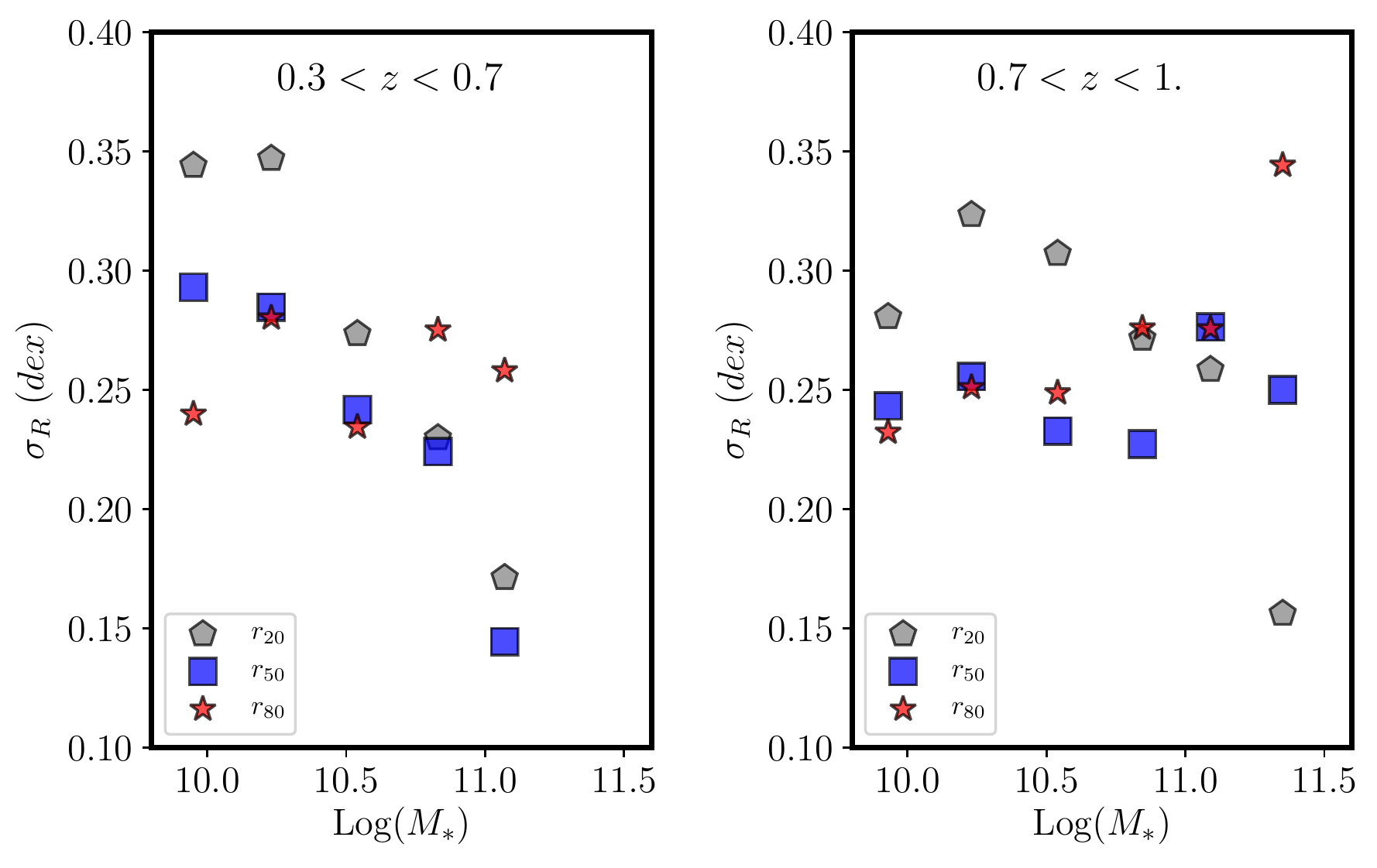}
\hspace{.2 mm}
\centering
\includegraphics[width=0.48\textwidth]{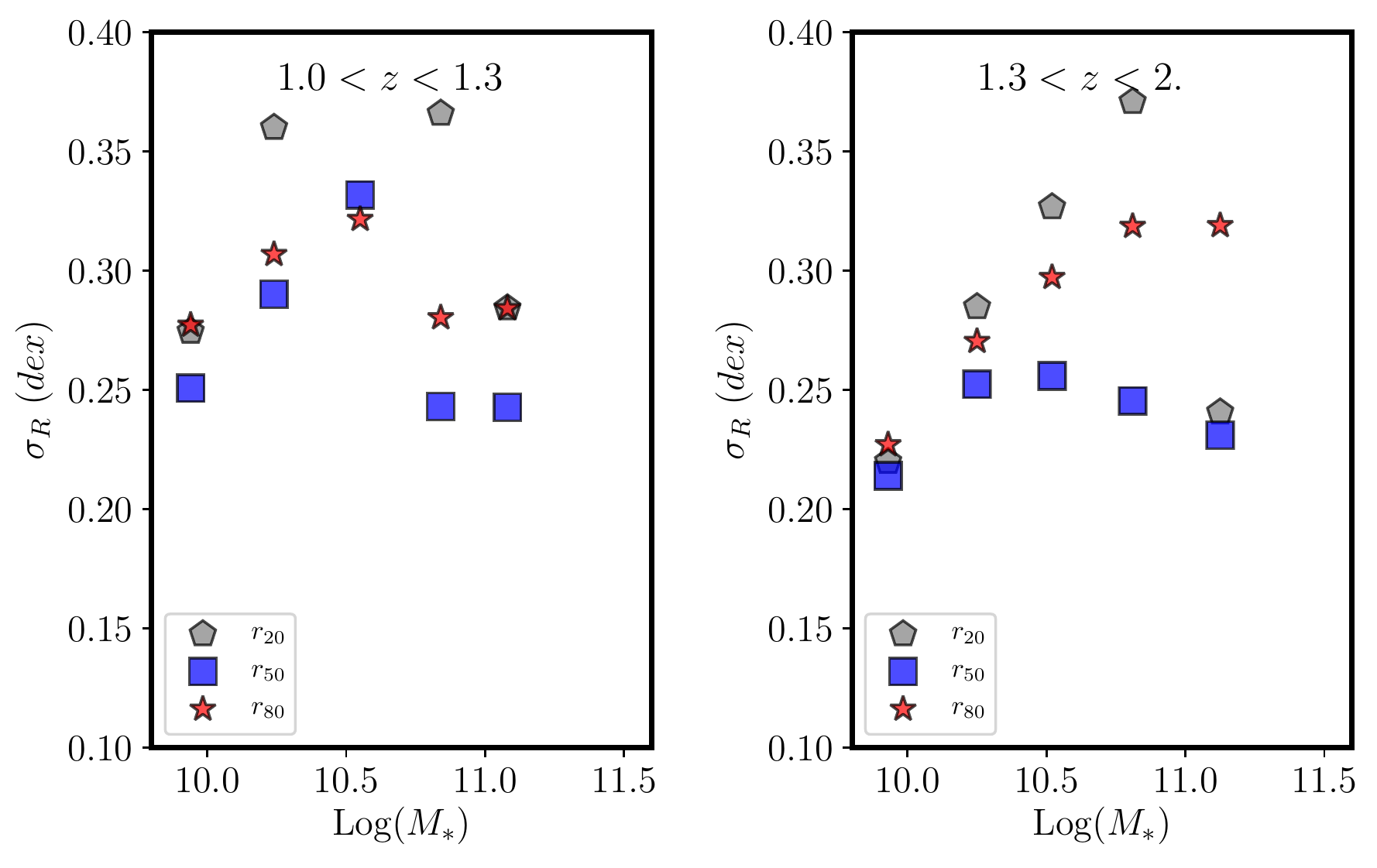}
\caption{Comparing the observed scatter of the size-mass relation as a function of stellar mass for different size definitions. The scatter is slightly reduced as a larger fraction of the stellar mass density profile is considered, i.e. moving from $r_{20}$ to $r_{80}$. The plot is shown for different redshift bins increasing from top left to bottom right panel.}
\label{fig15}
\end{figure}

\subsection{Evolution of the \ser Index at Fixed Stellar Mass}

We show in Figure \ref{fig14} the redshift evolution of the median \ser index for four different mass bins (same bins as previous figures). Star-forming galaxies and quiescent galaxies are shown in the top and bottom panel, respectively. By construction, the \ser index is tightly correlated with the previously shown sizes, i.e. $r_{20}$, $r_{50}$ and $r_{80}$ are related to $n$ via Equations \ref{eq:r20} and \ref{eq:r80}. Therefore, it is not surprising that Figure \ref{fig14} shows a consistent picture with Figures \ref{fig8}-\ref{fig13}.

Specifically, Figure \ref{fig14} shows that the median \ser index for low-mass star-forming galaxies does not evolve significantly between redshift 0.3 to 2.0. The median \ser index is $n\approx2$, consistent with what is expected for a disk-dominated galaxy with a small bulge component. The median \ser index for higher-mass, star-forming galaxies ($\log (\mstar/\msun)>10.5$) increases with cosmic time: above $z\sim1$, $n$ is roughly 2 (consistent with lower-mass galaxies), and increase at $z<1$ to about $4$.

The median \ser index for quiescent galaxies is consistent with no evolution, with $n\approx4-6$ at all masses and redshifts. There is weak trend that higher mass galaxies on average have a higher \ser index. 


\subsection{Scatter of the Size-Mass Relation}

In this section, we briefly investigate the scatter of the size-mass relation. From a physical point of view, the distribution of sizes at fixed stellar mass is of interest because it can be related to the angular momentum, velocity dispersion, stellar age and metallicity of galaxies \citep{scott2017, li2018, wu2018b, rosito2019, diaz2019, walo2020}. In addition this physical origin of the scatter, effects of projection and galaxy orientation can also produce scatter \citep[e.g.,][]{price2017}. Furthermore, the measured scatter in sizes at a given mass is a combination of the true intrinsic scatter of the underlying size-mass relation and of the measurement uncertainties of the sizes and stellar masses \citep[e.g.][]{vanderwel2014}. Here, we only consider the measured scatter and postpone a more detailed modeling of the scatter to a future publication. 

We show in Section 5.1 that the scatter of the size-mass relation of star-forming and quiescent galaxies depends on the choice of size definition. The scatter decreases toward using radii which encompass larger fractions of the total stellar mass (i.e., $r_{80}$). This has been also shown by \cite{miller2019} for sizes based on light profiles. In Figure \ref{fig15} we quantify the scatter of all galaxies on the size-mass relation for different redshift bins. In general, the scatter of all galaxies increases if $r_{20}$ sizes is used. The $r_{20}$ is more sensitive to the \ser index and hence using this radii, can be related to the formation of the central regions and star-formation activity. There is a hint that the scatter of sizes depends on the stellar mass, however, this is not conclusive based on the current analysis.


\section{Discussion}

We present in this paper detailed measurements of the stellar mass surface density distribution within galaxies. Our galaxy sample lies in the redshift range of $z=0.3-2.0$ and has stellar mass above $\log (\mstar/\msun)>9.8$. In the previous section, we quantify how the sizes $r_{20}$, $r_{50}$ and $r_{80}$ (containing 20\%, 50\%, and 80\% of the stellar mass) depend on stellar mass (size-mass relation) and redshift (size evolution at fixed stellar mass with cosmic time). In this section, we discuss the size evolution of individual galaxies and uncertainties in our measurements.

\subsection{Interpretation of the Size Evolution of Star-Forming Galaxies}

We now use $r_{20}$, $r_{50}$ and $r_{80}$ measurements to explore how galaxies change their stellar mass density profiles with cosmic time. From this analysis, we will shed light on the physical processes that could be responsible for driving the stellar mass growth on spatially resolved scales. We first focus on star-forming galaxies. Recently, several studies have measured how the star-formation rate density is distributed within galaxies at $z=0.3-2.5$ \citep[e.g.,][]{morselli2019, nelson2016, tacchella2015a, tacchella2018}. These studies found that galaxies with $\log (\mstar/\msun)<10.8$ on the star-forming main sequence have flat specific star-formation rate profiles, indicating that galaxies grow self-similarly. Only at higher masses ($\log (\mstar/\msun)>10.8$), galaxies have a reduced specific star-formation rate in their cores relative to their outskirts \citep{tacchella2015a}. 

These measurements of the spatial distribution of star formation is consistent with what we find for the evolution of the mass-based sizes (see Figure \ref{fig13}): star-forming galaxies have similar sizes (all size definitions) at all stellar masses and all redshifts. This is a direct consequence of the self-similar growth. Only for the highest mass bin (Figure \ref{fig13}), we find that $r_{50}$ and $r_{80}$ are increasing from $z\approx1$ to $z\sim0$. Again, this is consistent with inside-out growth, where specific star-formation rates in the outkirts are higher than in the centers as expected from inside-out growth.

The constancy for $r_{20}$ with mass and redshift does not mean that galaxies do not increase their stellar mass in their cores: galaxies actually building up dense cores while star forming, as we can see by looking at the evolution of the \ser index (see Figure \ref{fig14}). Massive, star-forming galaxies have typically a higher \ser index than lower-mass, star-forming galaxies ($n\approx3-4$, increasing towards $z\sim0$). This is consistent with previous studies that show that massive galaxies assemble dense cores already at early epochs \citep{dokkum2010, vandokkum2014, saracco2012, mosleh2017, barro2016} and bulge formation on the star-forming main sequence \citep{lang2014, tacchella2015a, tacchella2018, tadaki2017}. 

The pivot mass scale of the size-mass relation of star-forming galaxies, tracing where the low-mass slope $\alpha\approx0$ transitions to a high-mass slope $\beta\approx0.2-0.4$, decreases from $\log (\mstar/\msun)\approx10.8$ at $z\sim2$ to $\log (\mstar/\msun)\approx10.3$ at $z\sim0$ (see Figure~\ref{fig11}). This indicates that this transition from self-similar growth to inside-out growth takes place at low masses towards more recent epochs. 

Theoretical studies can be used to shed more light onto the physics itself from these observations. The size evolution of galaxies, and in particular the size-mass relation (normalization, slope and scatter) provide important constrain for numerical simulations \citep[e.g.][]{furlong2017, genel2018, rodriguez2019}.
Cosmological zoom-in simulations can reproduce the aforementioned, observed flat sSFR profiles at lower stellar masses on average \citep{tacchella2016b}. In more detail, \cite{tacchella2016} shows that in these simulations, galaxies actually oscillate about the main sequence ridgeline, where bulges are being built at the upper envelope of the main sequence (sSFR increase to the centers), while outskirts are being built when galaxies below the star-forming main sequence (sSFR decraese to the centers) -- on average, the sSFR are flat. This is consistent with tidal effects during mergers, misaligned accretion (i.e. counter-rotating gas accretion), and violent disk instabilities leading to gas compaction and the formation of a central spheroidal component \citep{Hernquist1989, Sales2012, Dekel2014, Zolotov2015}.

\subsection{Interpretation of the Size Evolution of Quiescent Galaxies}

Moving to quiescent galaxies, there are two interesting questions we want to focus on: (i) do galaxies change their stellar structure when they cease their SFRs, i.e. when moving through the ``green valley''? (ii) do galaxies grow in size once they are quiescent? As we discuss below, we cannot provide satisfying answers with our measurements alone. Future studies that include constraints on the stellar ages and SFHs of the galaxies will help with answering these questions. 

As we mention in the Introduction, although today's star-forming and quiescent galaxies have different morphologies, it does not necessarily imply that the morphology of a galaxy needs to change when ceasing its star formation since the progenitors of today's quiescent galaxies are star-forming galaxies at higher redshifts \citep[e.g.][]{tacchella2019, park2019}. When comparing the morphology of star-forming and quiescent galaxies, a complication arises due to M/L gradients, hence, effects such as disk fading need to take into account \citep[e.g.][]{carollo2016}. Since we work with mass-based quantities, our comparison is straight forward. Comparing the sizes of star-forming and quiescent galaxies, we find that $r_{20}$ is on average larger for star-forming galaxies than for quiescent galaxies at all masses and epochs. The average size $r_{50}$ is larger for star-forming than for quiescent galaxies at low masses, and reverses when going to higher masses. We find a similar trend for $r_{80}$. We note that at fixed mass and $r_{50}$ sizes, galaxies with higher Sersic index (n) is expected to have a smaller $r_{20}$  but larger $r_{80}$ sizes. However, this can not be the only reason why quiescent galaxies are larger than the star-forming galaxies at the high-mass end of $r_{50}$ and $r_{80}$-mass relations, as described in the following paragraphs.

Even though we have these observations at hand, we cannot conclude how galaxies' sizes evolve when they cease their star formation. This is because these are average statements over the whole population: the quiescent population at any epoch consists of galaxies that ceased their SFRs at a range of earlier cosmic epochs. Therefore, the evolution of quantity averaged over the quiescent galaxy population includes evolutionary effects of quiescent galaxies themselves (such as mergers) \textit{and} the addition of newly quenched galaxies \citep[``progenitor bias'',][]{vandokkum1996, carollo2013a, poggianti2013a, belli2015, fagioli2016}. These newly quenched galaxies stem from the star-forming galaxy population, which itself evolves, and during quenching, some morphological evolution might also take place. 

Nevertheless, we find that $r_{50}$ and $r_{80}$ is larger for quiescent galaxies than for star-forming galaxies at high masses $\log (\mstar/\msun)>11$. Under the assumption of negligible size growth when star-forming galaxies cease their star formation, individual quiescent galaxies indeed need to increase $r_{50}$ and $r_{80}$, i.e. they need to add mass in the outskirts. In addition, the pivot mass in the size-mass relation for quiescent galaxies decreases from $\log (\mstar/\msun)\approx10.8$ at $z\sim2$ to $\log (\mstar/\msun)\approx10.2$ at $z\sim0$ (see Figure~\ref{fig11}, similar to star-forming galaxies), which is similar to the mass where the fraction of star-forming galaxies is $50\%$, indicating that the pivot mass reflects a transition from in-situ, dissipational to ex-situ, dissipationaless growth \citep[see also][]{mowla2019a, zahid2019}. This is also consistent with the decline in the scatter of sizes toward high stellar masses (see Section 5.4). Finally, also the IllustrisTNG simulations show that the ex-situ mass fraction of $\approx0.2$ decreases from $\log (\mstar/\msun)\approx10.9$ at $z=1.5$ to $\log (\mstar/\msun)\approx10.6$ at $z=0$ \citep[see also][]{Pillepich2018, tacchella2019}. In summary, there are indications that merger play a role in setting the size for the most massive systems, which is also consistent with the abundance of slow rotators at the massive end of the mass function \citep[e.g.][]{bezanson2009, cappellari2013, dokkum2015, Cappellari2016, vandesande2017}. 

Furthermore, we show in Section 5 that quiescent galaxies have smaller $r_{20}$ sizes compared to the star-forming ones at all stellar masses, reflecting the higher concentration of these galaxies. For quiescent galaxies, the relation is steeper at high masses above the pivot mass ($M_p$). The existence of the $r_{20}$-mass relation with a larger slope for massive quiescent galaxies suggests that as the total stellar mass increases, these objects have larger cores and higher stellar mass concentrations. Below the pivot stellar mass, the relation tends to be flattened, at least for the redshift bins of $z<1$. For the higher redshift bins ($z>1$), the trends seem to be reversed, though this might be affected by the incompleteness in our sample (see Section 4). Nevertheless, if this is real, then the contribution of recently quenched galaxies might affect the trend and cause this relation to be anti-correlated. If different quenching mechanisms can be assumed for galaxies with different stellar masses \citep[e.g., mass and environmental quenching described in ][]{penglilly2010}, then the low-mass galaxies -- which quenched via environmental effects -- could have been least affected by structural reshaping. Hence, the central regions of the low-mass quiescent galaxies would be similar to star-forming galaxies. Studying the environment and age of these systems can possibly help with testing this scenario. 

\subsection{Uncertainties in Parameter Measurements}

In this study, we created the stellar-mass maps of the galaxies with $\log(\mstar/\msun) \geqslant 9.8$ at $0.3 \leqslant z\leqslant 2$ to measure their mass-based structural parameters. Converting multi-wavelength images to the stellar mass maps and measuring the structural parameters is associated with several uncertainties. First, in our methodology, all pixels are treated independent during the SED fitting. Usually, pixels within central regions have sufficient S/N. However, the stellar mass estimates of pixels in the outer regions with low S/N will be associated with large uncertainties. Using the pixel binning methods used in some studies such as \citet{wuyts2012} degrades the resolution and introduces step-like profiles, which is not suitable for analyzing the morphology of galaxies. Some authors \citep[e.g.,][]{lang2014, morselli2019} tried to use $H_{160}$-band images and multiplied them to the estimated (pixel-binned and smoothed) $M/L$ maps. As discussed in \cite{sorba2015}, this can possibly reduce outshining effects and, hence, might change the shape of the true stellar mass profiles. The concerns on the effects of low S/N pixels on the results of this study is examined by means of the simulated galaxies (Appendix \ref{sec:appendixB}).  Although, the outer regions of the stellar mass maps are slightly noisy (see for instance Figure \ref{fig6}), the overall shape of these maps (in particular, the outer regions) are within the expected range of the stellar masses densities. In addition, conversion of the 2D stellar mass maps to the 1D density profiles, reduces the effect of these low S/N pixels in the outskirt of galaxies, while recovering the true profiles. Our approach of using a formalism which looks into a wide range of models also helps to mitigated issues regarding the fitting procedure. Therefore, we believe that the results of this paper are robust concerning low S/N pixels in the outskirts.

The second concern is related to the choice of the SFH model of the SED fitting. In this work, an exponentially declining SFH is assumed. Different assumptions such as additional random burst, constant or delayed SFH can alter the estimated stellar masses. This has been tested in \cite{mosleh2017} for the 1D stellar mass profiles. Although there might be some systematic on the stellar mass estimates \citep[e.g.,][]{madau2014, leja2019}, \cite{mosleh2017} could not find any effects on the mass-based size measurements. Testing different SFH models is beyond the scope of this work and this has to be examined in future studies. Nevertheless, we expect that this will not have significant effects on the general results of this paper, reminding the fact that the stellar masses used in this work is the integral of the SFH, which includes remnant and mass returned to the interstellar medium. It is worth noting that the variation of the initial mass function (IMF) as a function of radius or redshift  \citep[see e.g.,][ and references there in]{labarbera2016, eftekhari2019} can also affect the stellar mass maps. In this work, we only assume \cite{chabrier2003} for all radii and redshift and the possible consequences due to the changes of the IMF is not studied. Clear pictures of the dependence of the IMF on redshift and radius will be valuable for testing its effect on the structural analysis.

Another source of uncertainty is due to the effect of dust on the stellar mass maps, in particular for massive star-forming galaxies and also its dependence on inclination of galaxies \citep{hemmati2015}. Understanding radial variation of the dust attenuation \citep{wang2017, nelson2016b, tacchella2018} and its effect requires detail spectroscopic analysis for a large sample of galaxies. Hence, using a limited number of wavelength bands at short rest-frame filters can be problematic in estimating the dust attenuation. Perhaps, a better coverage at short rest-frame wavelengths can be a first step for testing the effects on our final results. We have examined this by exploiting the new ultra-violate (UV) imaging data from the Hubble Legacy Fields (HLF) on the GOODS-South field \citep{whitaker2019}. We apply our method for deriving the stellar mass maps for about 300 galaxies in common with 3D-HST at $0.5<z<1.3$, using 13 bandpass observations in HLF, including rest-frame UV. This allows us to estimate the stellar mass and $A_{\rm V}$ profiles of these galaxies and comparing the profiles with rest-frame UV and without rest-frame UV coverage. Overall, we could not see any significant differences between the stellar mass profiles of star-forming galaxies. The largest difference ($\sim0.2$ dex) is found for the central regions (within 1 kpc) of the most massive bin in the redshift range of $1<z<1.3$. This might affect $r_{20}$ sizes. Therefore, in general, we expect that the effect of dust might not have a notable consequences on our final results (except for the most massive star-forming galaxies at $z>1$). Nonetheless, future studies with a better wavelength coverage are required for testing this further.

Number of galaxies (especially massive ones) decline towards lower redshifts ($z\lesssim0.5$). Hence, in the lowest redshift bin, the analysis might suffer from low statistics. Using the five fields of 3D-HST has reduced this issue, but still a large sample of galaxies is required for tracing the structural evolution at fixed mass at these low redshifts \citep[such as][]{damjanov2019}. 

In addition, galaxies are observed to consist of bulge and disk components out to high-$z$ \citep{margalef2016}. It has been shown that fitting profiles with a single component models can change the slope of size-mass relation \citep{mosleh2013, bernardi2014}. However, using simulated data, \cite{mosleh2013} showed that for high redshift galaxies ($z\sim1$), single \ser model is sufficient for recovering their true sizes \citep[see also][]{davari2016}. Therefore, the result of this paper is not expected to be affected assuming single \ser model for high-$z$ galaxies. In spite of that, sizes of galaxies at the lowest redshift range of $z\sim 0.2-0.4$, might be slightly under/over-estimated. Investigating this issue is beyond the scope of this paper, but needs to be considered in complementary studies. 

Finally, our strategy was to get the stellar mass density profiles as straight forward as possible without resorting to the light profile fitting in each filter prior to the SED fitting \citep{szomoru2013}, or any additional assumption on the $M/L$ profiles \citep[e.g.,][]{suess2019a}. This will reduce many sources of uncertainties on converting the light to mass profiles and measuring structural parameters. Our first and second methodology described in Section 4 and Appendix \ref{sec:appendixA} are complementary to each other, and they have been tested via creating a large sample of mock galaxies to ensure that the results are not biased. To the best of our knowledge, this is one of the most comprehensive tests for these sort of studies. The methodology also helps to estimated the PSF-corrected morphological parameters based on the stellar mass maps. However, as discussed in the text, using SED fitting method for deriving mass-based structural parameters of galaxies beyond  $z>1.3$ could be accompanied with another uncertainties, due to the combination of effects such as low S/N and lack of rest-frame NIR coverage. Hence, based on the currently available data, this redshift range should be treated with caution. From our simulated objects, the stellar mass limit of $\log(\mstar/\msun) \sim 9.8$ is obtained, below which the uncertainties on the parameter measurements increase significantly (see Appendix \ref{sec:appendixB}).

\section{Conclusions}

In this paper, we created the stellar mass maps of a sample of $\sim5557$ galaxies up to $z=2$ from the CANDELS/3D-HST observations to measure mass-based structural parameters. We utilized pixel-by-pixel SED fitting for constructing these maps. These maps are used to derive mass-based sizes ($r_{20}$, $r_{50}$, and $r_{80}$ that enclose 20\%, 50\%, and 80\% of the total stellar mass, respectively) and \ser indices. These measurements are made available to community as an online table (Table \ref{tableA1}). 

The methodologies used in this study for deriving the stellar mass maps and structural parameters are tested via creating a sample of $\sim3000$ mock galaxies. Based on the results of these simulations, we show that the analysis is robust for galaxies with stellar mass $>10^{9.8}\msun$ and up to the redshift of $z\sim 2.0$. Beyond this redshift, the robustness of mass maps and sizes are reduced. 

We constrain the size-mass relations for all mass-based $r_{20}$, $r_{50}$ and $r_{80}$ sizes. The scatter of size-mass relations depends on the size definition; it reduces considerably for $r_{80}$ sizes. We use the different size definition to understand how centers and outskirts are building up stellar mass as a function of cosmic time and redshift. 

The sizes of star-forming galaxies are similar at all mass and redshifts, regardless of the size definition ($r_{20}$, $r_{50}$ and $r_{80}$). Only $r_{50}$ and $r_{80}$ of the most massive star-forming galaxies increase weakly towards lower redshifts. This leads to a picture that below a pivot stellar mass of $\sim10^{10.5}\msun$, star-forming galaxies have similar stellar mass profiles and grow self-similarly. This is consistent with flat sSFR profiles from other observations. Above the pivot stellar mass of $\sim10^{10.5}\msun$, the outskirts of star-forming galaxies grow slightly faster than the inner region. This is consistent with sSFR profiles that are rising towards the outskirts. 

Quiescent galaxies at low masses also have similar sizes at all epochs. At high masses, the size-mass relation steepens. Furthermore, massive quiescent galaxies on the other hand show strong size evolution at fixed mass, regardless of the size definition. We argue that progenitor bias and accumulation of the stellar masses via minor/major mergers contribute to the evolution.

\begin{acknowledgments}
We thank the anonymous referee for the comments that helped to improve the manuscript. This work is based on observations taken by the 3D-HST Treasury Program (GO 12177 and 12328) with the NASA/ESA HST, which is operated by the Association of Universities for Research in Astronomy, Inc., under NASA contract NAS5-26555. S.T. is supported by the Smithsonian Astrophysical Observatory through the CfA Fellowship. 

\end{acknowledgments}

\appendix

\section{Size Measurements} \label{sec:appendixA}

As described in the text, we use two methods for measuring the mass-based sizes of galaxies. The first method is based on fitting the 1D stellar mass density profile. We extensively discuss the details of this methodology and the results in the main text. In the following, we describe the 2D fitting method of the stellar mass maps of galaxies. This complementary approach allows us to assess possible systematics in our final results of the stellar-mass based sizes.

\subsection{Estimation of the Background Level}

Before describing this method, it is important to mention that estimating the background level is crucial for this fitting approach. Therefore, we need to estimate the background level of the stellar mass maps. The 3D-HST mosaic images are background subtracted, however, these images have surface brightness limits down to which extended sources can be detected. We find the corresponding stellar mass density limit to those surface brightness limits around each source. We measure the 1$\sigma$ scatter of the background pixels in the $48'' \times 48''$ postage stamps for all filters. We perform SED fitting of the background assuming the same redshift as the galaxy. The estimated stellar masses of the background pixels are then perturbed randomly by their errors assuming a Gaussian distribution and are then added to the background pixels in the galaxies' stellar mass maps. We express that the method has been tested extensively using simulated objects as discussed in Appendix \ref{sec:appendixB}. 

\subsection{2D Profile Fitting Method ($M_{II}$)}

\begin{figure*}
\centering
\includegraphics[width=0.32\textwidth]{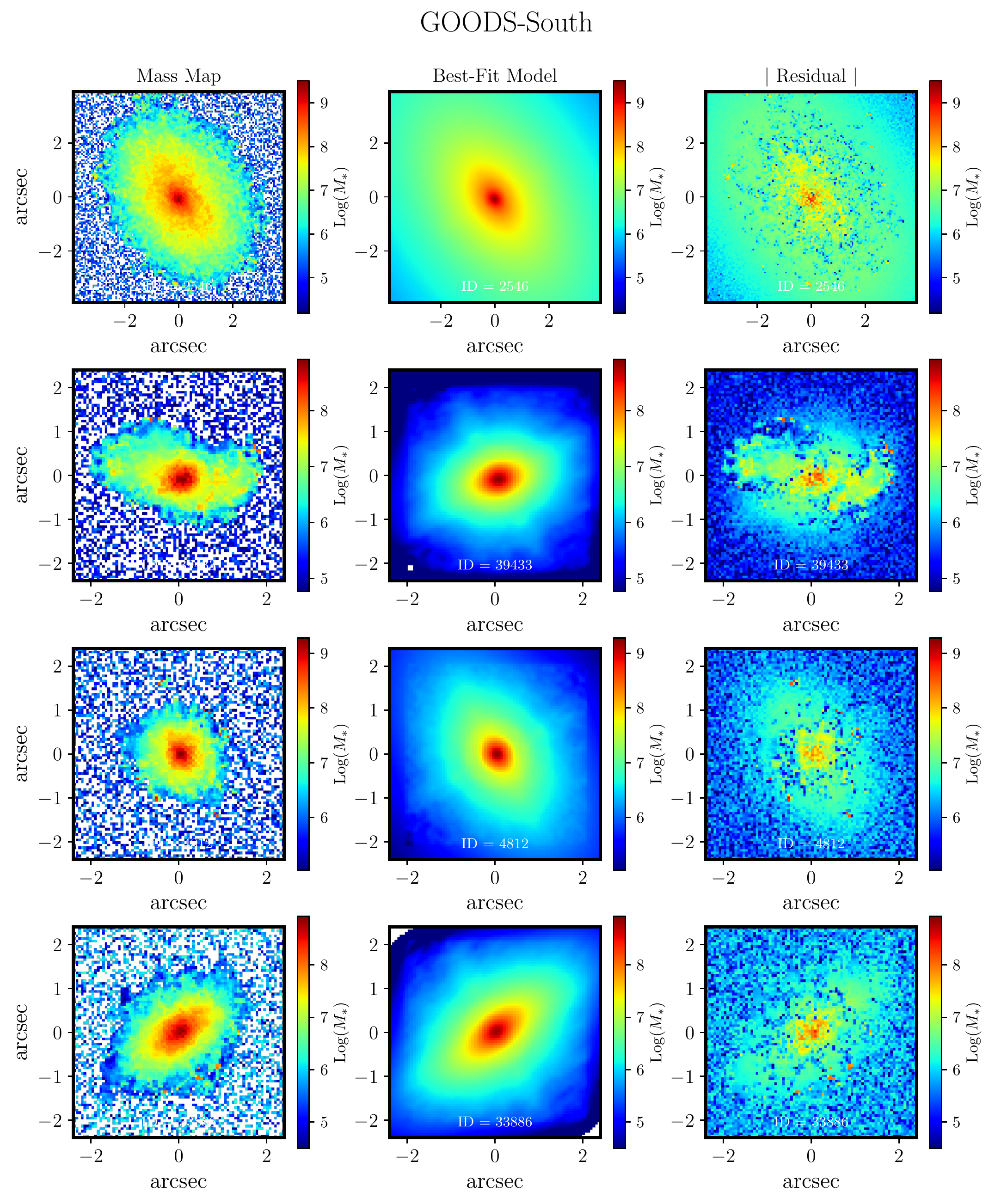}
\vspace{.1 mm}
\centering
\includegraphics[width=0.32\textwidth]{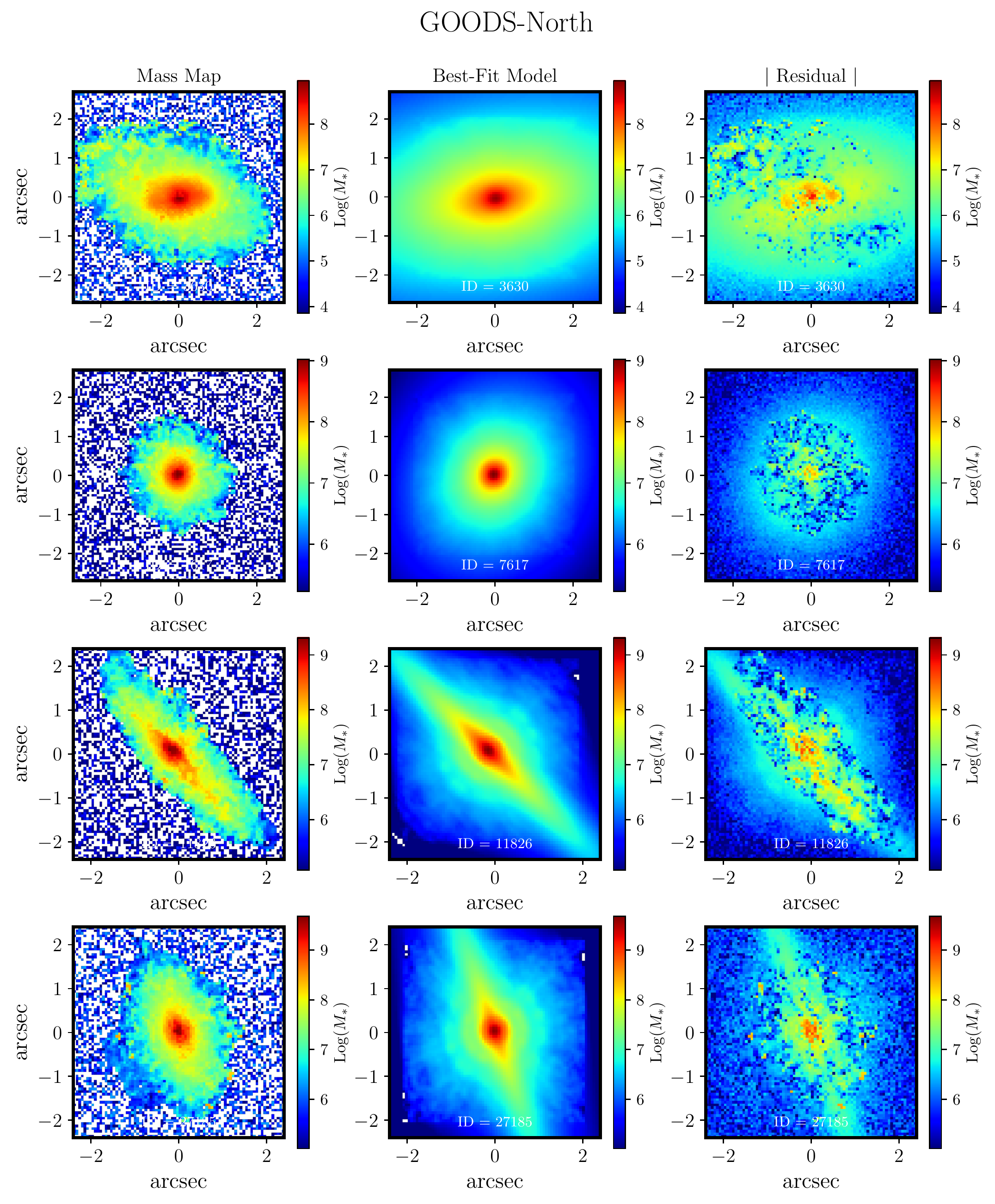}
\vspace{.1 mm}
\centering
\includegraphics[width=0.32\textwidth]{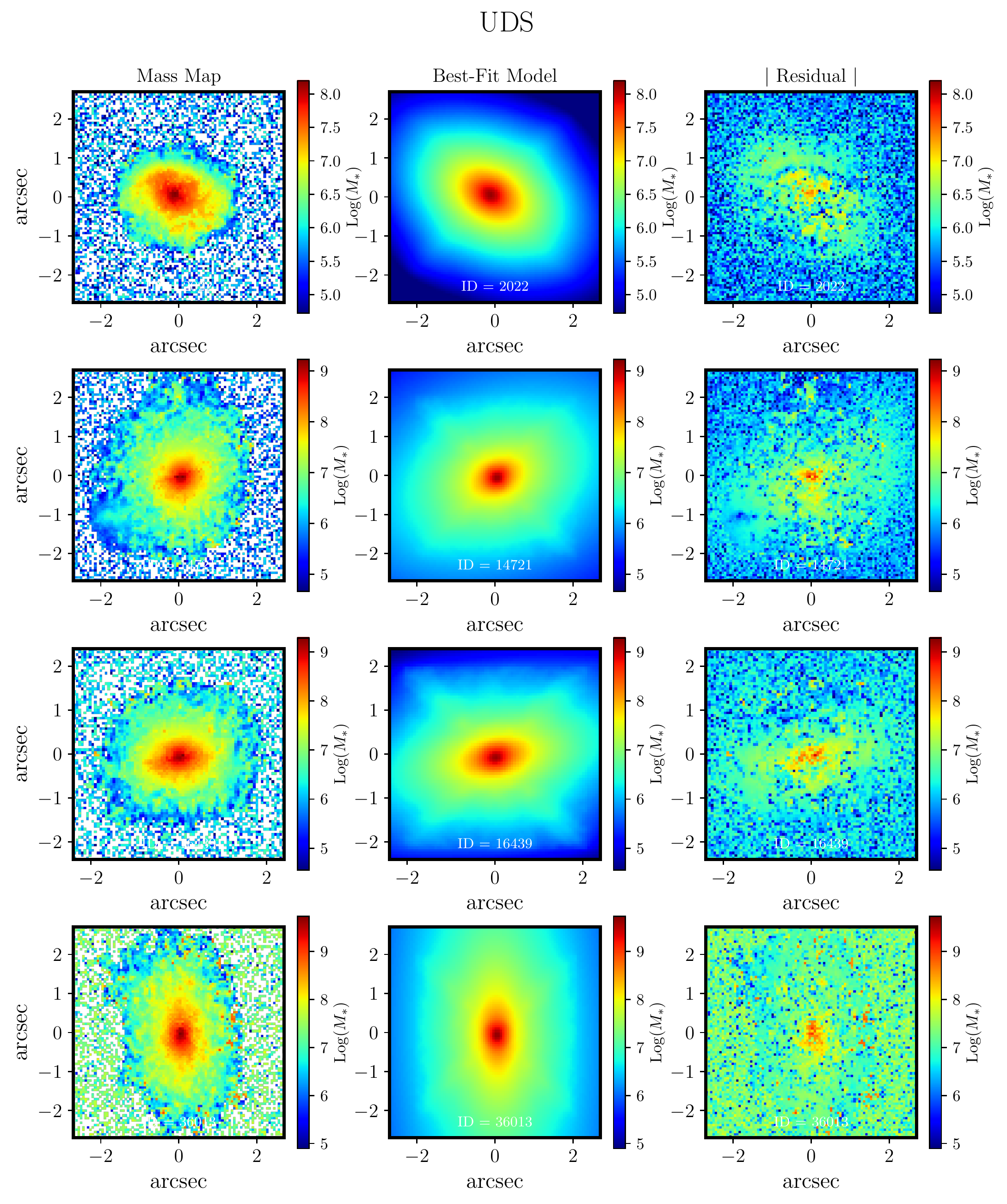}
\caption{The best-fit \ser models for all objects in Figure \ref{fig3} based on the 2D method described in Appendix \ref{sec:appendixA}. The mass map, the best fit model, and the absolute residual are shown from left to right for each object. In this method, the 2D stellar mass maps are fitted with a single \ser models using \texttt{GALFIT} and assuming the $H_{160}$-band PSF. The robustness of this technique has been tested via simulations in Appendix \ref{sec:appendixB}.} 
\label{figA1}
\end{figure*}

In this method, we use the 2D stellar mass maps and utilize \texttt{GALFIT} v3 \citep{peng2010} to perform two-dimensional $\chi^{2}$ fitting. We use the $H_{160}$-band PSFs since all images are convolved to the $H_{160}$ resolution. We therefore assume that the resolution has not been affected during the pixel-by-pixel SED fitting process. We follow a similar procedure as in \cite{mosleh2012} for treating the background as a free parameter during the fitting process. In Figure \ref{figA1}, we show for the same objects as in in Figure \ref{fig3} the mass maps, the best-fit 2D \ser models and the associated residuals. In order to test the reliability and limits of this method, a set of mock galaxies is required and we refer the readers to Appendix \ref{sec:appendixB} for more details. As explained and examined in that section, the fraction of galaxies with reliable sizes drops significantly (by $\sim 20\%$) beyond redshift of $z\sim1.3$ as the uncertainties of the stellar mass maps increase. For these sources, \texttt{GALFIT} fails to converge or it converges to the set boundaries. Below this redshift, on average only $\sim6$ percent of the sources are returned with unreliable size estimates. We should note that in order to be consistent with the first method and reduce the effects of ellipticity, the sizes are also circularized \citep[e.g.,][]{trujillo2006a}.

\begin{figure*}
\includegraphics[width=0.48\textwidth]{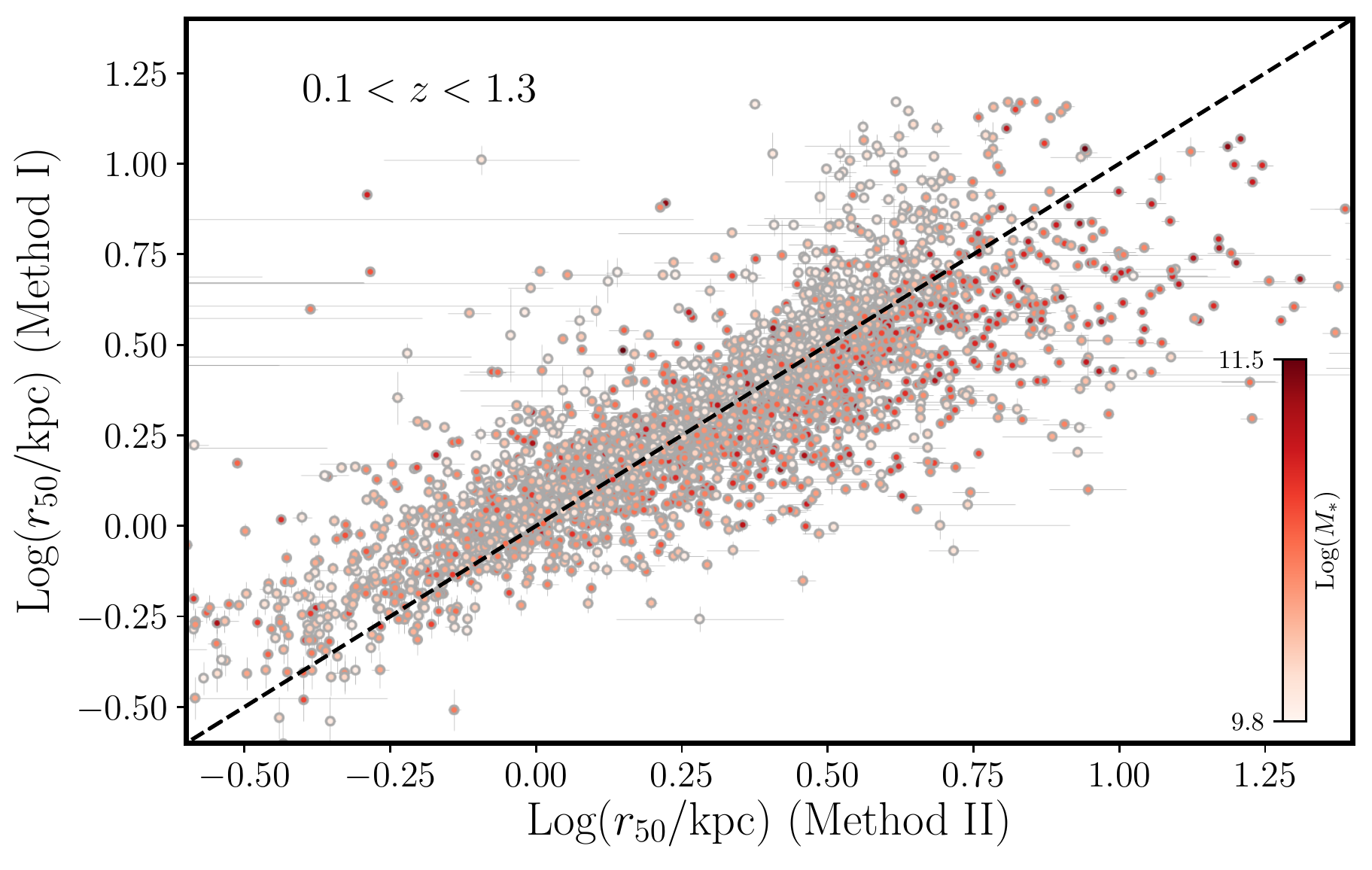}
\hspace{2. mm}
\includegraphics[width=0.48\textwidth]{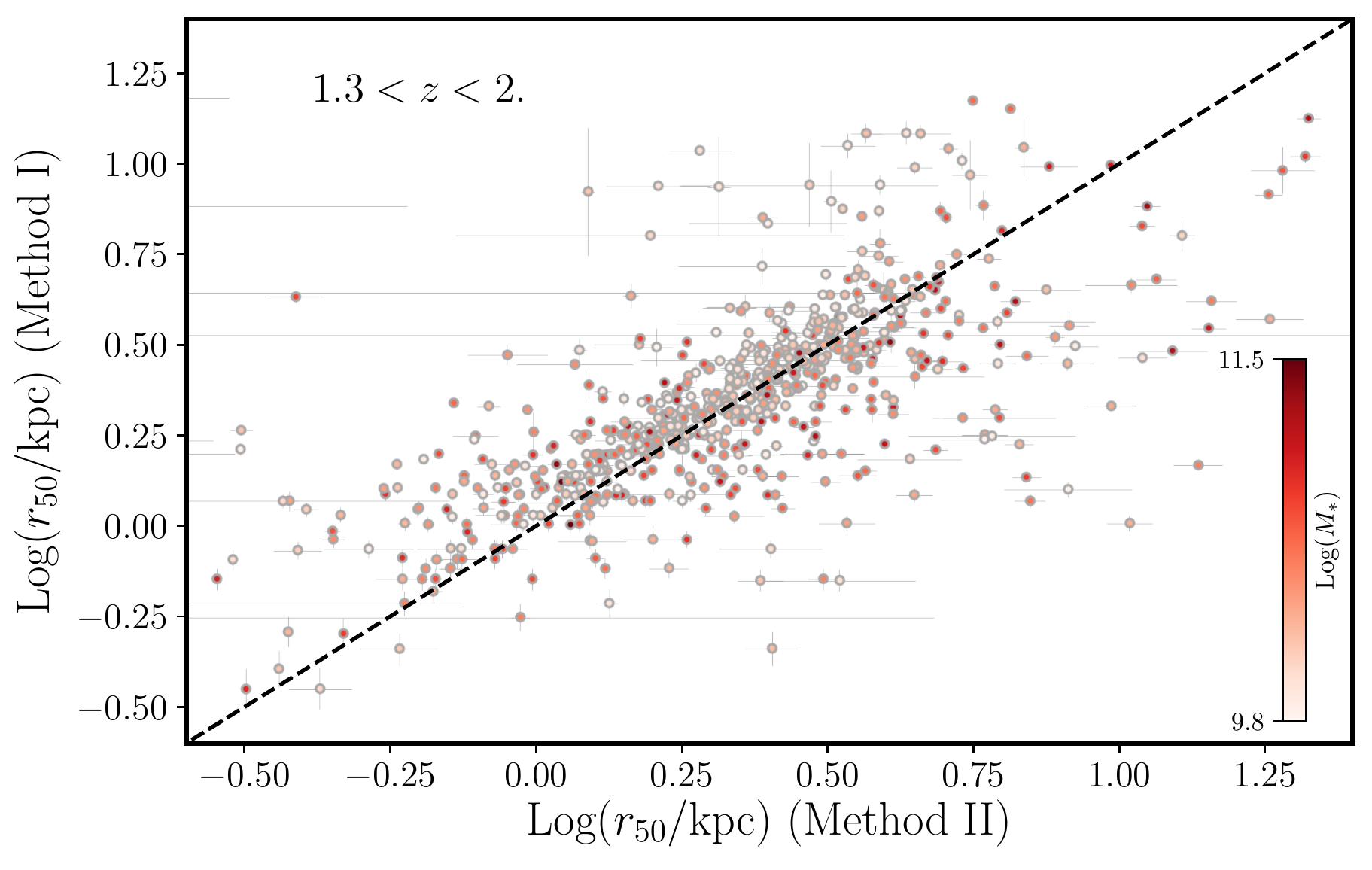}
\caption{Comparison between half-mass sizes ($r_{50}$) from the our first and second method for two different redshift bins (left and right panels). The differences between sizes estimated based on these methods are small ($\sim 0.01$ dex) and sizes from 2D and 1D method are consistent with each other.}
\label{figA2}
\end{figure*}

\begin{figure}
\centering
\includegraphics[width=0.48\textwidth]{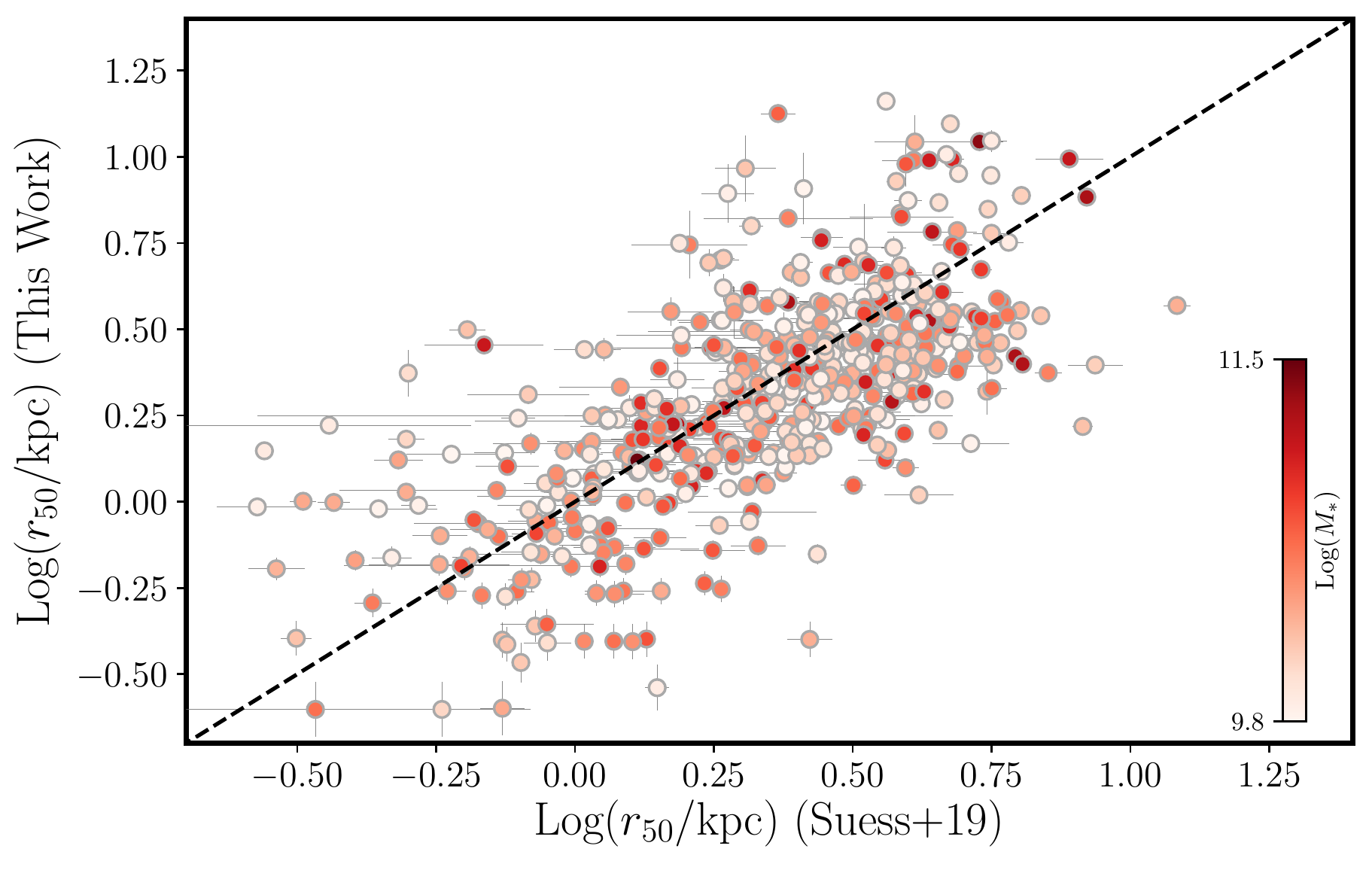}
\caption{Comparison between half-mass radii of $\sim700$ galaxies in common between \cite{suess2019a} and this work. The overall difference between both measurements is small ($0.04$ dex), but the scatter with $\sim0.2$ dex is substantial.}
\label{figA3}
\end{figure}

A comparison between the $r_{50}$ sizes estimated from the first and the second method is illustrated for the two redshift ranges in Figure \ref{figA2}. In general, the results from both methods are consistent with a median offset of $\Delta (r_{50}) < 0.02$ dex for both redshift ranges and a scatter of $\leqslant 0.11$ and $\leqslant 0.13$ dex for the low- and high-redshift bin, respectively. We have also compared the half-mass sizes for $\sim700$ galaxies that are in common with \cite{suess2019a} in Figure \ref{figA3}. There is a tendency that the sizes from \cite{suess2019a} are larger than the ones from this work, though, the overall differences between both measurements is small ($0.04$ dex), but the scatter with $\sim0.2$ dex is substantial.

\section{Simulations} \label{sec:appendixB}

We perform extensive simulations to test the reliability of our methods for deriving stellar mass maps and sizes of galaxies. The results of this Section will help to constrain the stellar mass and redshift limits to which our methods are robust. We first describe how the models are created and then we test the size measurement methods using these simulated data.

\begin{figure*}
\centering
\includegraphics[width=\textwidth]{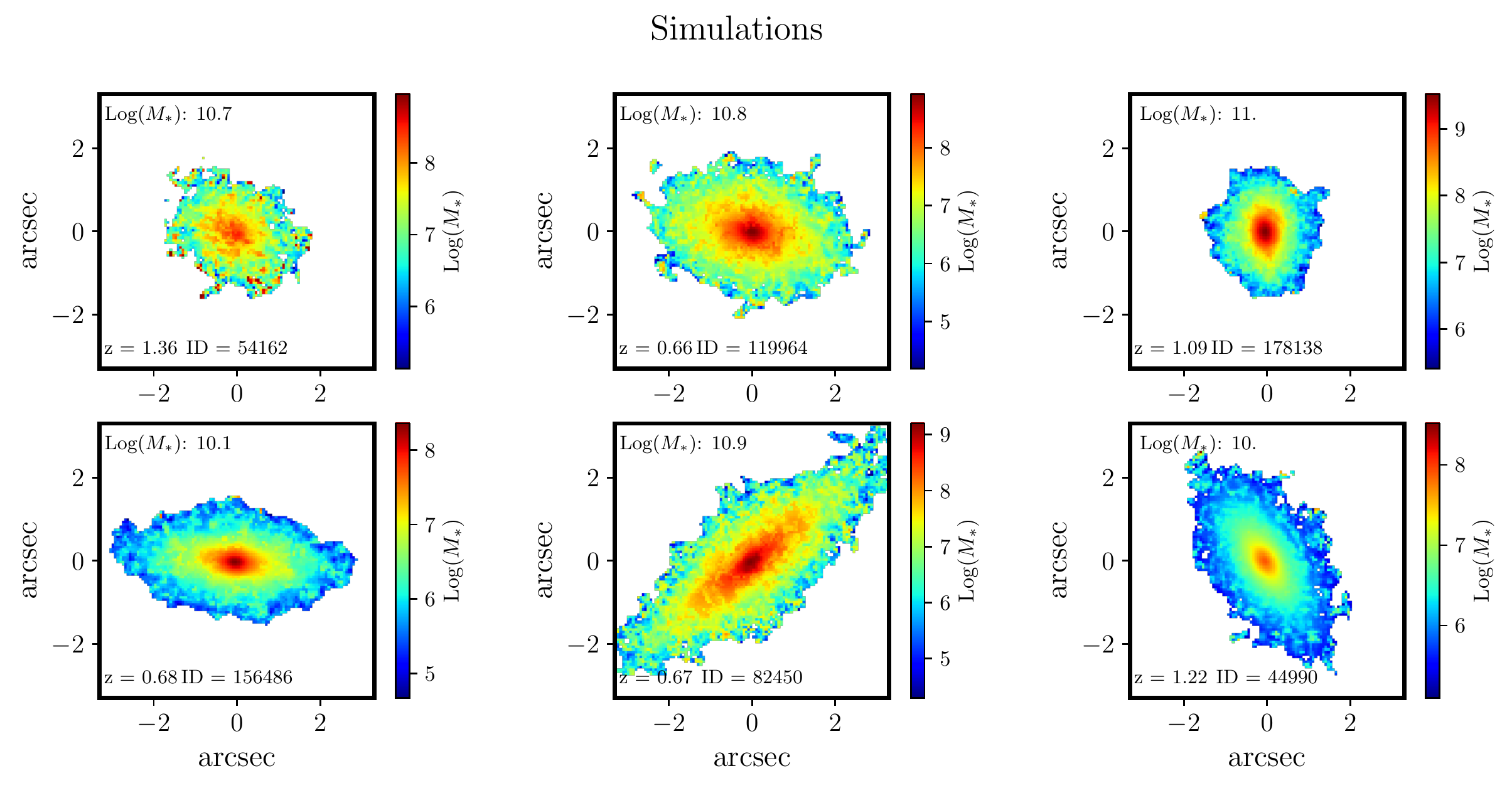}
\caption{The stellar mass maps for a few simulated galaxies. As described in text (Appendix \ref{sec:appendixB}), these galaxies are created assuming single \ser models and without any color gradients in their light profiles. This will help to check the reliability of the methods for constructing the stellar mass maps and of the recovery of the mass-based sizes. }
\label{figB1}
\end{figure*}

\begin{figure*}
\centering
\includegraphics[width=0.48\textwidth]{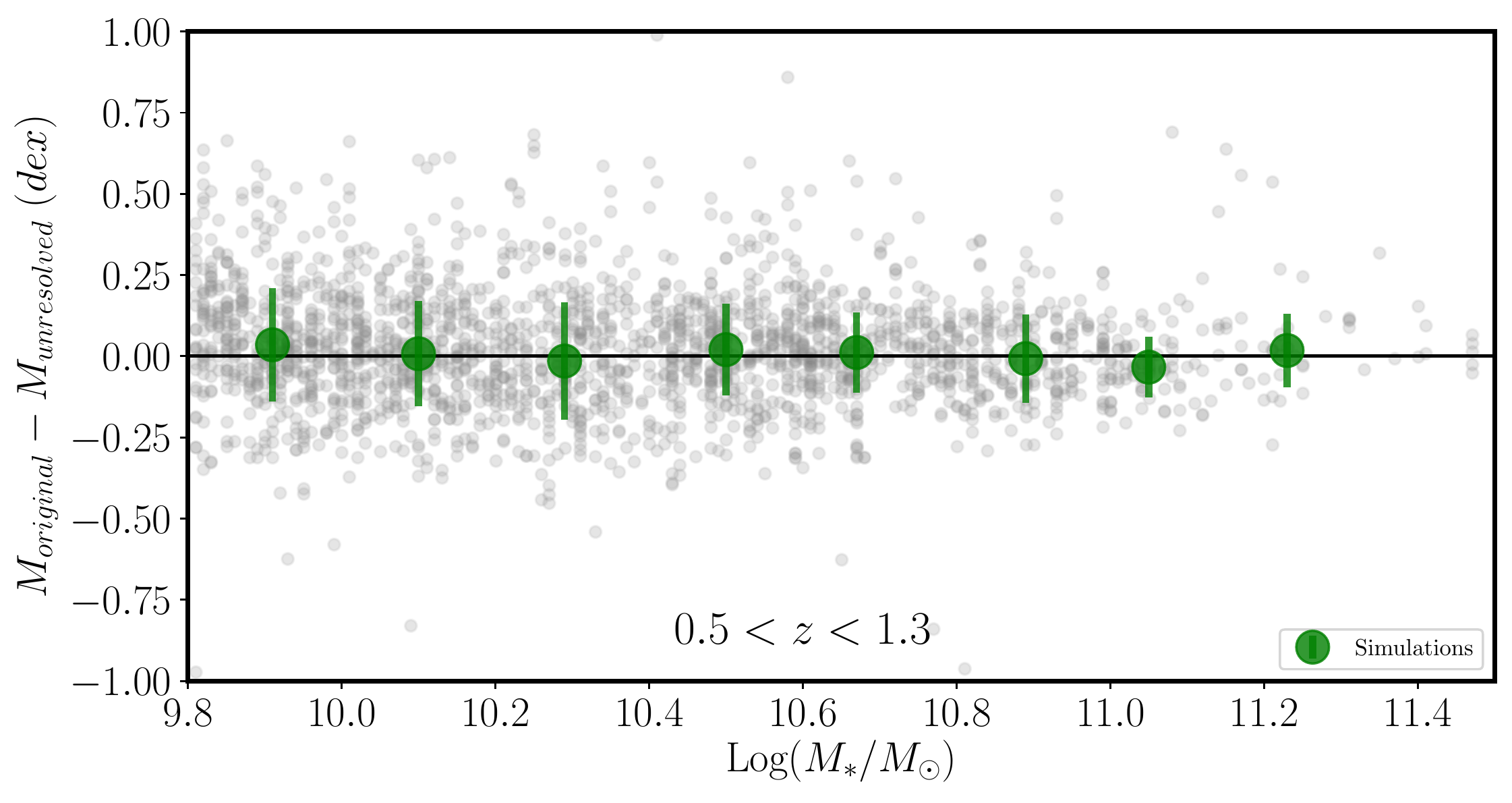}
\hspace{2. mm}
\centering
\includegraphics[width=0.48\textwidth]{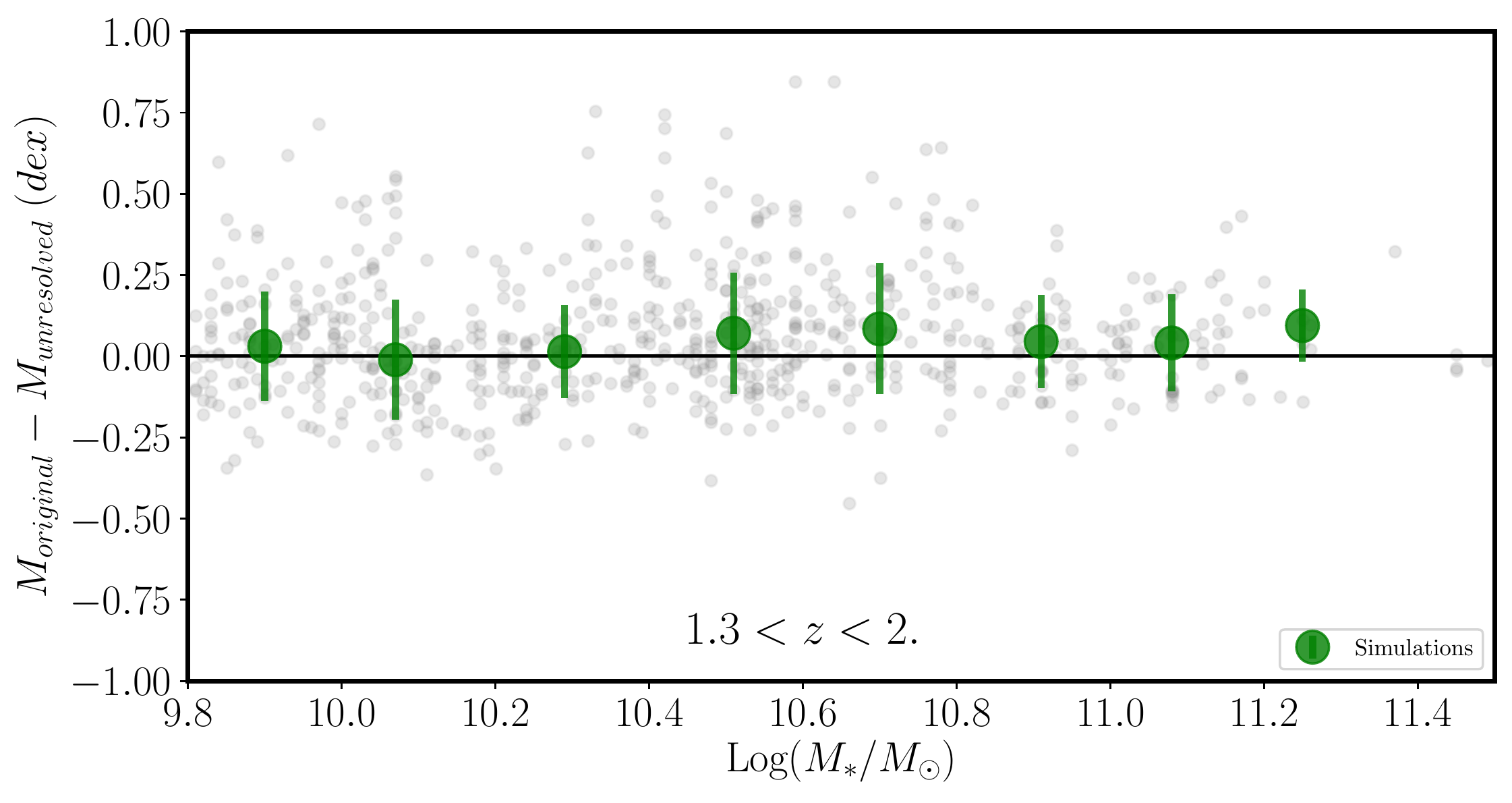}
\vspace{2. mm}
\centering
\includegraphics[width=0.48\textwidth]{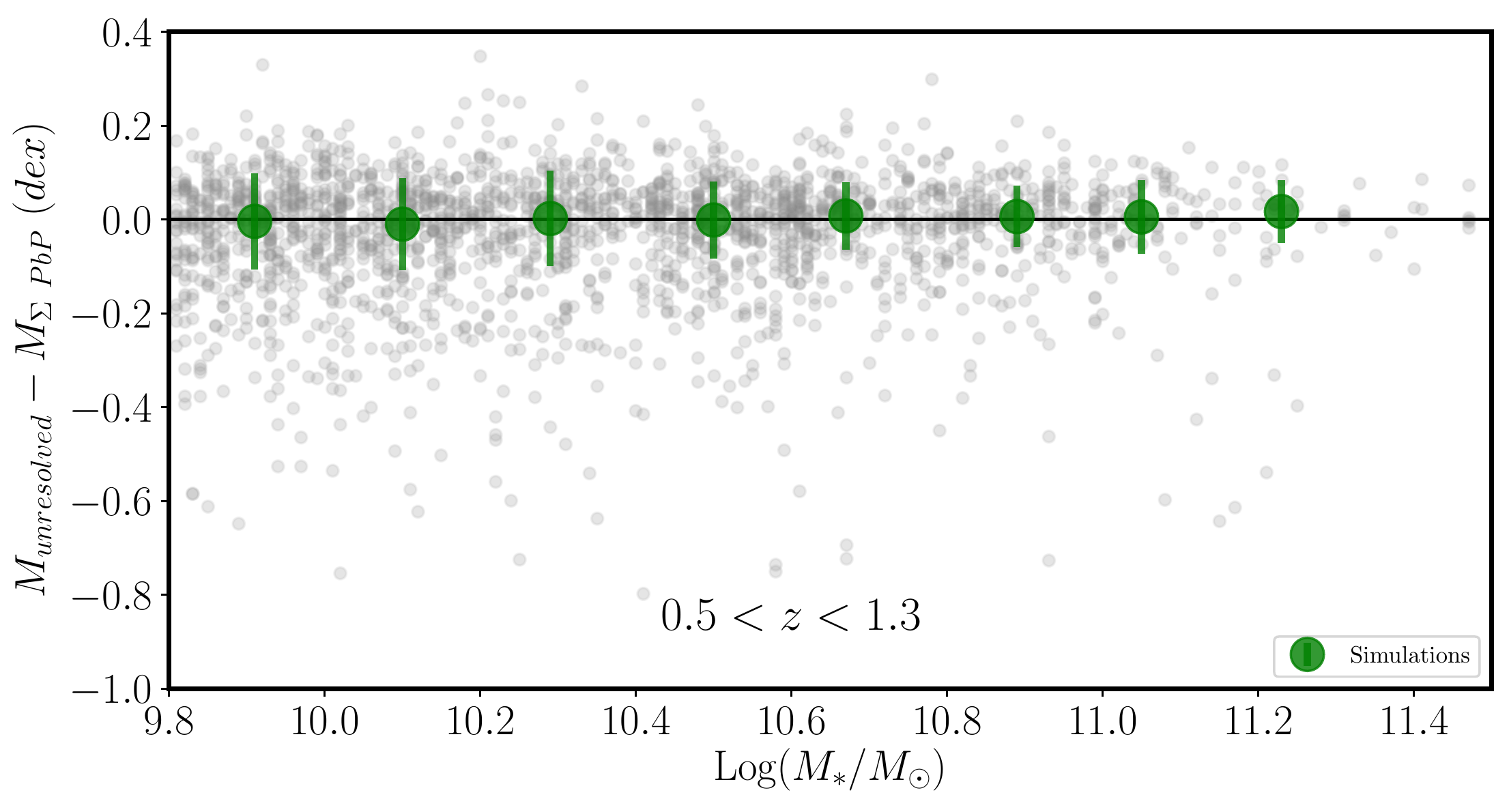}
\hspace{2. mm}
\centering
\includegraphics[width=0.48\textwidth]{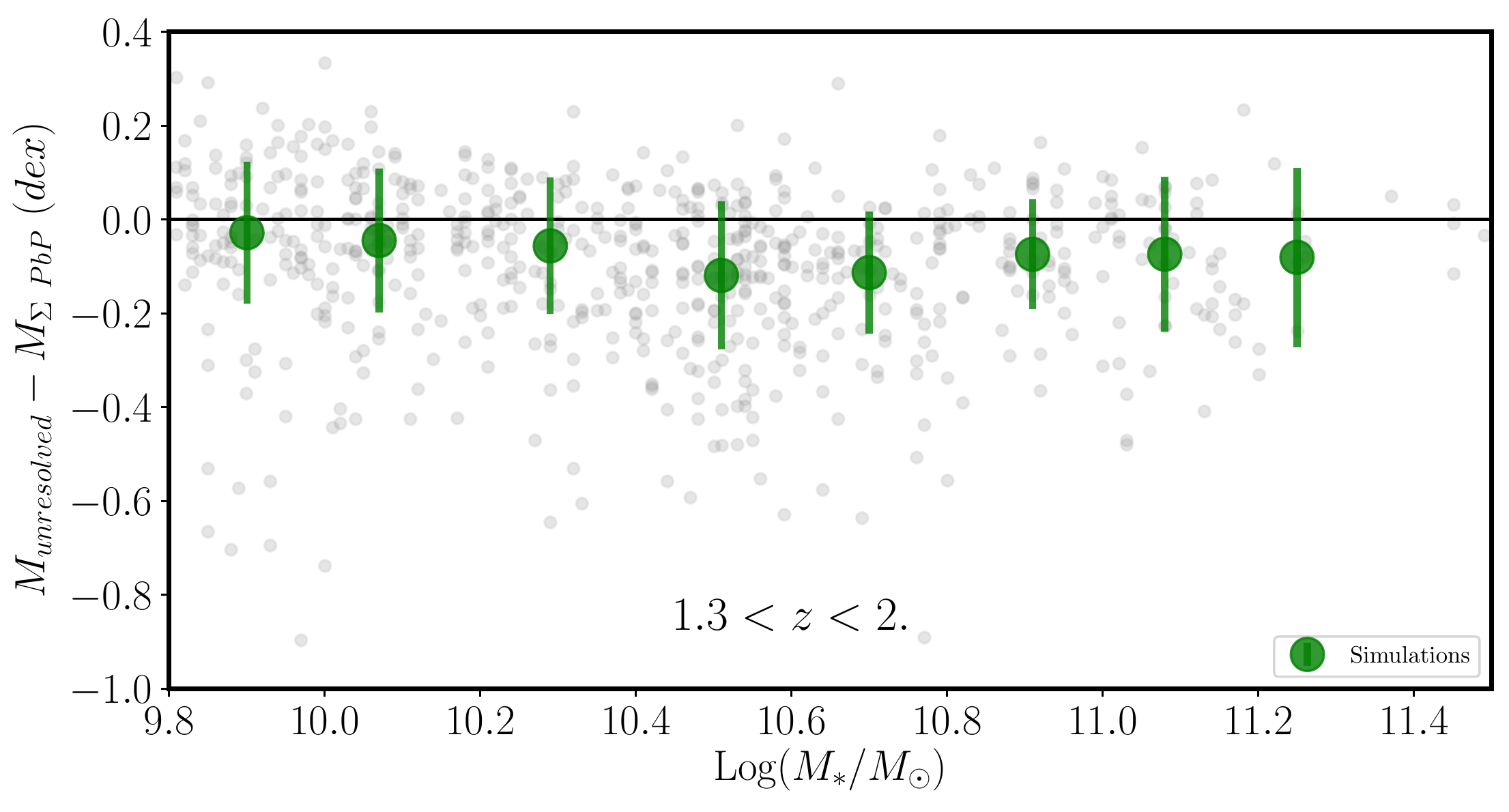}
\caption{\textit{Top panels:} Differences between unresolved total stellar mass estimation (sum of the total fluxes for the pixels detected to belong the simulated objects) and the total stellar mass from the original catalog (based on wide range of wavelength coverage) for two redshift ranges (left and right panels). This result indicates that the methodology for constructing 2D stellar mass maps is robust.  \textit{Bottom panels:}  Same as Figure \ref{fig4} for the differences between resolved and unresolved total stellar masses for two redshift ranges (left and right panels). An offset exists for sources at $z>1.3$, although this is not as significant as for real objects. }
\label{figB2}
\end{figure*}

\begin{figure*}
\centering
\includegraphics[width=\textwidth]{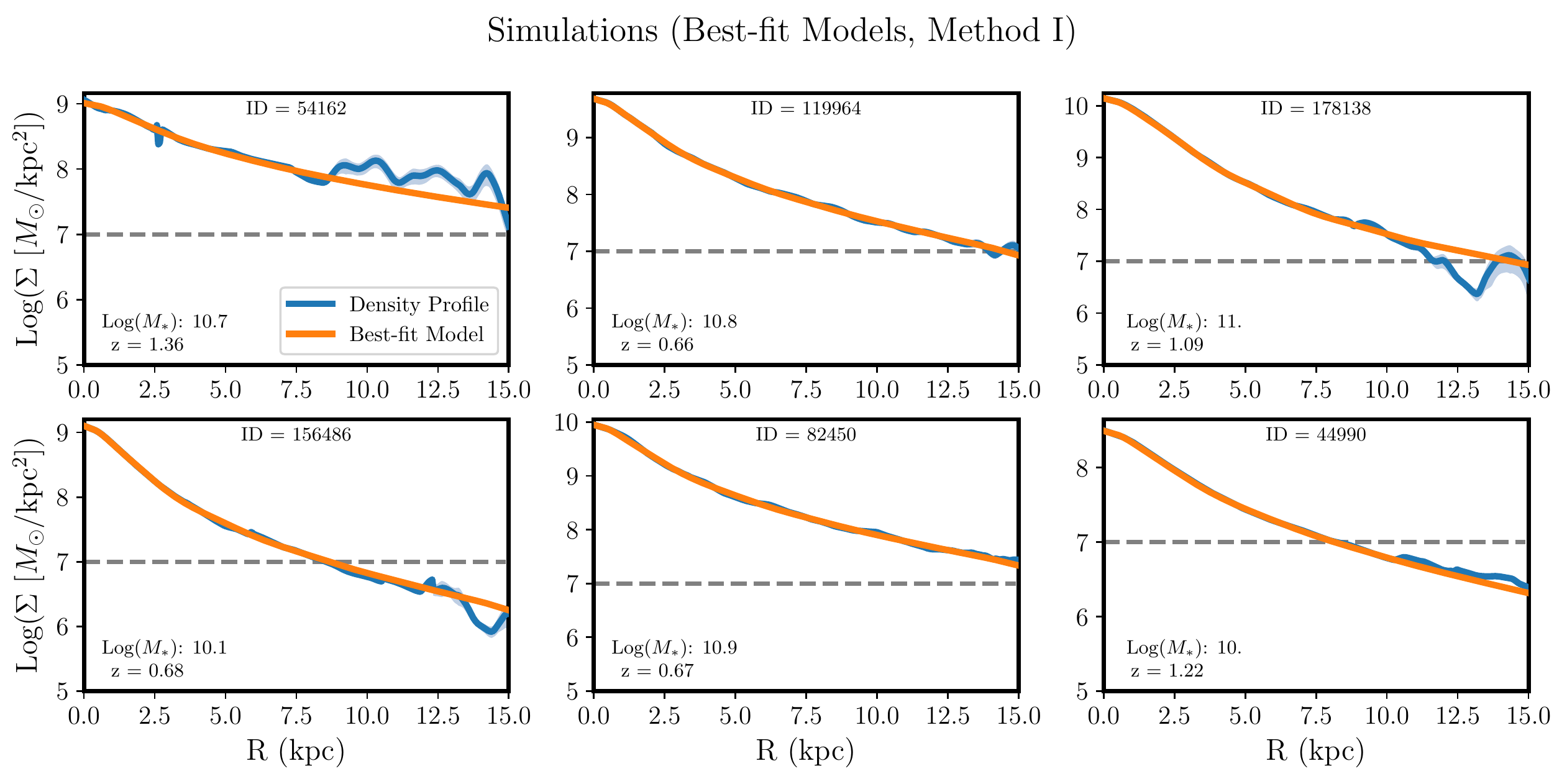}
\caption{Comparison between the 1D stellar mass surface density profiles and the associated best-fit models from the 1D method for our simulated galaxies in Figure \ref{figB1}. }
\label{figB3}
\end{figure*}

\begin{figure*}
\centering
\includegraphics[width=0.48\textwidth]{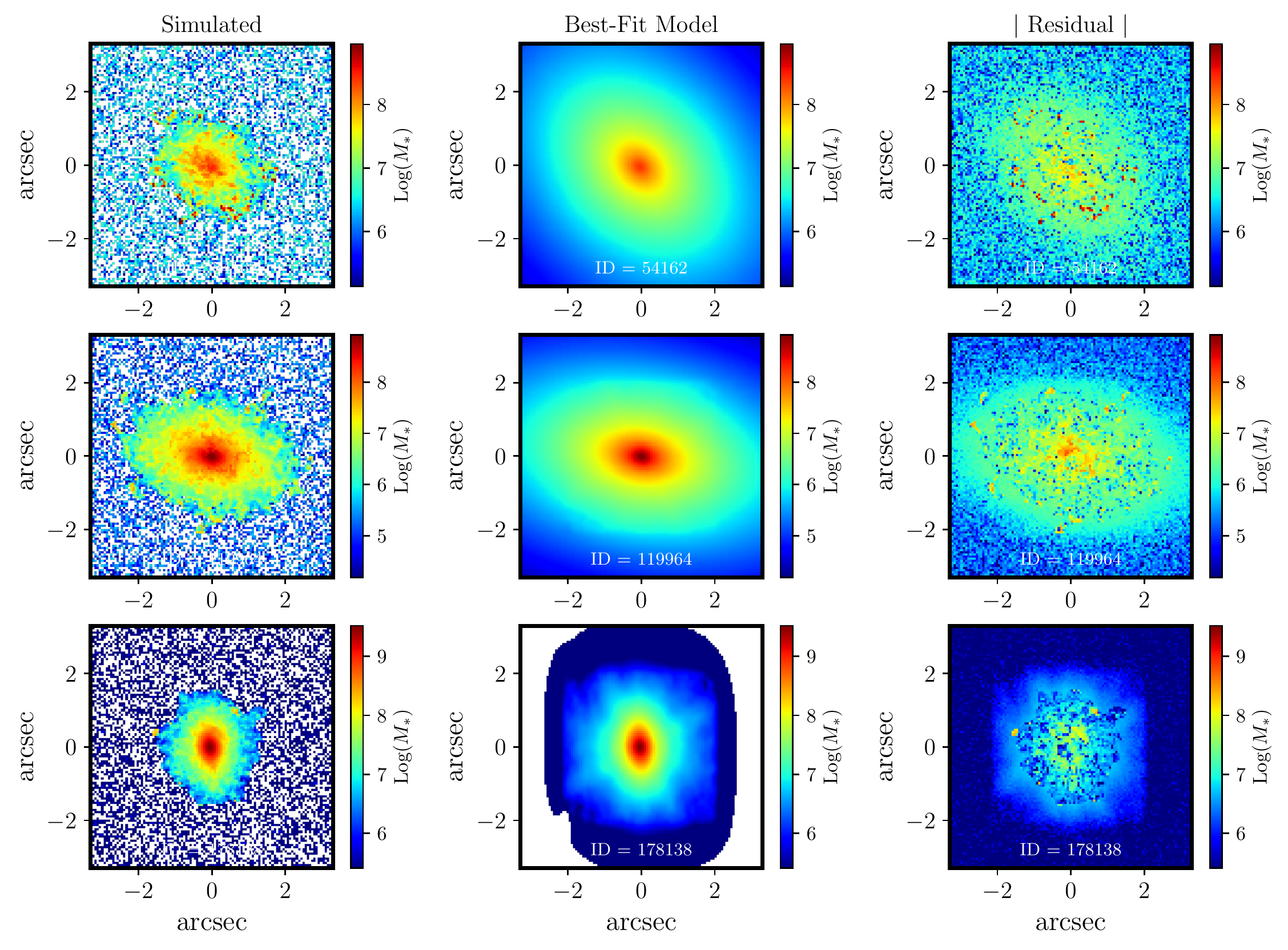}
\hspace{1. mm}
\centering
\includegraphics[width=0.48\textwidth]{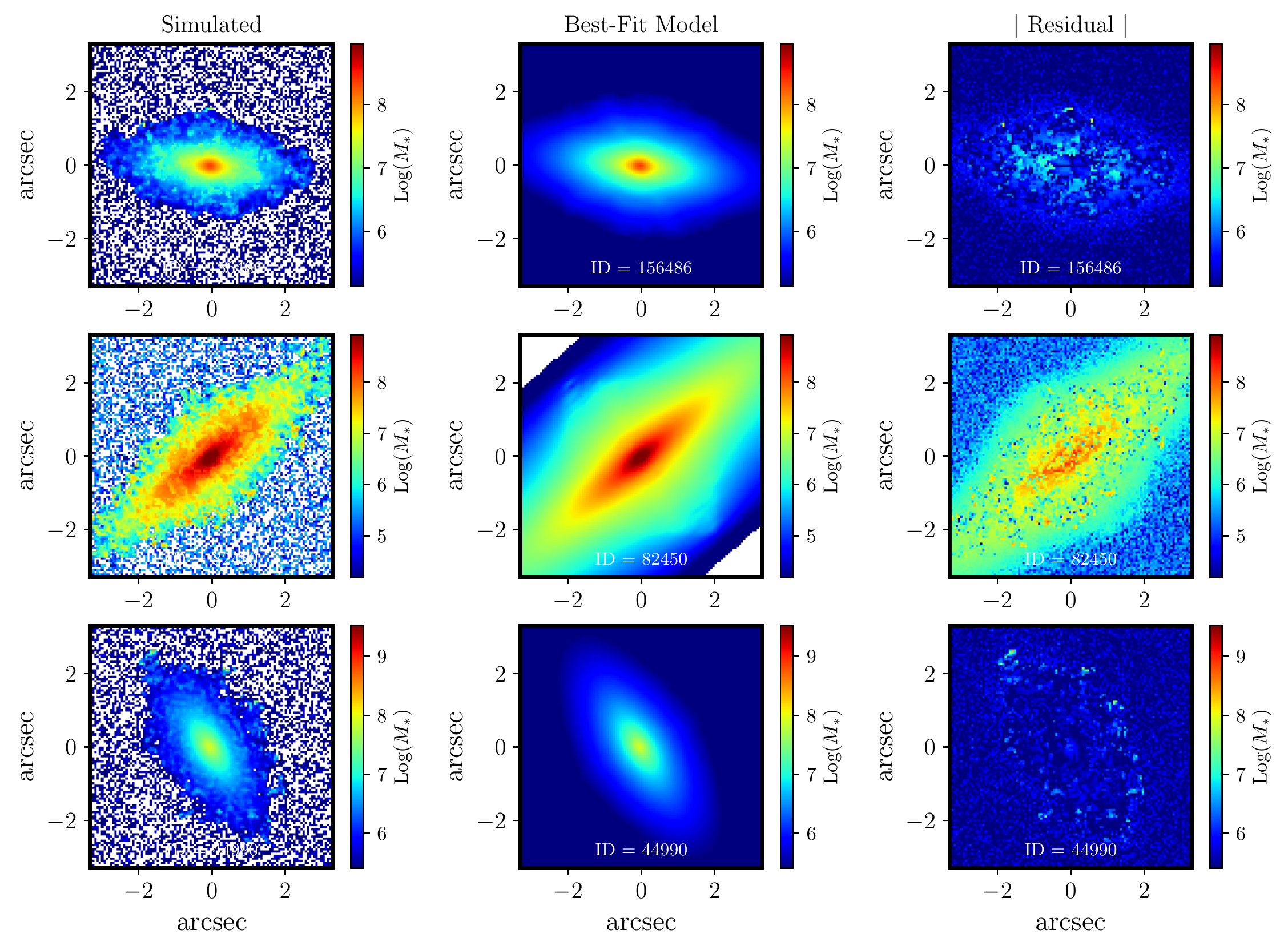}
\caption{The best-fit 2D models of the stellar mass maps for the simulated galaxies in Figure \ref{figB1} using our 2D method.}
\label{figB4}
\end{figure*}

\subsection{Models}

\begin{figure*}
\includegraphics[width=0.48\textwidth]{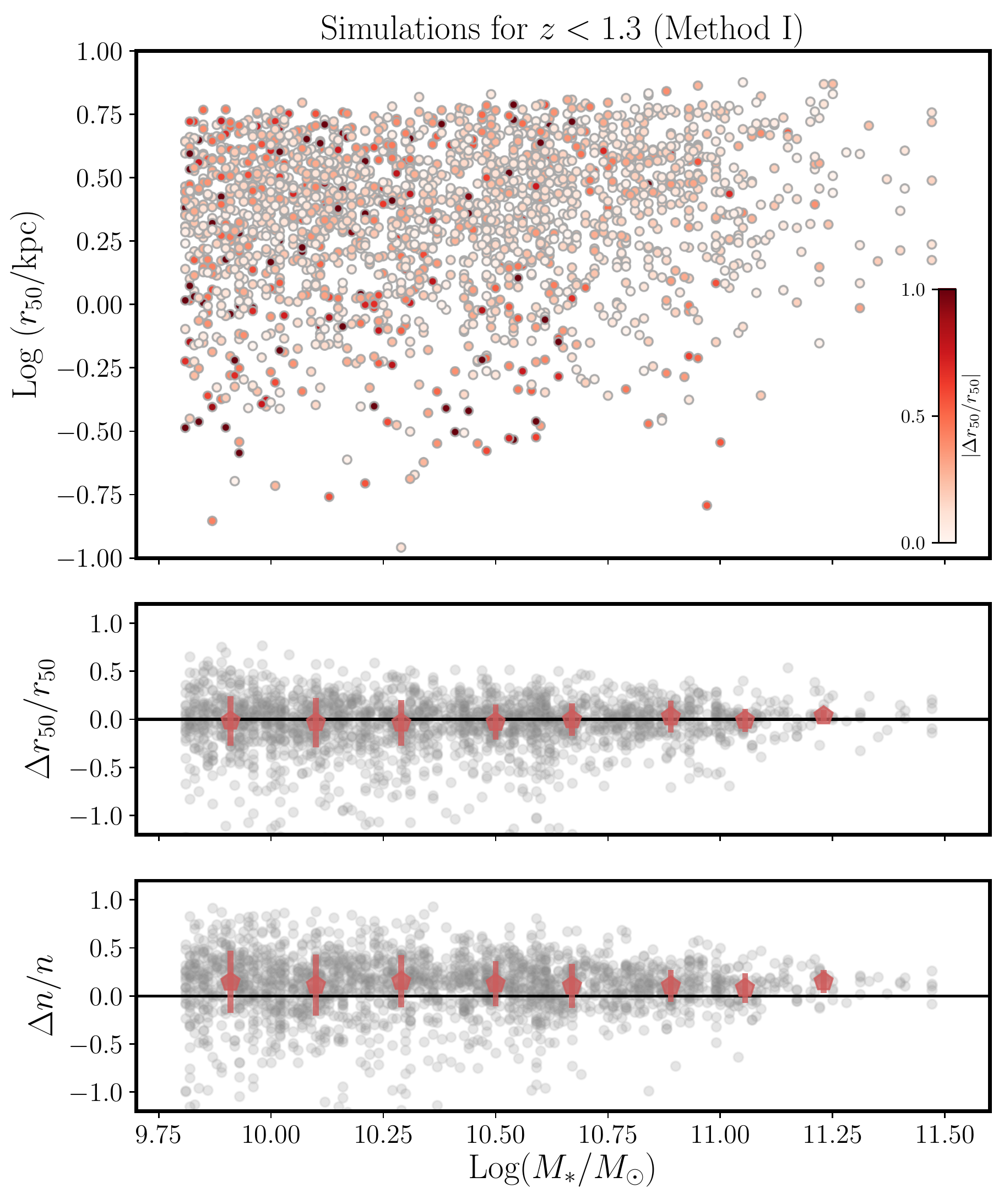}
\hspace{2. mm}
\includegraphics[width=0.48\textwidth]{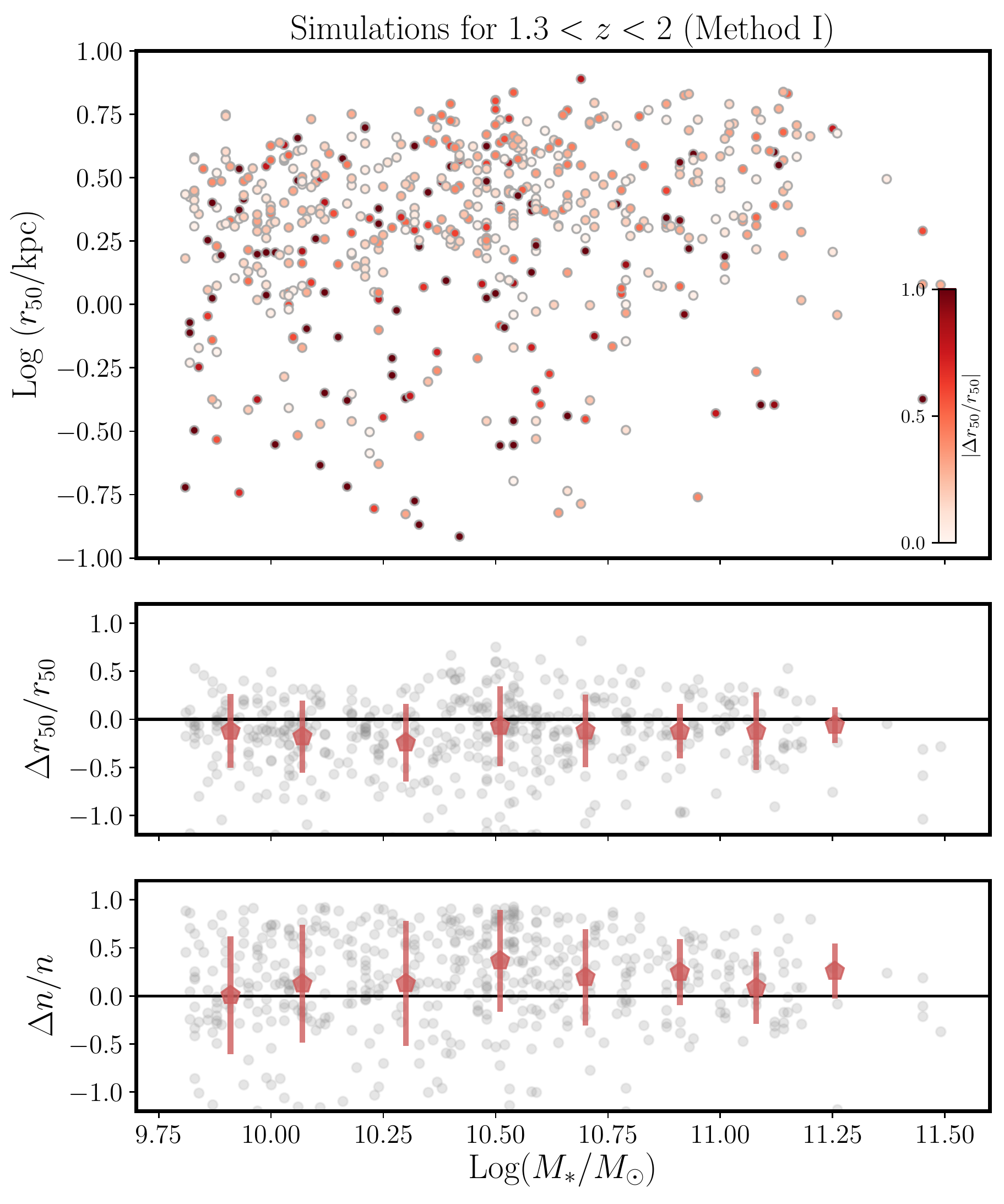}
\caption{\textit{Top Left}: The half-mass size-stellar mass plane of the simulated galaxies at $z<1.3$, color-coded based on their relative recovery ($\Delta r_{50}/r_{50}$) using our primary method (1D). The relative differences for sizes and \ser indices for objects above the $0.5$ kpc are shown in the middle and bottom left panels. There is no systematic offset for the studied stellar mass range, but the random uncertainties increase towards low mass galaxies. \textit{Right Panels:} The same as the left ones, but for the redshift range of $1.3<z<2$. At these redshifts, the uncertainties increase significantly, and the number of sources that failed to be fitted increase to $\sim 9\%$. There is a weak systematic bias in the recovery of \ser parameter with this method at all redshifts, but the uncertainties are relatively small in low redshift bin (bottom left panel).}
\label{figB5}
\end{figure*}

First, we create a catalog of mock galaxies by randomly selecting objects from the 3D-HST catalog. We then create single \ser models (using \texttt{GALFIT}) with the same structural parameters ($n$, $r_{50}$, $b/a$ and PA) in each filter. The range of properties is selected to be within a similar range of the observed galaxies on the stellar mass-size plane (see Figure \ref{figB5}). The models are all convolved to the same $H_{160}$-band PSF. Assuming the same shape in all filters will ensure that the shape of simulated galaxies at different wavelengths is the same (i.e., removing any color gradients) and, hence, there will not be any inherent bias between the light- and mass-based sizes for these mock galaxies. To ensure that the total stellar mass is conserved, the total flux of the model in each filter corresponds to flux from that randomly chosen real object. The models are then added to the empty regions of the mosaic images. We then run the SExtractor \citep{bertin1996} to create segmentation maps for each mock galaxy. We use the same procedure described in Section 3.1 to derive the stellar mass maps of these simulated galaxies. The stellar mass maps of a few simulated galaxies are shown in Figure \ref{figB1}. The simulated mass maps are similar to real objects (see Figure \ref{fig3}) with higher surface densities in the center and a relatively smooth distribution. 

For these simulations, we created about 3000 mock galaxies, in which two-thirds of them are added to the GOODS-South field (7 HST filter coverage) and the rest to the COSMOS and UDS fields (5 HST filter coverage). The redshift range of the simulated object in the GOODS-South field is between $0.5 \leqslant z \leqslant 2.$, and for the other fields are chosen to be $0.5 \leqslant z \leqslant 1.3$. 

\begin{figure*}
\includegraphics[width=0.48\textwidth]{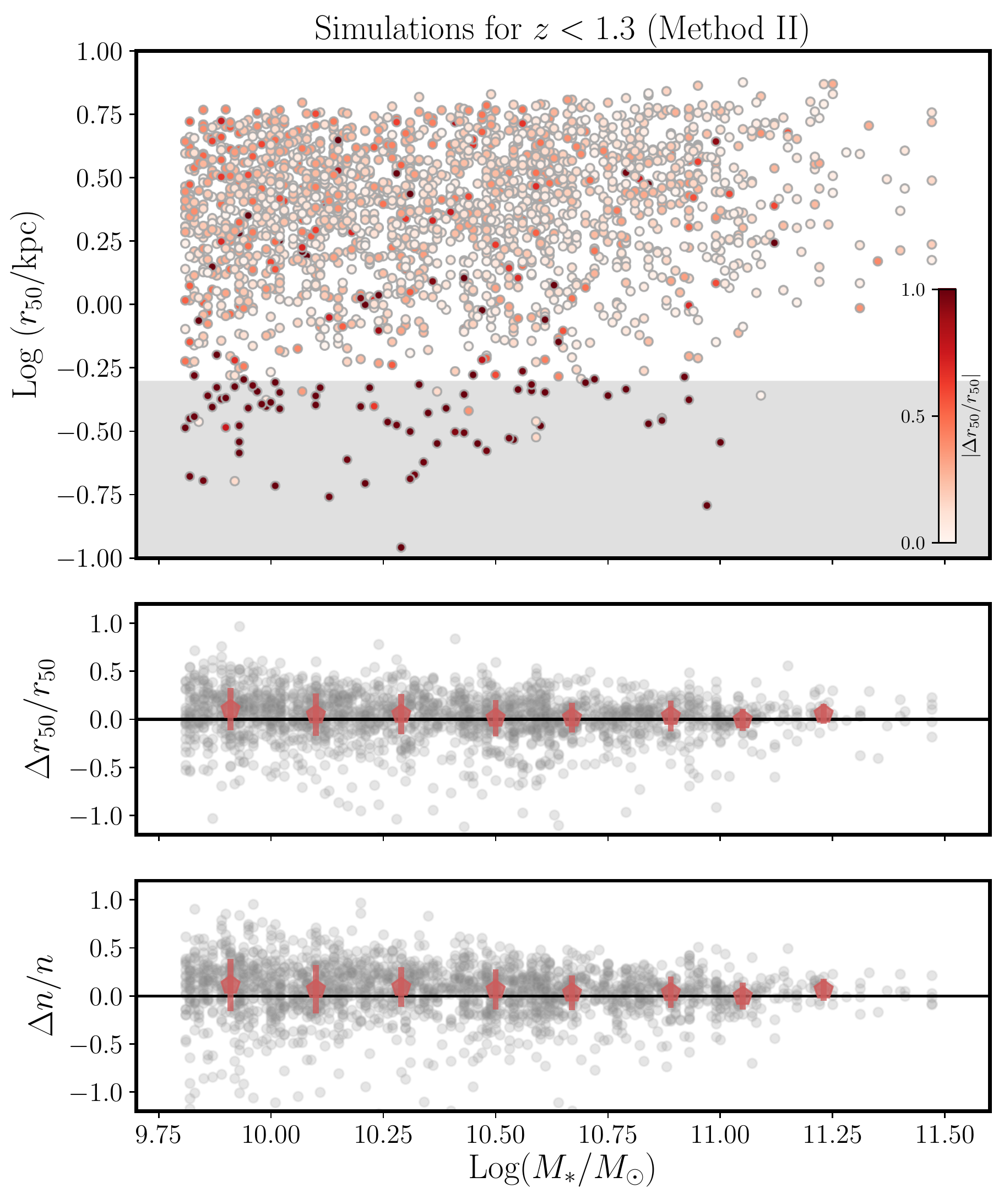}
\hspace{2. mm}
\includegraphics[width=0.48\textwidth]{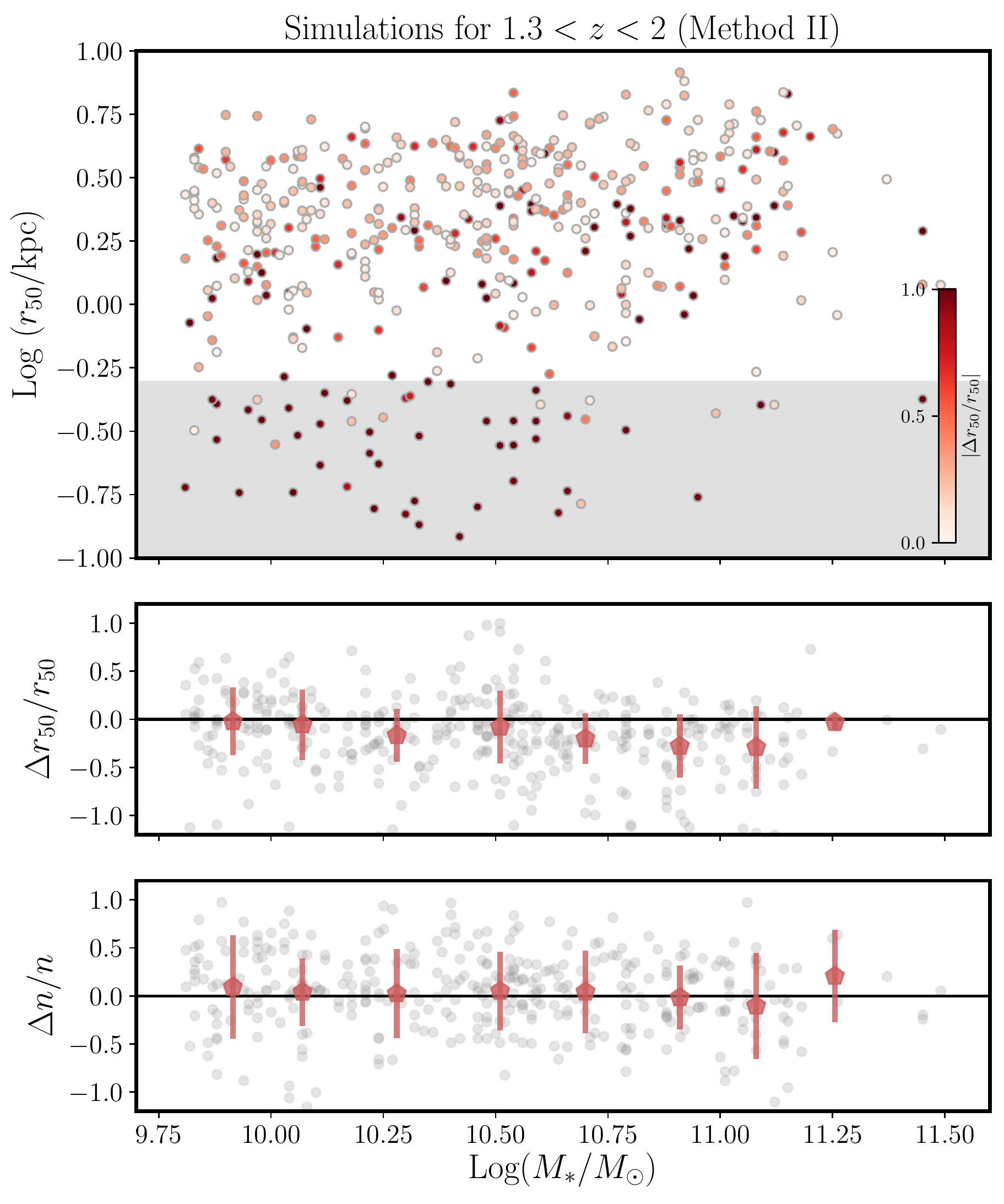}
\caption{The same as Figure \ref{figB5}, but for the second method (2D fitting approach). There is no systematic offset on sizes for the studied stellar mass range at $z<1.3$, but the random uncertainties increase towards low mass systems at this redshift (middle left panel). The uncertainties on sizes also increases for sources at $z>1.3$ (middle right panel). The results of the simulation show that the half-mass sizes of objects with $r_{50} \lesssim 0.5$ kpc are not reliable.}
\label{figB6}
\end{figure*}

\subsection{Testing Total Stellar Masses}

As a first step we check the differences between the input (original) stellar mass and the output total unresolved stellar mass for these simulated galaxies. Overall, there is a very good consistency with less than $\sim 0.04$ dex median differences at all redshifts (see top panels of Figure \ref{figB2}). This indicates that the total stellar masses (regardless of the number of filters) are recovered without any systematic offset. Therefore, the methodology for estimating the stellar mass maps are robust.

Next, we compare the total stellar mass from the unresolved and resolved (pixel-by-pixel) methods for the two redshift ranges (below and above $ z \sim 1.3$). The differences are shown in the bottom panels of Figure \ref{figB2}. For simulated objects at $z<1.3$, there is no systematic offset in the median values (green points) at all stellar masses above $\log \msun = 9.8$. Above $z \sim 1.3$, the uncertainties and biases increase, similar to the real objects (see Figure \ref{fig4}), though, the systematic differences are less ($\sim0.07$ dex) compared to the real objects ($\sim0.17$ dex). This might be due to our choice of rather simplistic models without any color gradients for these mock galaxies. Nevertheless, at redshifts beyond $z=1.3$, the resolved total stellar masses start to deviate from the resolved ones. As discusses in Section 3, the origin of this deviation has been attributed to several effects such as dusty regions, SFHs, outshining effects or signal to noise issues. The next generation of (space) telescopes and high-resolution instruments will assist to resolve this issue by looking into the rest-frame NIR of these high-$z$ galaxies.

\subsection{Testing Mass Profile Parameters}

We used two methods for measuring the sizes and \ser parameters of galaxies' stellar mass maps, which are based on 1D and 2D, as described in Section 4. We apply the same techniques to the simulated stellar mass maps. In Figure \ref{figB3}, the best-fit 1D \ser models are shown for the objects in Figure \ref{figB1}. The results for the 2D fitting method are also illustrated in Figure \ref{figB4} for the same objects. For both methods, the true shapes of the profiles have been recovered by their best-fit models. 

We examine this further by looking into the relative difference (between input and measured) of the $r_{50}$ sizes on the stellar mass-size plane. The top panels of Figure \ref{figB5} show the distributions of the simulated galaxies on the size-mass plane for the two redshift bins of $0.5 \leqslant  z \leqslant 1.3$ (left panel) and $1.3\leqslant z \leqslant 2.$ (right panel). The symbols are color-coded according to the relative difference of input and output sizes ($\Delta r_{50}/r_{50}$) from the first method (1D fitting). For the redshift range of $0.5<z<1.3$, the relative differences of the sizes are small (middle left panel). In spite of that, the median relative differences of the \ser indices at this redshift range, show small systematic offset of $12\%$, i.e., the output values of \ser indices are smaller than the inputs. This might be due to limitations of ellipse fitting in the very central regions of the stellar mass maps caused by the pixel resolution. The systematic offset of the median relative sizes in the high redshift sample ($z>1.3$) is about $12\%$ (middle-right panel of Figure \ref{figB5}), but the scatter is also considerable. Moreover, a systematic offset on the recovery of the \ser parameter exists at this redshift bin ($16\%$) and the scatter is also large.

Identical tests for the second method (2D profile fitting) are presented in Figure \ref{figB6}. From this figure, it is clear that for objects with $r_{50}$ smaller than 0.5 kpc (gray shaded regions), the size measurement is not reliable. The median relative differences of the sizes and \ser index are robust for the studied stellar mass range at $z<1.3$ and objects with mass-based sizes $> 0.5$ kpc. However, the scatter increase for the galaxies below $\log \msun \sim 10$ (see middle and lower panels of Figure \ref{figB6}). For the sources at the redshift of $1.3<z<2.$, the uncertainties of the output parameters are large, and the systematic differences in sizes can be seen for massive sources, but we do not see a significant systematic difference in the recovery of the \ser parameters. 

\begin{figure*}
\includegraphics[width=0.48\textwidth]{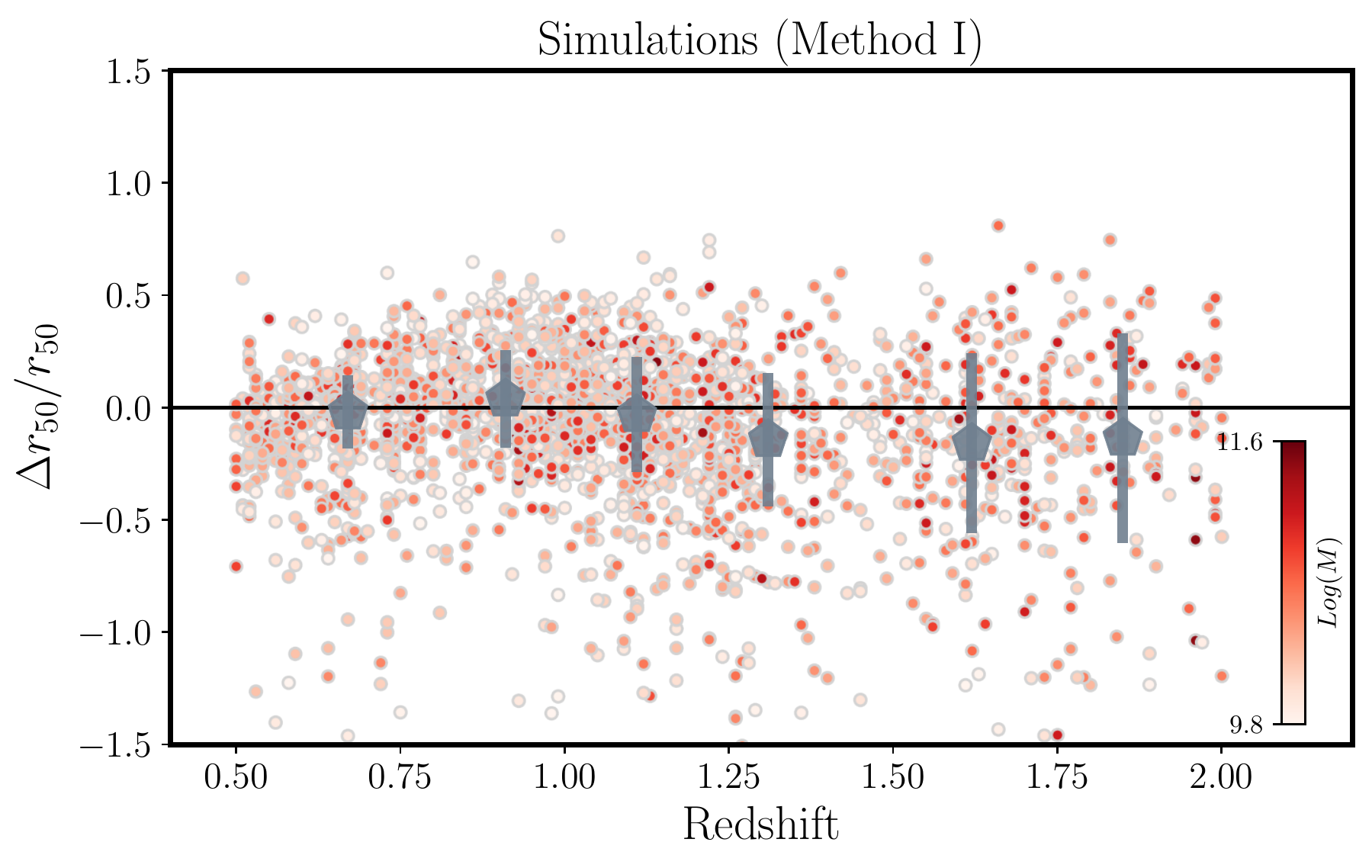}
\hspace{2. mm}
\includegraphics[width=0.48\textwidth]{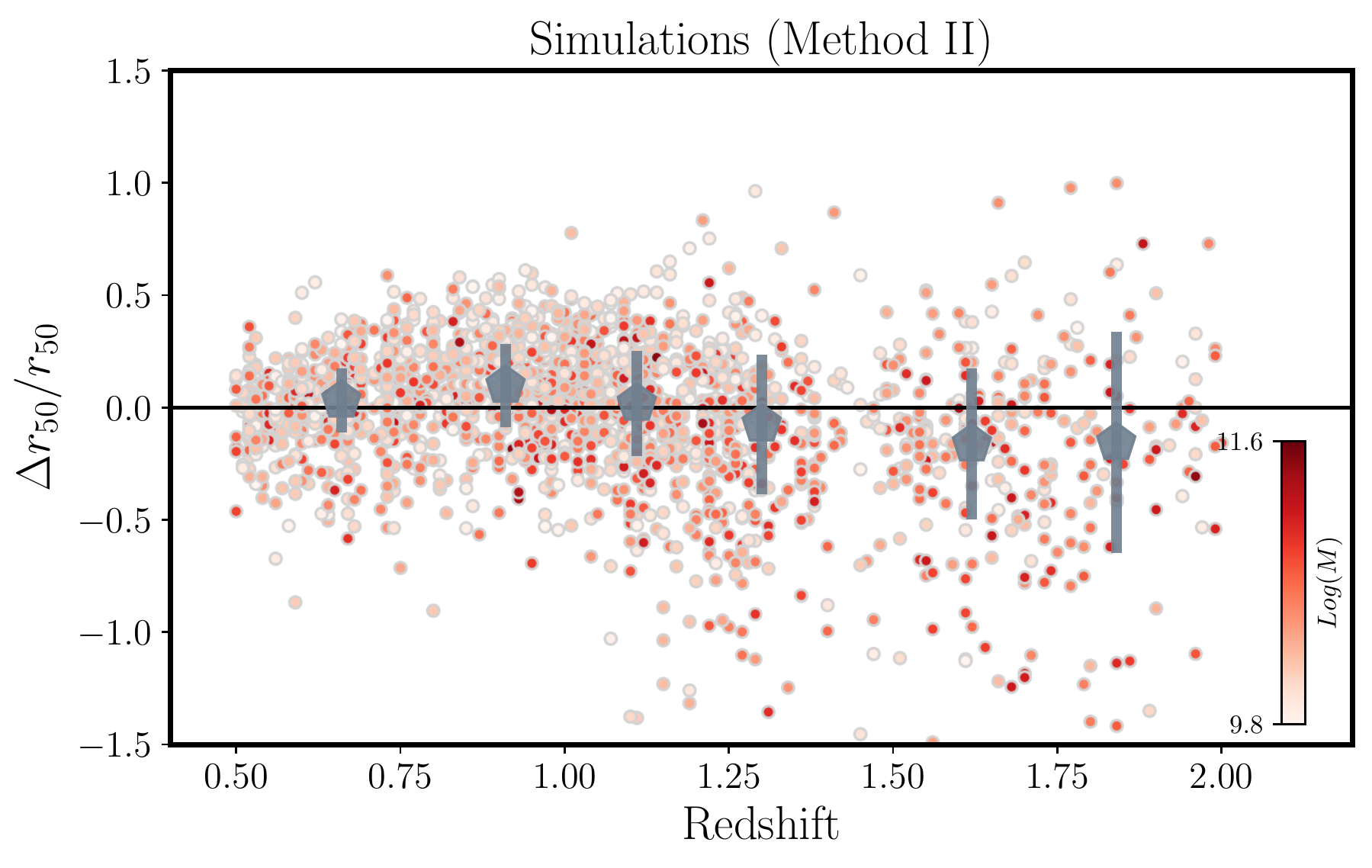}
\caption{The relative size recovery error for all simulated objects as a function of redshift. This shows that for galaxies below $z \sim 1.3$, the results are relatively robust compared to the higher redshift range. The behavior is almost the same for both methods (left and right panels), though the systematic and the rate of recovery is better for the first method and hence selected to be our fiducial method for mass-size analysis.}
\label{figB7}
\end{figure*}

The relative size ($r_{50}$) differences for the simulated galaxies as a function of redshift are shown in Figure \ref{figB7}. In this figure, the symbols are color-coded according to their stellar masses. For both methods, the scatter increase significantly beyond $z\sim1.3$. In addition, the fraction of sources that have been fitted without any failure in their fitting procedure drops significantly beyond $z>1.3$, but the systematic differences are smaller for the 1D method. For the first and second methods, 98\% and 95\% of the simulated galaxies within $0.5<z<1.3$ have been recovered without failure, respectively. However, this drops to about 92\% and 73\% for objects at $1.3<z<2$ mainly due to having less robust and noisier stellar mass maps. From this analysis, we conclude that our methods for deriving the stellar mass sizes and \ser indices are reliable for stellar masses beyond $10^{9.8} \msun$ and up to the redshift of $z\sim1.3$ (for the 2D method) and $z\sim2.0$ (for the 1D method). Beyond these redshift limits, the uncertainties of the size estimates increase, although the systematics are small. Moreover, as the first method have better coverage over the full size-mass plane and redshift range. We therefore treat this as our fiducial size measurement method for the main analysis of this work. The results based on the second method are presented in Appendix \ref{sec:appendixC}.

\section{Results Based on Sizes from Method ($M_{II})$} \label{sec:appendixC}

\begin{figure*}
\centering
\includegraphics[width=0.32\textwidth]{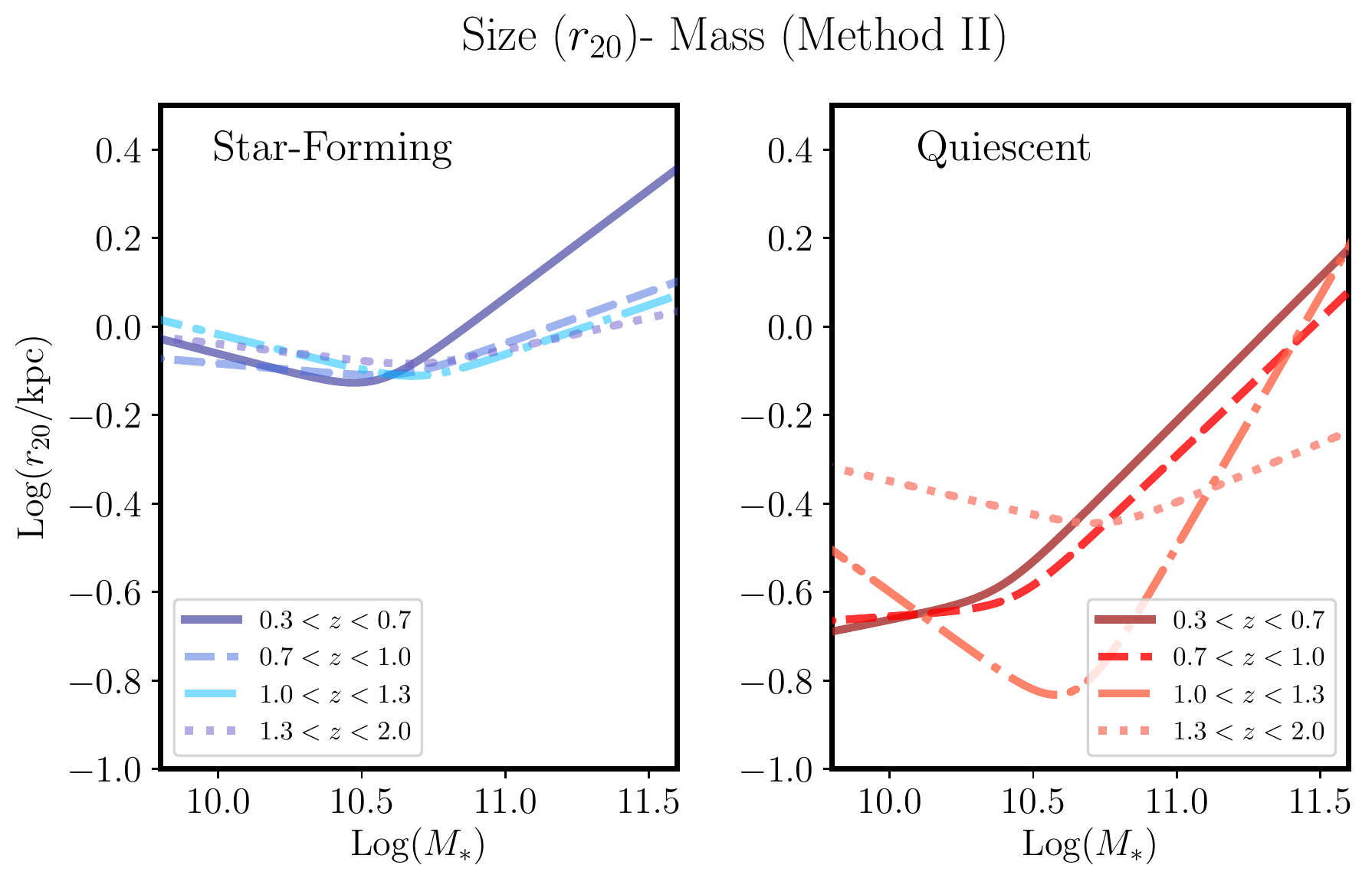}
\hspace{.2 mm}
\centering
\includegraphics[width=0.32\textwidth]{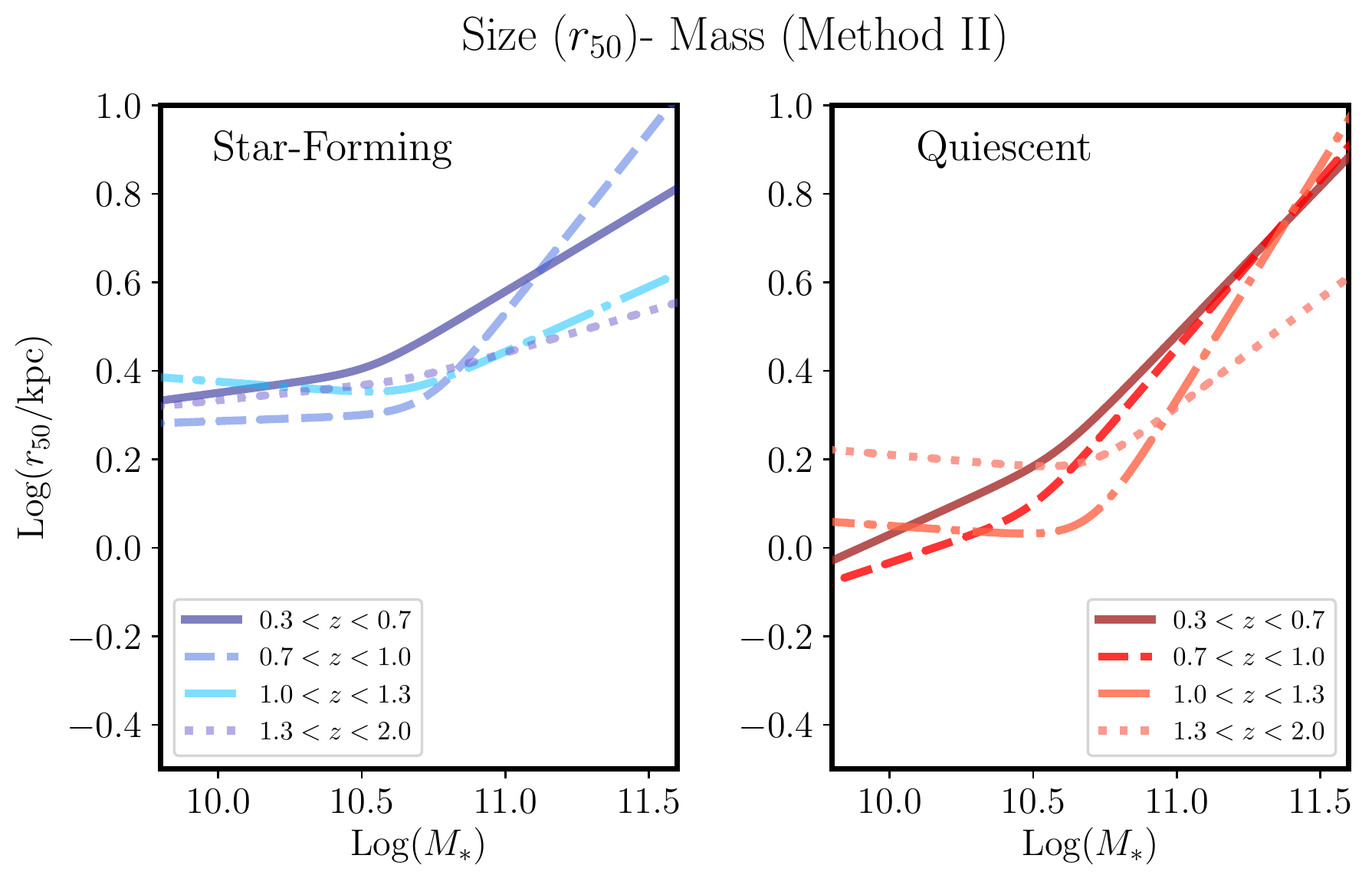}
\hspace{.2 mm}
\centering
\includegraphics[width=0.32\textwidth]{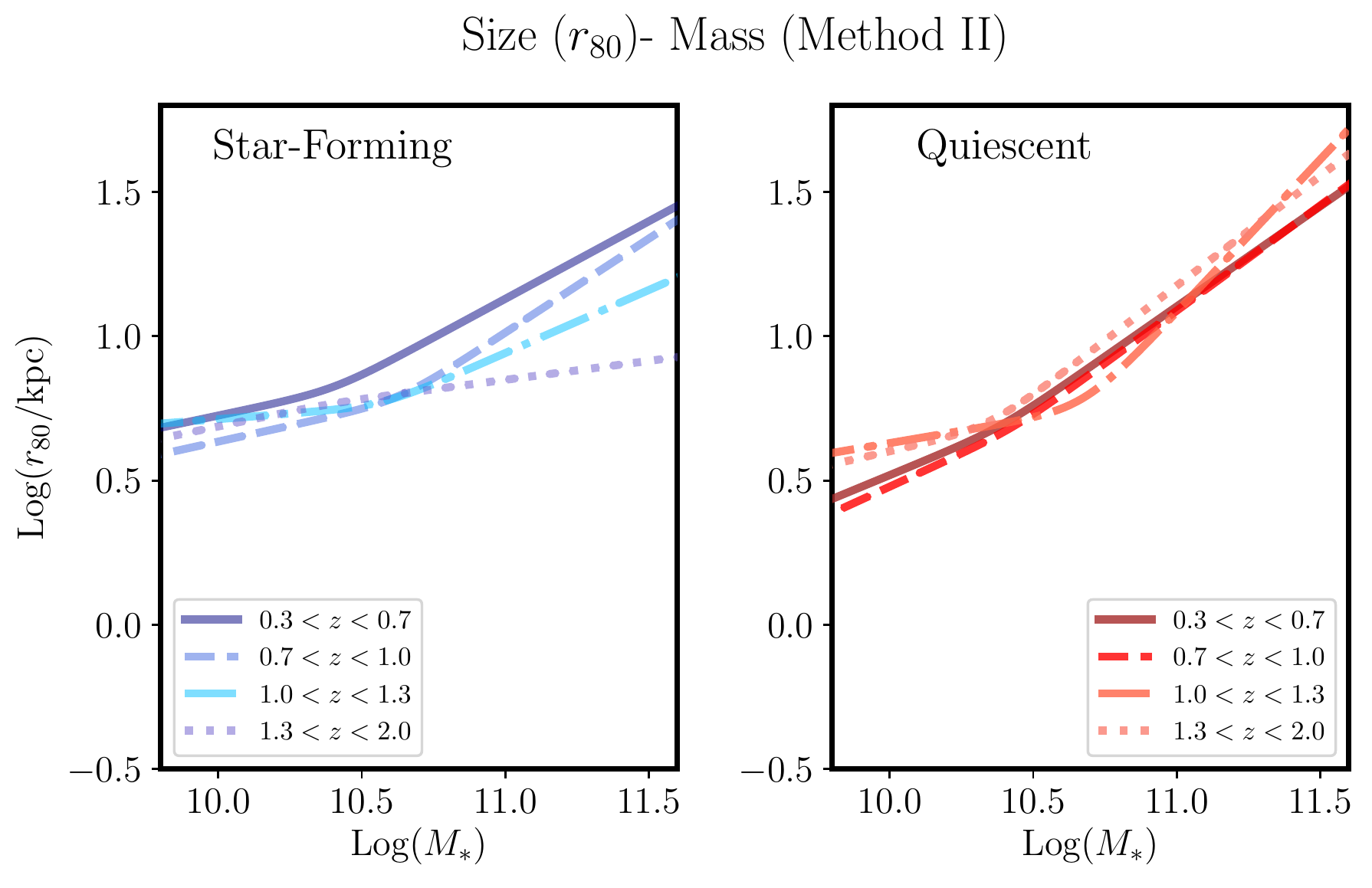}
\caption{Size-mass relation of star-forming and quiescent galaxies at different redshift similar to Figure \ref{fig9}, but based on sizes obtained from the second method. In general, there is a good agreement between the results of the two methods.}
\label{figC1}
\end{figure*}

\begin{figure*}
\centering
\includegraphics[width=0.32\textwidth]{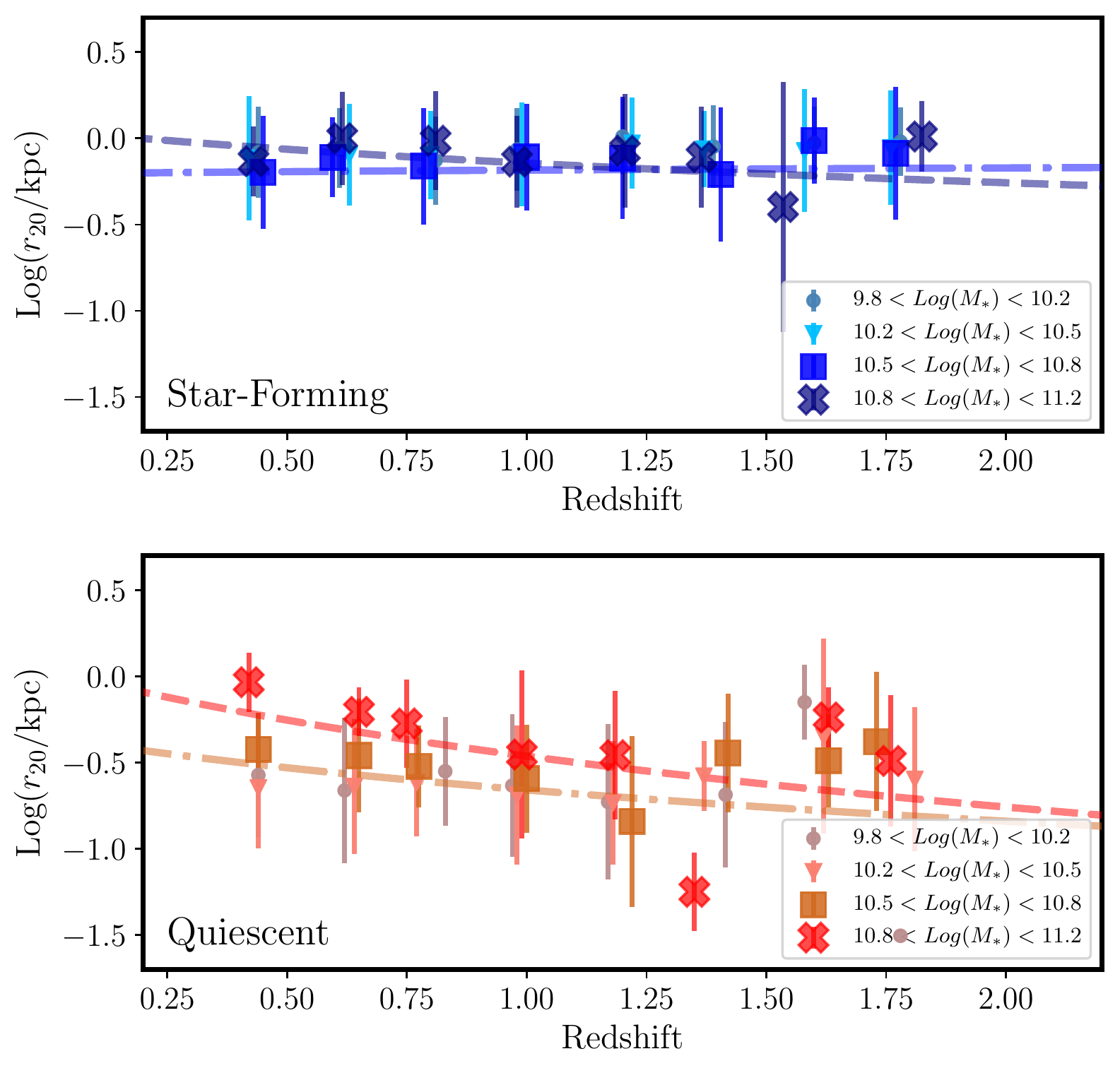}
\hspace{.2 mm}
\centering
\includegraphics[width=0.32\textwidth]{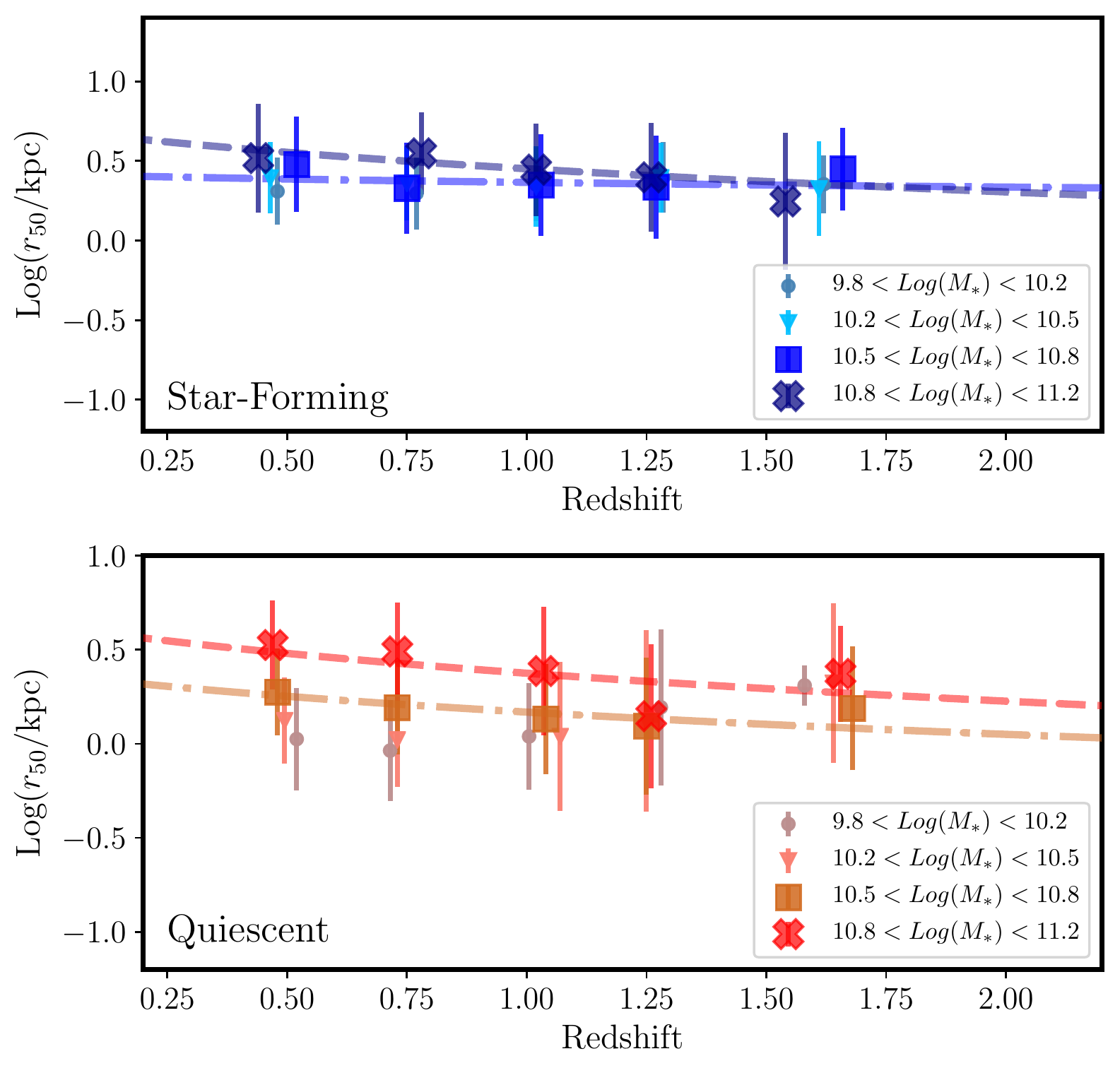}
\hspace{.2 mm}
\centering
\includegraphics[width=0.32\textwidth]{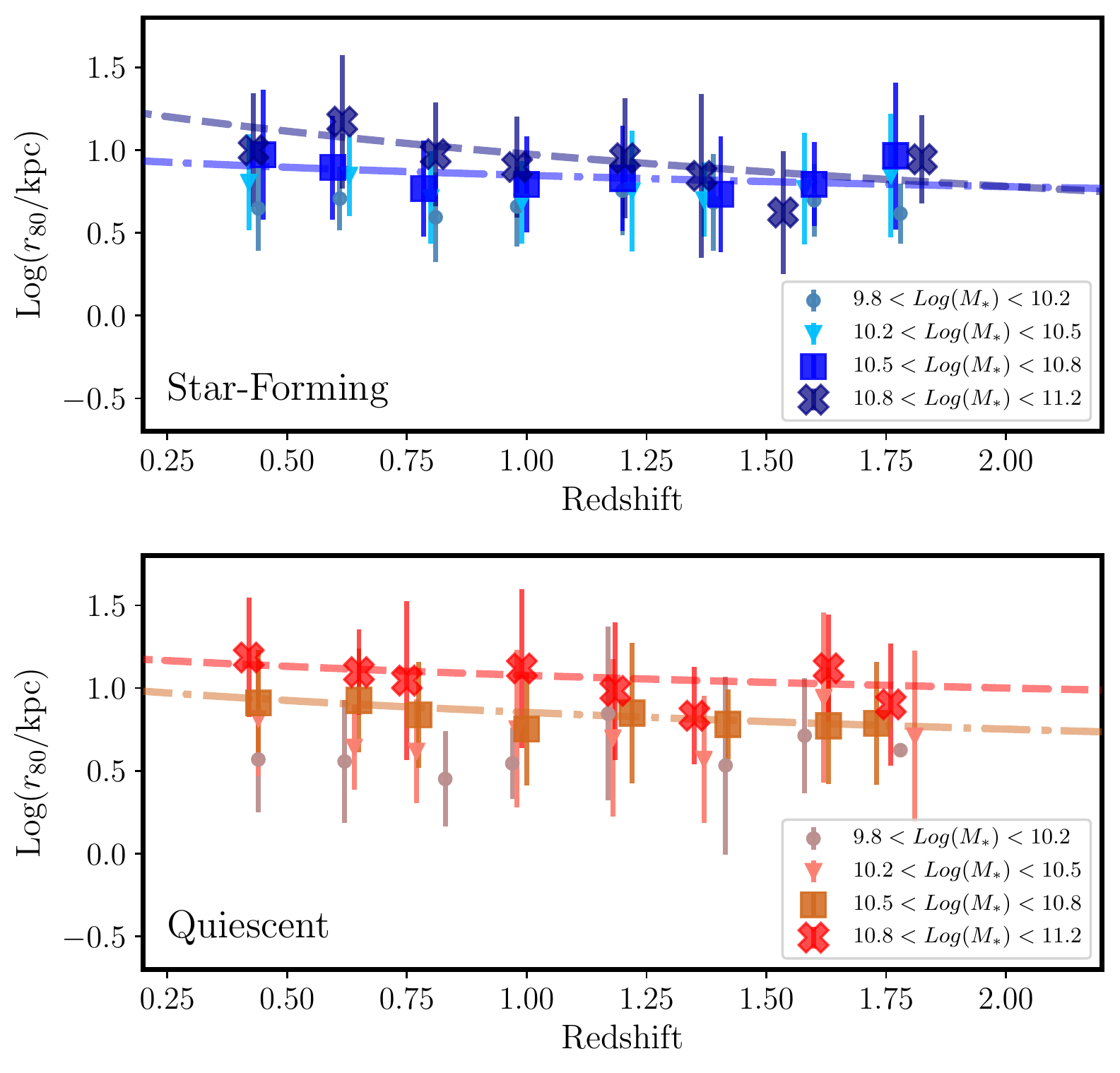}
\caption{Size evolution of galaxies with redshift for different size definitions and galaxy types, similar to Figure \ref{fig13}. The slopes are reported in Tables 3, 4, and 5.}
\label{figC2}
\end{figure*}

We used our second method (2D fitting approach) for exploring the size-mass relations, similar to the first method. The results are shown in Figure \ref{figC1}. There is a general agreement between size-mass relations comparing this with Figure \ref{fig9} from the 1D method. Star-forming galaxies have shallower slopes compared to quiescent galaxies in all redshift bins, regardless of size definition. The strong size evolution at fixed mass takes place for those above the pivot stellar mass. The evolution of sizes is also illustrated in Figure \ref{figC2}. We note that the results based on this second method ($M_{II}$) are noisy at $z>1.3$, as also shown in the simulations above.

\begin{deluxetable*}{cll}
\tablecolumns{3}
\tablewidth{\textwidth}
\tabletypesize \footnotesize
\tablecaption{Structural Properties for $ 5557$ Galaxies on CANDELS Fields}
\tablehead{
\colhead{Column Name} &
\colhead{Descriptions} &
\colhead{Comments}
}
\startdata
 ID & Identification Number for each source & from the v4.1 3D-HST catalogs \\
 Field & Name of the CANDELS Field \\
 RA & Right Ascension  (degree) & from the v4.1 3D-HST catalogs \\
 Dec & Declination   (degree) & from the v4.1 3D-HST catalogs\\
 $z$ & Redshift & from the v4.1 3D-HST catalogs \\
 LMASS & Total stellar mass of sources in $\log (\mstar/\msun)$ & from the v4.1 3D-HST catalogs\\ 
 $n_{MI}$ & best-fit \ser parameter and associated errors from the first method  & See Appendix \ref{sec:appendixB} for corrections. \\
$r50_{MI}$  & best-fit half-mass ($r50$) parameter and associated errors from the first  method \\
$r20_{MI}$ & best-fit $r20$ parameter and associated errors from the first method \\
$r80_{MI}$ & best-fit $r80$ parameter and associated errors from the first method \\
$flag_{MI}$  & Quality flag for the first method & use $flag_{MI} = 0$ for selecting reliable ones \\
$n_{MII}$ & best-fit \ser parameter and associated errors from the second method \\
$q$        & best-fit $q$ parameter and associated errors from the second method \\
$r50_{MII}$  & best-fit half-mass ($r50$) parameter and associated errors from the second method \\
$r20_{MII}$ & best-fit $r20$ parameter and associated errors from the second method \\
$r80_{MII}$ & best-fit $r80$ parameter and associated errors from the second method \\
$flag_{MII}$  & Quality flag for the second method & use $flag_{MII} = 0$ for selecting reliable ones \\
 \enddata
\tablecomments{The \ser parameters from the first approach ($n_{MI}$) should be corrected based on the results from the simulations (see Appendix \ref{sec:appendixB}). The full table is available online.}
\label{tableA1}
\end{deluxetable*}

\bibliographystyle{apj}


\end{document}